\documentclass[a4paper,11pt]{article}	
\usepackage{jheppub} 
\usepackage{makeidx}

\usepackage[table]{xcolor}                  
\usepackage{colortbl}
\usepackage{booktabs}                       
\usepackage{multirow,bigdelim}              
\usepackage{longtable}                      
\usepackage{arydshln}                       

\usepackage{units}                          
\usepackage{amsmath,amssymb,mathtools}      
\usepackage{slashed}                        
\usepackage{amsthm}
\usepackage{bbm,dsfont}                     

\usepackage{bm}                             
\usepackage{upgreek}                        
\usepackage[vcentermath]{youngtab}          
\usepackage{empheq}                         
\usepackage{suffix}	                        
\usepackage{lscape}	                        

\usepackage{graphicx}                       
\usepackage{subfig}	                        
\usepackage{float}
\usepackage[export]{adjustbox}              

\usepackage{url}                            
\usepackage[numbers,sort&compress]{natbib}  
\usepackage[colorlinks=true,urlcolor=blue,anchorcolor=blue,citecolor=blue,filecolor=blue,linkcolor=blue,menucolor=blue,pagecolor=blue,linktocpage=true,pdfproducer=medialab,pdfa=true]{hyperref}	                    

\numberwithin{equation}{section}        
\numberwithin{table}{section}
\numberwithin{figure}{section}

\def\md{\mathbf}
\def\mf{\mathfrak}

\def\IC{\mathbb{C}}
\def\IE{\mathbb{E}}
\def\IF{\mathbb{F}}

\def\IP{\mathbb{P}}

\def\IR{\mathbb{R}}
\def\IZ{\mathbb{Z}}

\def\cN{\mathcal{N}}
\def\cO{\mathcal{O}}

\def\fg{\mathfrak{g}}

\def\vev#1{\left\langle #1 \right\rangle}

\def\({\left(}
\def\){\right)}
\def\[{\left[}
\def\]{\right]}

\newcommand{\re}{{\rm e}}
\newcommand{\ri}{{\mathsf{i}}}

\newcommand{\nn}{\nonumber \\}

\makeindex

\newcommand{\nD}{g}
\newcommand{\nC}{b}
\newcommand{\BPS}{N^{\bf d}_{j_L,j_R}}
\newcommand{\hF}{\widehat{F}}
\newcommand{\hZ}{\widehat{Z}}
\newcommand{\fq}{\mathfrak q}
\newcommand{\PE}[1]{\text{PE}\[#1\]}

\newcommand{\aroot}{\vev{\boldsymbol \alpha,\md a}}
\newcommand{\arooti}[1]{\vev{\boldsymbol\alpha_{#1},\md a}}
\newcommand{\avec}{\md a}
\newcommand{\alvec}{\boldsymbol\alpha}
\newcommand{\hG}{h^\vee_G}
\newcommand{\ucol}[1]{\textcolor{blue}{#1}}
\newcommand{\vcol}[1]{\textcolor{green}{#1}}
\newcommand{\limprime}{\lim\nolimits'}

\newcommand{\Zloop}{Z^\text{1-loop}}

\def\eq{{\epsilon_1}}
\def\et{{\epsilon_2}}

\newcommand{\br}{{\md r}}

\newcommand{\Qtau}{Q_{\tau}}

\newcommand{\be}{\begin{equation}}
\newcommand{\ee}{\end{equation}}
\newcommand{\ba}{\begin{aligned}}
\newcommand{\ea}{\end{aligned}}

\title{\boldmath Blowup Equations for 6d SCFTs. I}
\author{Jie Gu${}^{a,b}$, Babak Haghighat${}^c$, Kaiwen Sun${^d}$, Xin
  Wang${}^{e,f}$}
\affiliation{
  ${}^a$Laboratoire de Physique Th\'eorique, \'Ecole Normale Sup\'erieure\\
  CNRS, PSL Research University, Sorbonne Universit\'es, UPMC, 75005,
  Paris, France\\
  \\
  ${}^b$ D\'epartement de Physique Th\'eorique et Section de
  Math\'ematiques\\
  Universit\'e de Gen\`eve, Gen\`eve, CH-1211 Switzerland\\
  \\
  ${}^c$ Yau Mathematical Sciences Center, Tsinghua University, Beijing, 100084, China\\
  \\
  ${}^d$ Scuola Internazionale Superiore di Studi Avanzati (SISSA), via Bonomea 265, 34136, Trieste, Italy\\
  \\
 ${}^e${Bethe Center for Theoretical Physics, Physikalisches Institut, Universit\"{a}t Bonn, 53115 Bonn, Germany}\\
\\
 ${}^f${Max Planck Institute for Mathematics,
Vivatsgasse 7, D-53111 Bonn, Germany}
\\}

\emailAdd{jie.gu@unige.ch,\ babak@math.tsinghua.edu.cn,\ ksun@sissa.it,\ wxin@mpim-bonn.mpg.de}

\abstract{We propose novel functional equations for the BPS partition
  functions of 6d $(1,0)$ SCFTs, which can be regarded as an elliptic
  version of G\"{o}ttsche-Nakajima-Yoshioka's K-theoretic blowup
  equations. From the viewpoint of geometric engineering, these are
  the generalized blowup equations for refined topological strings on
  certain local elliptic Calabi-Yau threefolds. We derive recursion
  formulas for elliptic genera of self-dual strings on the tensor
  branch from these functional equations and in this way obtain a
  universal approach for determining refined BPS invariants. As
  examples, we study in detail the minimal 6d SCFTs with $SU(3)$ and
  $SO(8)$ gauge symmetry. In companion papers, we will study the
  elliptic blowup equations for all other non-Higgsable clusters.}

\begin{document}
\maketitle
\flushbottom

\section{Introduction}
\label{sc:introduction}

Quantum field theories with the highest amount of symmetries, namely
supersymmetry as well as conformal symmetry, in the highest possible
dimension are 6d superconformal field theories. The classification of
such theories subdivides between two main classes of theories, namely
the $\mathcal{N}=(2,0)$ theories and the $\mathcal{N}=(1,0)$
theories. The former have been studied intensively for a long time now
and there are powerful techniques available for constructing their
partition functions on various manifolds. The subject of
$\mathcal{N}=(1,0)$ theories, which preserve half of the supercharges
of the $(2,0)$ theories, has recently enjoyed a resurgence due to a
proposed classification of such theories in terms of F-theory
compactifications on non-compact elliptic Calabi-Yau three-folds
\cite{Heckman:2013pva,Heckman:2015bfa}.

In this classification, the geometry of the base $B$ of the Calabi-Yau
manifold directly translates into the tensor multiplet sector of the
6d SCFTs where the number of tensor multiplets is given by the
dimension of $H^{1,1}(B,\mathbb{Z})$ and the intersection form on $B$
gives the couplings of these tensor multiplets to each other. Note
that an action is not available as field strengths of tensor
multiplets are constrained to be self-dual. Nevertheless, it is useful
to write down a ``formal" action on the tensor branch from which many
properties of the theory and its compactifications can be deduced, see
for example \cite{Blum:1997mm,Ohmori:2015pia}. Furthermore, the base
$B$ of the Calabi-Yau is non-compact and all curve classes inside it
are required to be simultaneously shrinkable to zero volume in order
to restore conformal invariance of the resulting 6d theory at its
tension-less limit. This gives strong constraints on the geometry and
in particular forces all curves to be $\mathbb{P}^1$'s which have
negative self-intersection number. Moreover, for self-intersection
numbers $-n$ lower than $-2$ the elliptic fiber above the
corresponding curve $\Sigma$ becomes singular with a singularity type
determined by Kodaira's classification of elliptic fibers
\cite{Morrison:1996pp,Morrison:2012np}. The singularity becomes worse
when $n$ increases such that beyond $n=12$ it becomes too bad for a
smooth description of the Calabi-Yau threefold. The physical
interpretation of these singularities is the emergence of a bulk gauge
group (whose Lie Algebra $\mf g_{\Sigma}$ is determined by the
intersection form of the resolved singularity) in the 6d SCFT on its
tensor branch. If we have two curves $\Sigma_1$ and $\Sigma_2$ with
non-trivial intersection number and gauge groups, then the
corresponding 6d theory will also have bi-fundamental matter in
suitable representations of the arising gauge groups. In the current
paper we want to focus on the cases where the base $B$ contains only
one curve with self-intersection $-n$, i.e. it is a certain
decompactification limit of a Hirzebruch surface $\mathbb{F}_n$. Then
the possible gauge groups which arise as a function of $n$ are as
follows\footnote{The cases of $n=9,10,11$ involve points of enhanced
  singularities on the base curve which must be resolved giving rise
  to additional curves in the base.}:
\begin{center}
  \begin{tabular}{|c|c|c|c|c|c|c|c|}
    \hline
    \textrm{$n$} & 3 & 4 & 5 & 6 & 7 & 8 & 12 \\
    \hline
    $G_{\Sigma}$ & $SU(3)$ & $SO(8)$ & $F_4$ & $E_6$ & $E_7$ & $E_7$ & $E_8$ \\
    \hline
  \end{tabular}
\end{center}
\vspace{-0.4ex} These theories are known as the Non-Higgsable Clusters
\cite{Morrison:2012np}.\footnote{The Non-Higgsable
  Clusters also include three examples of intersecting chains of two
  or three curves, which we do not cover here.} Note that the $E_7$
Lie algebra appears twice, namely for the self-intersections $-7$ and
$-8$. The difference is that in the $-7$ case there is also
fundamental matter which is non-Higgsable. These theories are the
subject of the current paper where we focus on the cases $n=3$ and
$n=4$ and aim at testing a novel method for computing
BPS partition functions of the corresponding minimal 6d SCFTs. Let us
describe our procedure for computing such partition functions in the
following.

In order to be able to compute a partition function for our 6d
theories we first need to make a choice for a background geometry. As
it turns out, an appropriate choice for 6d SCFTs is the Omega
background $\IR^4 \times_{\epsilon_1,\epsilon_2}T^2$
\cite{Haghighat:2013gba}. This background not only regularizes the
infinities arising from non-compactness of $\IR^4$ but also serves as
a building block for computing partition functions for other
backgrounds like superconformal indices and $T^2 \times S^4$ partition
functions \cite{Lockhart:2012vp,Kim:2012qf}. As was observed in
\cite{Haghighat:2013gba}, instantons on this background arise from
self-dual strings wrapping the $T^2$ and localized at a point on
$\IR^4$. From the F-theory point of view, such strings arise from D3
branes wrapping a curve $\Sigma$ in the base which in our case is a
$(-n)$-curve. The bulk gauge group will descend to a flavor symmetry
on the worldvolume theory of these strings which is a 2d
$\mathcal{N}=(0,4)$ supersymmetric theory with $R$-symmetry
$SU(2)_{\epsilon_+} \times SU(2)_R$ where $SU(2)_R$ is the R-symmetry
of the 6d SCFT. The partition function on the tensor branch
$Z^{\textrm{6d}}$, i.e. when the volume $t_b$ of the $(-n)$-curve in
the base is non-zero, is the generating function of the elliptic
genera $\IE_k$ of $k$ strings up to a prefactor
\cite{Haghighat:2013gba,Haghighat:2013tka,Haghighat:2014vxa}.

For 6d theories corresponding to the particular choices $n=3$ and
$n=4$ the worldvolume theory of the strings is known in terms of a
quiver gauge theory whose single gauge node is of rank $k$
\cite{Haghighat:2014vxa,Kim:2016foj}. For the other cases, references
\cite{DelZotto:2016pvm,DelZotto:2018tcj} give some descriptions for
the $k=1$ subsector of a single string but a complete description
including a computation scheme for all the $\IE_k$ is still lacking. In
this paper we want to remedy this gap by providing a novel computation
scheme for $Z^{\textrm{6d}}$ which allows us to derive expressions for
the $\IE_k$ recursively. This is done by using the so-called
\textit{blow-up equations}.

The blowup equations have their origin in the studies of Donaldson
invariants
\cite{MR1394968,Moore:1997pc,Marino:1998bm,Edelstein:2000aj}. But the
version we are most interested in is the generalized version proposed
and later proved by G\"{o}ttsche, Nakajima, and Yoshioka. Nakajima
and Yoshioka \cite{Nakajima:2003pg} first considered the 4d $\cN=2$
$SU(N)$ supersymmetric Yang-Mills theory on the Omega background. The
idea is to view $\IR^4 \cong \IC^2$ as the limit of its blow-up at the
origin \cite{Nakajima:2003pg}, denoted by
$\widehat{\mathbb{C}^2}$, when one sends the size of the exceptional
divisor $\mathbb{P}^1$ to zero. Then
$U(1)_{\epsilon_1} \times U(1)_{\epsilon_2}$ has a natural action on
$\widehat{\mathbb{C}^2}$ with two fixed points, one at the north pole
and one at the south pole of the exceptional $\mathbb{P}^1$. Computing
the partition function on the background
$\widehat{\mathbb{C}^2} \times_{\epsilon_1,\epsilon_2} T^2$ through
localization then contributes a product of two copies of the partition
function on our original background while one has
to sum now over non-trivial fluxes of the $B$-field through the
exceptional divisor. This idea can be put into functional equations for
the Nekrasov partition function \cite{Nakajima:2003pg} (see also
\cite{Nakajima:2003uh}), and they were instrumental for Nakajima and
Yoshioka to prove Nekrasov's conjecture \cite{Nekrasov:2002qd}.  Later
together with G\"{o}ttsche they generalized and then proved the blowup
equations for 5d $\cN=1$ $SU(N)$ super-Yang-Mills theories on the Omega
background $\IR^4 \times_{\epsilon_1,\epsilon_2}S^1$ with a possible
Chern-Simons term of level $m$
\cite{Nakajima:2005fg,Gottsche:2006bm,Nakajima:2009qjc}. On the other
hand, the Nekrasov partition function of such a 5d theory can be
computed by the refined topological string theory with target space
the local toric Calabi-Yau threefold $X_{N,m}$, which is the resolution of the
cone over the $Y^{N,m}$ singularity
\cite{Iqbal:2003zz,Taki:2007dh}. This is an example of the geometric
engineering \cite{Katz:1996fh}. Inspired by the consistency study of
the exact quantization program of mirror curves of local Calabi-Yau
threefolds
\cite{Grassi:2014zfa,Codesido:2015dia,Wang:2015wdy,Sun:2016obh,Grassi:2016nnt},
the G\"{o}ttsche-Nakajima-Yoshioka blowup equations were reformulated
completely in terms of the geometric data of the Calabi-Yau threefold
$X_{N,m}$ \cite{Gu:2017ccq}. The reformulation, however, was not
complete, and the complete set of equations were provided in
\cite{Huang:2017mis}.

These geometrically reformulated or generalized blowup equations prove
to be very powerful. First of all, just as in the case of the
G\"{o}ttsche-Nakajima-Yoshioka blowup equations \cite{Keller:2012da},
they can be used to compute the Nekrasov instanton partition functions
\cite{Huang:2017mis}. Second, the generalized blowup equations open up
possibilities for various directions of generalization. As we will see
in the next subsection, the form of the generalized blowup equations
is simple and universal, and it does not put any constraints on the
target space of the topological string except that it has to be
non-compact to allow for $U(1)$ isometry crucial for the preservation
of supersymmetry in the presence of the Omega background. This
naturally poses the question of the validity of the generalized blowup
equations beyond 5d $SU(N)$ SYM engineered by the $X_{N,m}$
geometries. Indeed it was checked in \cite{Huang:2017mis} that the
generalized blowup equations are satisfied by some toric Calabi-Yau
threefolds which engineer 5d $SU(N)$ gauge theories with
matter. Moreover, what is fascinating is that the generalized blowup
equations may even be valid for 6d SCFTs as the topological string
theory on non-compact elliptic Calabi-Yau threefolds used in F-theory
compactifications computes precisely $Z^{\textrm{6d}}$ of these 6d
SCFTs on the Omega background. As a first step it was checked in
\cite{Gu:2017ccq,Huang:2017mis} that the simplest 6d SCFT, the
E-string theory, respects half of the generalized blowup
equations. The verification of the other half is a bit trickier, and
it will be discussed in our upcoming work. In this paper, we
demonstrate the validity of the generalized blowup equations through
the already well-studied cases of $n=3$ and $n=4$ minimal SCFTs in the
present paper, and illustrate their power by computing the elliptic
genera as well as the BPS invariants with them. Furthermore, by
reducing the $n=4$ model down to the 5d $SO(8)$ SYM, we verify the
validity of the generalized blowup equations for the latter theory as
well, which is also new. We will cover all the remaining minimal SCFTs
in companion papers. In the next subsection we give a quick overview
of the generalized blowup equations.

\subsection{Overview of geometric blowup equations}
\label{sc:overview}

Consider putting the refined topological string theory on a
non-compact Calabi-Yau threefold $X$.  Let $H_{2i}(X,\IZ)$ be the
homology groups of compact $2i$-cycles. In particular $H_2(X,\IZ)$
includes compact curve classes $\{\Sigma_i\}$, and $H_4(X,\IZ)$
compact divisor classes $\{D_j\}$. We denote the complexified K\"ahler
moduli of the compact curve classes by $t_i$ with
$\text{Vol}(\Sigma_i) =-\text{Re}(t_i)$, and the dimensions of the two
homology groups by
\begin{equation}
  \nC:=\dim H_2(X,\IZ) \ , \quad \nD:=\dim H_4(X,\IZ) \ .
\end{equation}
Since $X$ is not compact, these two numbers are not necessarily
identical. We encode the intersection numbers of curve classes and
divisor classes in a matrix
\begin{equation}\label{eq:C-def}
  \md C = (C_{ij})\ ,\quad \text{with}\;
  C_{ij} = \Sigma_i . D_j \ ,\quad \Sigma_i\in H_2(X,\IZ),\;D_j\in
  H_4(X,\IZ) \ .
\end{equation}
It is always possible to find $\nC-\nD$ independent linear
combinations of the curve classes so that they have zero intersection
number with any compact divisor. We call the corresponding K\"ahler
moduli mass paramters and sometimes denote them by $t_{m_i}$, as they
are interpreted as masses of hypermultiplets or instanton fugacity in
5d or 4d theories.

It was first observed in \cite{Choi:2012jz} and later confirmed in
many examples that the non-vanishing BPS invariants $\BPS$ on a
noncompact Calabi-Yau threefold respect a checkerboard pattern: there
exists a $\nC$ dimensional vector $\md B$ with entries in $\IZ_2$
such that\footnote{The entries of the $\md B$ field may be
  fractional if it is not expanded in integral curve classes.}
\begin{equation}\label{eq:checker}
  2j_L + 2j_R + 1 \equiv \md B \cdot\md d \quad \text{mod}\; 2
\end{equation}
holds for any non-vanishing $\BPS$. Then we could define the twisted
refined free energy via
\begin{equation}\label{eq:hF}
  \hF(\md t, \epsilon_1,\epsilon_2)
  = F^{\rm pert}(\md t, \epsilon_1,\epsilon_2)
  + F^{\rm inst}(\md t+\pi\ri \md B, \epsilon_1,\epsilon_2) \ ,
\end{equation}
where only in the worldsheet instanton contributions are the K\"ahler
moduli $\md t$ shifted by the vector $\md B$. We call $\md B$ the
$\md B$ field as it combines into the Kalb-Ramond part of the
complexified $\md t$. The twisted free energy appears prominently in
the geometric engineering of Nekrasov partition functions
\cite{Iqbal:2003zz,Taki:2007dh,Grassi:2016nnt} as well as in the
program of exact quantum mirror curves
\cite{Grassi:2014zfa,Codesido:2015dia}. We also define the twisted
partition function
\begin{align}
  \hZ(\md t, \epsilon_1,\epsilon_2) =
  \exp(\hF(\md t,\epsilon_1,\epsilon_2)) =
  & Z^{\rm pert}(\md t,\epsilon_1,\epsilon_2)
    \hZ^{\rm inst}(\md t,\epsilon_1,\epsilon_2) \ .
\end{align}
Note we do not put hat on
$Z^{\rm pert}$ since the K\"ahler moduli are not shifted there.

In terms of these quantities, the blowup equations can be reformulated
in the following way
\begin{equation}\label{eq:geom-blowup}
  \nn
  \sum_{\md n\in \IZ^g}(-1)^{|\md n|}
    \hZ\(\md t+\epsilon_1\md R,
    \epsilon_1,\epsilon_2-\epsilon_1\)
    \hZ\(\md t+\epsilon_2\md R,
    \epsilon_1-\epsilon_2,\epsilon_2\)=\Lambda
  (\md t_m,\epsilon_1,\epsilon_2,\md r)
    \hZ\(\md t,\epsilon_1,\epsilon_2\)
 \ ,
\end{equation}
with $|\md n| = \sum_i n_i$ and
\begin{equation}
  \md R = \md C\cdot \md n + {\md r}/2\ .
\end{equation}
Here $\md r = (r_i)$, which we call a $\md r$ field, is a $\nC$
dimensional vector with entries in $\IZ$ satisfying
\begin{equation}\label{eq:rB}
  r_i \equiv B_i \quad\text{mod}\; 2 \ .
\end{equation}
Two $\md r$ fields $\md r, \md r'$ are equivalent if
\begin{equation}\label{eq:r-equiv}
  \md r - \md r' = 2\md C\cdot \md n' \ ,\quad \md n' \in \IZ^g
\end{equation}
as the corresponding blowup equations can be identified by the shift
$\md n \to \md n + \md n'$.  The prefactor $\Lambda$ is trivial in the
sense that it only depends on the mass paramters $\md t_m$ but not on
the true moduli \cite{Huang:2017mis}. It also depends on the choice of
the $\md r$ field, thus it gives rise to different blowup equations
with different choices of the $\md r$ field. For some choices of the
$\md r$ field, $\Lambda$ vanishes all together, and we call the
corresponding equations the \emph{vanishing} blowup equations, while
the other equations with non-vanishing $\Lambda$ are called the
\emph{unity} blowup equations. Note that since the row vector of the
$\md C$-matrix corresponding to a mass parameter is null, a
multiplicative factor of $\hZ$ which depends only on mass parameters
but no other K\"ahler parameters decouples as its contributions to the
blowup equations can be factored out of the summation in $\md n$ and
be absorbed in $\Lambda$. We will thus discard this type of components
in $\hZ$.

It was conjectured in \cite{Gu:2017ccq,Huang:2017mis} that for any
non-compact Calabi-Yau threefold there is a finite but non-empty set
of $\md r$ fields so that the blowup equations \eqref{eq:geom-blowup}
hold. It was further conjectured and verified for some toric
Calabi-Yaus in \cite{Huang:2017mis} that the BPS invariants could be
computed from the blowup equations using classical geometric data of
$X$ as input. Furthermore, $\Lambda$ is modular invariant
with respect to the monodromy group of the topological string moduli
space. In this paper we demonstrate the validity of these statements
for the partition functions of the minimal 6d $n=3,4$ SCFTs.

The partition function of a 6d SCFT on the Omega background
$\IR^4\times_{\epsilon_1,\epsilon_2} T^2$ can be split to
three components
\begin{equation}
  Z(t_b,\tau,\md a,\epsilon_1,\epsilon_2) =
  Z^{\rm pert}(t_b,\tau,\md a,\epsilon_1,\epsilon_2)
  Z^{\text{1-loop}}(\tau,\md a,\epsilon_1,\epsilon_2)
  Z^{\rm ell}(t_b,\tau,\md a,\epsilon_1,\epsilon_2) \ .
\end{equation}
Here $t_b$, $\tau$ and $\md a$ are tensor modulus, complex structure
of $T^2$, and gauge fugacities (Wilson lines on $T^2$)
respectively. $Z^{\rm pert}$ contains perturbative
contributions. $Z^{\text{1-loop}}$ comes from Kaluza-Klein modes of 6d
particle multiplets on $T^2$. We denote it by the superscript 1-loop
because it descends to 1-loop contributions in 4d when we shrink $T^2$
to a point. Finally
$Z^{\rm ell}(t_b,\tau,\md a,\epsilon_1,\epsilon_2)$ splits by 

\begin{equation}
  Z^{\rm ell}(t_b,\tau,\md a,\epsilon_1,\epsilon_2)
  = 1 + \sum_{k=1}^\infty Q_{\rm ell}^k
  \IE_k(\tau,\md a,\epsilon_1,\epsilon_2)
\end{equation}
with $Q_{\rm ell} =\re^{t_{\rm ell}}$ the counting parameter, and
$\IE_k(\tau,\md a,\epsilon_1,\epsilon_2)$ the $k$ string elliptic
genus.

By the F-/M-theory duality and the relation of the BPS sector of the
M-theory with the refined topological string theory, the partition
function of a 6d SCFT on the tensor branch is computed by the
partition function of the refined topological string theory on the
same Calabi-Yau threefold $X$ encoding the BPS invariants on $X$, and
the moduli $t_{\rm ell},\tau,\md a$ are identified with
linear combinations of the K\"ahler moduli of compact curve classes in
$X$. In particular, $Z^{\rm ell}$ includes the BPS states of M2 branes
wrapping the base curve, while $Z^{\text{1-loop}}$ the BPS states of
M2 branes not wrapping the base curve at all. They combine into the
component $Z^{\text{inst}}$ that encodes all the BPS invariants.
$Z^{\rm pert}$ basically encodes the intersection numbers of divisors
in $X$. If we further decompactify $X$ along the direction of the
elliptic fiber in the M-theory picture, the 6d $(1,0)$ SCFT reduces to
a 5d $N=1$ SYM with the same gauge group on the Omega background,
where the tensor modulus $t_b$ becomes the gauge coupling.

The organization of the rest of the paper is as follows.
In Section \ref{sc:initial} we compute the initial data for blowup
equations. These include the curve-divisor intersection
$\md C$-matrix, the B-field for the checkerboard pattern,
$Z^{\rm pert}$ and $Z^{\text{1-loop}}$.
We give explicit expressions for these initial data for the cases of
6d SCFTs with $SU(3)$ and $SO(8)$ bulk gauge groups. In Section
\ref{sc:expansion} we put everything together and first demonstrate
the validity of the blowup equations order by order in terms of
$Q_{\rm ell}$ expansion and then proceed to recursively compute the
elliptic genera of multiple strings as well as the corresponding BPS
invariants. In Section \ref{sc:reduction} we study reductions of the
blowup equations in the 5d limit, that is when one of the circles of
the $T^2$ which is wrapped by the strings shrinks to zero
radius. Finally, in Section \ref{sc:conclusion} we present our
conclusions and give an overview of applications and open problems.

\section{Initial data for blowup equations}
\label{sc:initial}

We explain here how to compute the initial data for the blowup
equations: the curve-divisor intersection $\md C$-matrix, the $\md B$
fields, as well as the perturbative and 1-loop partition functions
$Z^{\rm pert}$, $Z^{\text{1-loop}}$, for 6d minimal SCFTs with no
matter. As we note in subsequent subsections, these data
are necessary if we wish to derive compact formulas of elliptic genera
from the blowup equations, while the piece $Z^{\text{1-loop}}$ is not
needed if we wish to directly compute BPS invariants from the blowup
equations.

\subsection{Curve-divisor intersection matrix}
\label{sc:C}

\begin{figure}
  \centering
  \subfloat[$n=3$]{\includegraphics[width=0.20\linewidth]{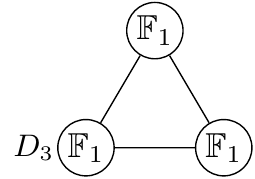}}\hspace{2ex}
  \subfloat[$n=4$]{\includegraphics[width=0.24\linewidth]{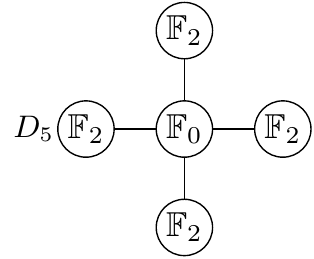}}\hspace{2ex}
  \subfloat[$n=6$]{\includegraphics[width=0.36\linewidth]{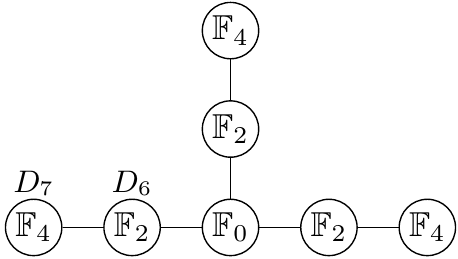}}
  \caption{Schematic structure of compact divisors in elliptic
    fibrations over $\cO(-n)\to\IP^1$ as affine Dynkin diagrams for
    $n=3,4,6$. The divisor $D_{r+1}$ corresponds to the affine
    node.}\label{fg:dynkins}
\end{figure}

The structures of the elliptic non-compact Calabi-Yau threefolds
underlying the 6d minimal SCFTs are for instance discussed in
\cite{DelZotto:2017pti}. Let the gauge group $G$ be of rank $r$. There
are $\nD = r+1$ compact divisors. They result from the resolution of
the singular elliptic fiber and they intersect with each other like
the nodes of the Dynkin diagram of $\widehat{\fg}$. One of the
divisors is special as it intersects with the base $B$ and it
corresponds to the affine node in the Dynkin diagram. We label the
special divisor by $D_{r+1}$ and the subsequent divisors
$D_{r}, D_{r-1},\ldots$. All these divisors are Hirzebruch surfaces
$\IF_{n_i}$. The indices $n_i$ of these divisors in different 6d
$(1,0)$ minimal SCFTs can be found in \cite{DelZotto:2017pti}, and we
give some examples in Figs.~\ref{fg:dynkins}. The number of
irreducible compact curves is $\nC = r+2$. Of these $r+1$ curves are
the $\IP^1$ fibers of the Hirzebruch surfaces $\Sigma_i$,
$i=1,\ldots,r+1$ and they stretch in the vertical direction. Their
labelling follows the labelling of the underlying divisors. These
curves satisfy
\begin{equation}
  \sum_i a_i [\Sigma_i] = [\delta] \ ,
\end{equation}
where $a_i$ are marks of the affine Lie algebra $\widehat{\fg}$, and
$\delta$ is the elliptic fiber. The last compact curve $\Sigma_b$ is in
the horizontal direction, and it projects down to the compact $-n$
curve $\Sigma_B$ in the base. In accord with topological string
calculations we choose it to be a Mori cone generator. It is always
the $\IP^1$ base of the Hirzebruch surface in the center of the chain
of $\IF_{n_i}$ with the lowest index. It is therefore related to the
base curve $\Sigma_{B}$ by
\begin{equation}
  [\Sigma_b] = [\Sigma_B] -
  \sum_{i=0}^{\lfloor \begin{small}{(n-3)/2}\end{small}\rfloor}(n-2-2i)[\Sigma_{r+1-i}] \ .
\end{equation}
In the case of $n=3,4$, we have
\begin{equation}
  [\Sigma_b] = [\Sigma_B] - [\Sigma_{r+1}] \ .
\end{equation}
We denote the volumes of these irreducible curves by $t_i$
($i=1,\ldots,r+1$) and $t_{r+2} = t_b$.

This geometric picture allows us to
write down the intersection $\md C$-matrix
\begin{equation}\label{eq:C-mat}
  \md C = \(\Sigma_i.D_j\) =
  \begin{pmatrix}
    \huge -\md A \\
    * \; \ldots \; *
  \end{pmatrix}
\end{equation}
where the $(r+1)\times (r+1)$ submatrix $-\md A$ is minus the Cartan
matrix of the affine Lie algebra $\widehat{\fg}$, and the last row
depends on the indices of the Hirzebruch surfaces $D_i =
\IF_{n_i}$. It is then easy to see that the only mass parameter is
\begin{equation}\label{eq:tau}
  \tau = \sum_{i=1}^{r+1}a_i t_i \ ,
\end{equation}
which is the comlexified volume of the elliptic fiber $\delta$.  We
will give the concrete expressions of the $\md C$-matrix of the
$SU(3)$ and the $SO(8)$ theories in the example subsetions.

Note that in most of the paper we expand $Z^{\rm ell}$ in terms of
$Q_b = e^{t_b}$
\begin{equation}
  Z^{\rm ell} = 1+\sum_{k=1}^\infty Q_b^k Z_k\ ,
\end{equation}
instead of $Q_{\rm ell} = e^{t_{\rm ell}}$ as the curve associated to
$t_{\rm ell}$ may not be in an integral class. These parameters are
related by\footnote{The K\"ahler modulus $t_b$ coincides with the
  volume of the curve class $l_b$ in \cite{DelZotto:2017mee}, but only
  coincides with the $t_b$ defined in \cite{DelZotto:2017mee} for
  $n=3,4$.}
\begin{equation}
  t_{\rm ell} = t_B - \frac{n-2}{2}\tau
  = t_b - \frac{n-2}{2}\tau
  + \sum_{i=0}^{\lfloor \begin{small}{(n-3)/2}\end{small}\rfloor}(n-2-2i)t_{r+1-i} \ .
\end{equation}

\subsection{The $\md B$ field}
\label{sc:B}


We would like to compute the $\nC$ dimensional $\IZ_2$ $\md B$ field which
characterizes the checkerboard pattern of non-vanishing BPS invariants
$N^{\md d}_{j_L,j_R}$ with identity
\begin{equation}
  2j_L + 2j_R + 1 \equiv   \md B\cdot \md d\quad \text{mod}\; 2 \ .
\end{equation}
Since the r.h.s.\ is linear in the curve class $\md d$, we only need
to know the entries of the $\md B$ field corresponding to each
individual irreducible curves. Each irreducible curve can be embedded
in an algebraic surface in the Calabi-Yau threefold $X$. Let us denote
the curve and the surface where it is embedded by $C$ and $S$
respectively. The non-vanishing BPS invariants associated to this
curve must have \cite{Hatsuda:2013oxa,Gu:2017ccq}
\begin{equation}
  2j_L^{\rm max} = \frac{C^2 + K_S\cdot C}{2} + 1 \ ,\quad 2j_R^{\rm max} =
  \frac{C^2 - K_S\cdot C}{2} \ ,
\end{equation}
where $C^2$ is the self-intersection number of the curve in the
surface $S$, and $K_S$ the canonical class of $S$. We have then
\begin{equation}\label{eq:B-C2}
  2j_L+2j_R + 1 = C^2 \quad\text{mod}\; 2 \ .
\end{equation}
Thus the entry of the $\md B$ field corresponding to $C$ is its
self-intersection number in the surface $S$ modulo two. On the other
hand, the self-intersection number $C^2$ is identified with the degree
of the normal bundle of $C$ perpendicular to $S$. Recall that the
normal bundle of a curve in a Calabi-Yau threefold has the form
\begin{equation}
  \cO(n)\oplus \cO(-2+n) \to C \ , \quad n\in \IZ \ .
\end{equation}
Since the two degrees $n$ and $-2+n$ are equivalent modulo two, we can
take either of them to be the entry of the $\md B$ field corresponding
to the curve $C$. Since we have a good understanding of irreducible
curves and surfaces in the Calabi-Yau threefold underlying the 6d
minimal SCFTs as we discussed in the beginning of the section, these
numbers can be easily computed for each irreducible compact curve.

\subsection{Perturbative partition function}
\label{sc:pert}


The perturbative contribution
$Z^{\rm pert}(\md t,\epsilon_1,\epsilon_2) =
\exp\(F^{\rm pert}(\md t,\epsilon_1,\epsilon_2)\)$ has the following
form
\begin{align}
  F^{\rm pert}(\epsilon_1,\epsilon_2;\md t) =
  & \frac{1}{\epsilon_1\epsilon_2}F^{{\rm pert}}_{(0,0)}(\md t,\epsilon_1,\epsilon_2)
    + F^{{\rm pert}}_{(1,0)}(\md t,\epsilon_1,\epsilon_2) -
    \frac{(\epsilon_1+\epsilon_2)^2}{\epsilon_1\epsilon_2}
    F^{{\rm pert}}_{(0,1)}(\md t,\epsilon_1,\epsilon_2) \nn =
  & \frac{1}{\epsilon_1\epsilon_2}\bigg(\frac{1}{6}\sum_{i,j,k=1}^{\nC}\kappa_{ijk}
    t_it_jt_k\bigg)
    + \sum_{i=1}^\nC b_i^{\rm GV}t_i
    -\frac{(\epsilon_1+\epsilon_2)^2}{\epsilon_1\epsilon_2}\sum_{i=1}^\nC
    b_i^{\rm NS} t_i \ .
\end{align}
The perturbative prepotential $F^{{\rm pert},(0,0)}$ is
decided\footnote{The perturbative prepotential can also include terms
  linear in $\md t$. But they decouple from the blowup equations.} by
the intersection numbers of divisors Poincar\'e dual to the curve
classes $\Sigma_i$ associated to $t_i$. Since the Poincar\'e duality
is only rigorously defined in a compact manifold, we should compute
$F^{{\rm pert},(0,0)}_{\rm cmp}$ in a compact Calabi-Yau threefold
where the non-compact Calabi-Yau threefold $X$ is embedded and take an
appropriate decompactification limit. Fortunately the compactification
of the Calabi-Yau threefolds underlying the 6d $SU(3)$ and $SO(8)$
gauge theories have been constructed in \cite{Haghighat:2014vxa}, and
we use the compact models there to compute
$F^{{\rm pert},(0,0)}_{\rm cmp}$ which is subsequently reduced to
$F^{{\rm pert},(0,0)}$ in the decompactification limit. On the other
hand, once $F^{{\rm pert},(0,0)}$ is computed for the 6d gauge theory,
we could obtain $F^{{\rm pert}, (0,0)}_{\rm 5d}$ for the 5d gauge
theory by further decompactifying the Calabi-Yau threefold $X$ along
the direction of the elliptic fiber while keeping the volumes of
$\Sigma_{i}$ $(i=1,\ldots,r)$ finite.\footnote{We send the volume of
  the curve class $\Sigma_{r+1}$ which intersects with the base to
  infinity.} The latter is also given by the perturbative Nekrasov
partition function \cite{Nakajima:2005fg}
\begin{align}
  F^{\rm Nek,pert}(\md a,\fq,\epsilon_1,\epsilon_2)
  =&
     -\frac{1}{\epsilon_1\epsilon_2}\sum_{\alvec \in
     \Delta_+}\bigg(\frac{\aroot^3}{6} - \frac{\log(\re^{-\hG
     \pi\ri}\fq)}{2\hG}\aroot^2 \bigg)\nn
  &-
    \frac{(\epsilon_1+\epsilon_2)^2+\epsilon_1\epsilon_2}{\epsilon_1\epsilon_2}
    \sum_{\alvec\in\Delta_+} \bigg(\frac{\aroot}{12}
    - \frac{\log(\re^{-\hG\pi\ri}\fq)}{24\hG}\bigg)  \label{eq:Nek-pert}
\end{align}
where $\avec$ is the vector of Coulomb moduli, $\fq$ is the instanton
counting parameter, $\Delta_+$ is the set of positive roots of the Lie
algebra $\fg$, and $\hG$ the dual Coxeter number of
$G$. $\vev{\bullet,\bullet}$ is the invariant bilinear
form\footnote{Here we normalize it so that the longest root has norm
  square 2.} in the Lie algebra $\fg$. The dictionary between field
theory parameters and geometric K\"ahler moduli is (see for instance
\cite{Grassi:2016nnt})
\begin{equation}\label{eq:dict}
  \left\{\begin{aligned}
      & t_i = \arooti{i} \\
      & t_m = - \log(\re^{\hG\pi \ri }\fq)
    \end{aligned} \right. \ , 
\end{equation}
where $\alvec_i$ are simple roots, and $t_m$ the mass parameter whose
associated curve, we recall, that does not intersect with compact
divisors. Once we could identify the first line of \eqref{eq:Nek-pert}
with $F^{{\rm pert}, (0,0)}_{\rm 5d}$, we could uplift the second line
of \eqref{eq:Nek-pert} to obtain the perturbative genus one free
energies $F^{{\rm},(1,0)}, F^{{\rm},(0,1)}$ for the 6d theories, as we
will do in example subsetions. In particular, we find in the examples
of the $n=3,4$ theories
\begin{equation}\label{eq:bp0}
  b_i^{\rm GV} + b_i^{\rm NS} = 0 \ ,\quad i=1,\ldots,\nC \ .
\end{equation}

\subsection{One-loop partition function}
\label{sc:1-loop}


$Z^{\text{1-loop}}$ has the contribution of the Kaluza-Klein modes on
the 6d $S^1$ of the 6d particle multiplets. The contribution of a
single supermultiplet of various types reads as follows
\cite{Hayashi:2016abm}\footnote{To be in line with the refined
  Gopakumar-Vafa formula of topological string theory
  \cite{Huang:2010kf,Iqbal:2007ii}, we suppress a term of $1/2$ in
  \cite{Hayashi:2016abm}.}
\begin{equation}\label{eq:multiplets}
  \begin{aligned}
    &Z_{\rm tensor}={\rm PE}\bigg[- \frac{q_L + q_L}{\big(q_1^{1/2} -
        q_1^{-1/2}\big)\big(q_2^{1/2} -
        q_2^{-1/2}\big)}\Big(\frac{Q_\tau}{1-Q_\tau}\Big)\bigg]\ ,\\
    &Z_{\rm vector}={\rm PE}\bigg[- \frac{q_R + q_R}{\big(q_1^{1/2} -
        q_1^{-1/2}\big)\big(q_2^{1/2} - q_2^{-1/2}\big)}
      Q_G^*\Big(\frac{Q_\tau}{1-Q_\tau}\Big)\bigg] \ , \\
    &Z_{\rm hyper}={\rm PE}\bigg[+ \frac{1}{\big(q_1^{1/2} - q_1^{-1/2}\big)\big(q_2^{1/2} -
        q_2^{-1/2}\big)}Q_G^*Q_F^*\Big(\frac{Q_\tau}{1-Q_\tau}\Big)\bigg] \ ,
  \end{aligned}
\end{equation}
where $Q_\tau = \re^{\tau}$\footnote{Note our convention here
  differs from the usual convention in the mathematics literature by a
  factor of $2\pi\ri$.}, while $Q_G = \re^{\md a}, Q_F=\re^{\md m_G}$
are gauge and flavor symmetry fugacities, with powers $*$ appropriate
charges of the supermultiplets. The plethystic exponential is defined
as
\begin{equation}
  \PE{f(\cdot)} = \exp\bigg[\sum_{i=1}^\infty \frac{1}{n}f(\cdot^n)\bigg]\ .
\end{equation}
In the case of 6d minimal SCFTs, there is no contributions from
charged hypermultiplets, while the contributions of tensor multiplets
only depend on the mass parameter $\tau$ and no other K\"ahler moduli
and can thus be factored out of the blowup equations. Therefore in
this paper we only consider contributions of 6d vector multiplets to
$Z^{\text{1-loop}}$. The spectrum of vector multiplets and thus their
total contribution to $Z^{\text{1-loop}}$ can be computed by the
refined topological string theory
\begin{align}
  Z^{\text{1-loop}}
  &(\md t,\epsilon_1,\epsilon_2) \nn =
  &\prod_{\Sigma\in H_2^{\rm vert}(X,\IZ)}
  \prod_{k_L=-j_L}^{j_L}\prod_{k_R =-j_R}^{j_R}
  \prod_{m_1,m_2=1}^\infty
  \Big(1-t^{k_L+k_R+m_1-\tfrac{1}{2}}q^{k_L-k_R+m_2-\tfrac{1}{2}}\md
  Q^{\Sigma} \Big)^{M_\Sigma^{(j_L,j_R)}}
\end{align}
where\footnote{Here $M_\Sigma^{(j_L,j_R)}$ differs from that in
  \cite{Iqbal:2007ii} by 1 in order for the contributions of vector
  multiplets to be in the denominator, as they should.}
\begin{equation}
  q = \re^{\epsilon_1}\ ,\quad t = \re^{-\epsilon_2} \ ,\quad
  M_\Sigma^{(j_L,j_R)} =(-1)^{2(j_L+j_R)}N_{j_L,j_R}^\Sigma \ ,
\end{equation}
with $(j_L,j_R) = (0,1/2)$ for vector multiplets. Here
$H_2^{\rm vert}(X,\IZ)$ is the homology group of compact curves
in the vertical direction, and it is generated by $\Sigma_{i}$
$(i=1,\ldots,r+1)$. Using $N^\Sigma_{0,1/2}=1$, we obtain
\begin{align}
  Z^{\text{1-loop}}(\md t,\epsilon_1,\epsilon_2) =
  &\prod_{\alvec\in\widehat{\Delta}_+}\prod_{i,j=0}^\infty
  \(1-t^iq^{j+1}Q^{\alvec}\)^{-1} \(1-t^{i+1}q^j Q^{\alvec}\)^{-1} \nn =\,
  &{\rm PE}\Bigg[-\frac{q_R+q_R^{-1}}
    {\big(q_1^{1/2}-q_1^{-1/2}\big)\big(q_2^{1/2}-q_2^{-1/2}\big)}
  \sum_{\alvec\in\widehat{\Delta}_+} \re^{\aroot}\Bigg]
\end{align}
where $\widehat{\Delta}_+$ is the set of positive roots of the affine
Lie algebra. By the identification of the imaginary root with the
elliptic fiber and \eqref{eq:tau}, the expression for
$Z^{\text{1-loop}}$ is equivalent to
\begin{equation}\label{eq:1-loop}
  Z^{\text{1-loop}}(\md t,\epsilon_1,\epsilon_2)
  =
  \,{\rm PE}\Bigg[-\frac{q_R+q_R^{-1}}
  {\big(q_1^{1/2}-q_1^{-1/2}\big)\big(q_2^{1/2}-q_2^{-1/2}\big)}
    \sum_{\alvec \in \Delta_+}  \(\re^{\aroot}
    +Q_\tau\re^{-\aroot}\) \frac{1}{1-Q_\tau}\Bigg] \ .
\end{equation}


\subsection{Examples}
\label{sc:init-examples}

\subsubsection{6d $SU(3)$ theory}

\label{sc:input-A2}

\begin{figure}
  \centering
  \includegraphics[width=0.6\linewidth]{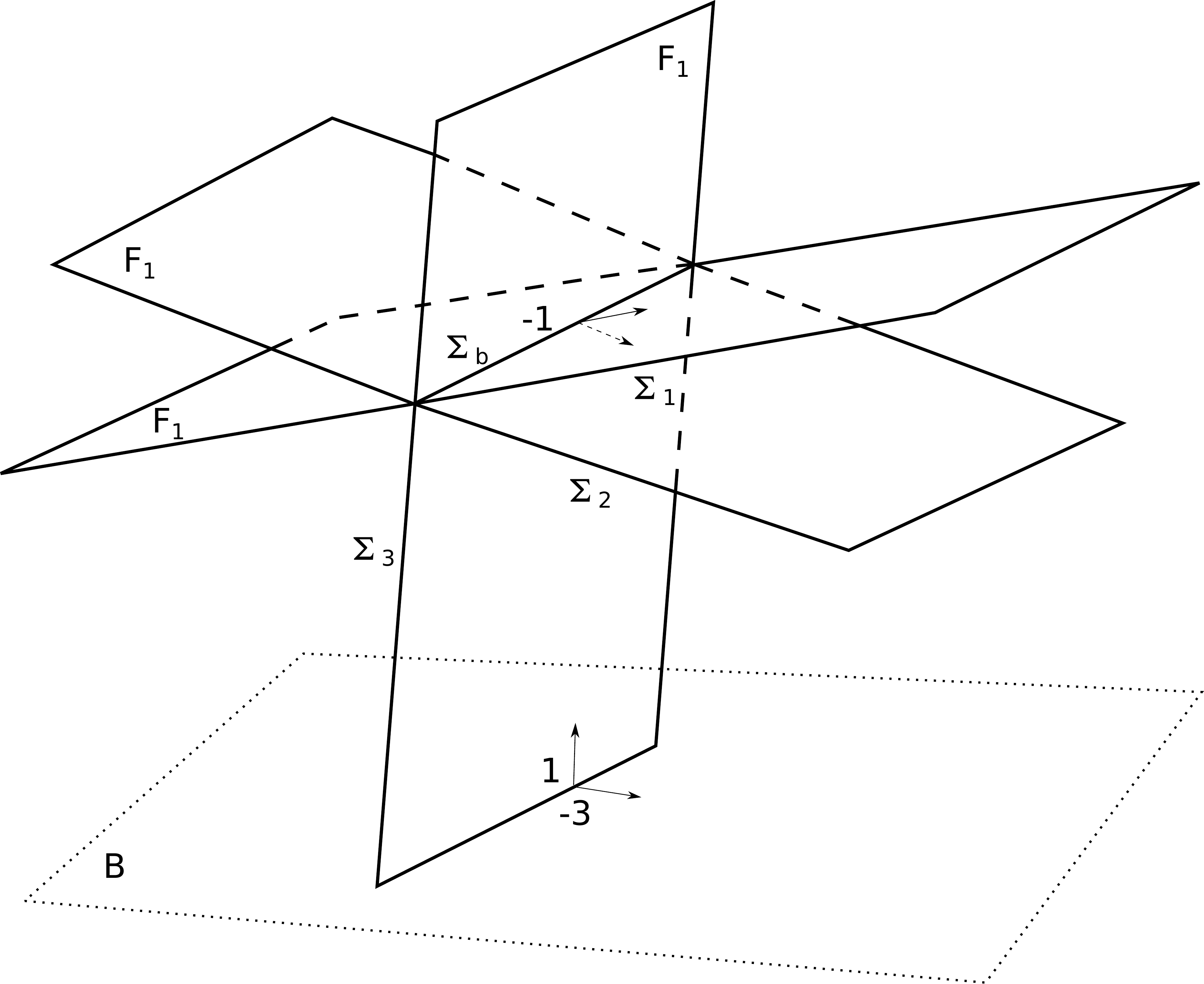}
  \caption{Compact curves and compact divisors in the elliptic
    fibration over $\cO(-3)\to\IP^1$.}
  \label{fg:A2}
\end{figure}

The non-compact Calabi-Yau threefold $X$ underlying the 6d $SU(3)$
model on the Omega background is the elliptic fibration over
$\cO(-3)\to\IP^1$ with the singular fiber resolved. As explained in
\cite{DelZotto:2017pti}, there are $\nC=4$ compact irreducible curves
and $\nD=3$ compact irreducible divisors. The latter $D_1,D_2,D_3$ are
three $\IF_1$ surfaces in the vertical direction intersecting at a
common $(-1)$-curve $\Sigma_4 = \Sigma_b$, which projects to the
$(-3)$-curve in the base. The other three curves $\Sigma_i$
($i=1,2,3$) are the $\IP^1$ fibers of the Hirzebruch surfaces. This
geometry is illustrated in Fig.~\ref{fg:A2}. The intersection matrix
of the curves $\Sigma_i$ ($i=1,\ldots,4$) and the divisors $D_j$
($j=1,\ldots,3$) is
\begin{equation}\label{eq:C-A2}
  C =
  \begin{pmatrix}
    -2 & 1 & 1 \\
    1 & -2 & 1 \\
    1 & 1 & -2 \\
    -1 & -1 & -1 
  \end{pmatrix} \ ,
\end{equation}
in accord with the general structure \eqref{eq:C-mat}.

The understanding of the embedding of the curves $\Sigma_i$ in
surfaces in $X$ allows us to write down the $\md B$ field
\begin{equation}\label{eq:B-A2}
  \md B = (0,0,0,1) \ ,
\end{equation}
following the discussion in section~\ref{sc:B}.

To compute the perturbative prepotential $F^{{\rm pert},(0,0)}$, we
follow \cite{Haghighat:2014vxa} and take $X$ as the decompactification
limit of the compact Calabi-Yau threefold $\widehat{X}$, the elliptic
fibration over $\IF_3$, along the horizontal direction perpendicular
to the $(-3)$-curve in the base. The compact model $\widehat{X}$ can
be realized as a hypersurface in a toric variety. Therefore its triple
intersection numbers and thus the perturbative prepotential can be
computed with the usual techniques in toric geometry (see
for instance \cite{Hosono:1993qy}). Then $F^{{\rm pert},(0,0)}$ of the
non-compact model is obtained by integrating over the periods which
remain finite in the decompactification limit \cite{Chiang:1999tz}. In
this way, we find
\begin{equation}\label{eq:F0d6-A2}
  F_{{\rm 6d},SU(3)}^{{\rm pert}, (0,0)}(\md t,\epsilon_1,\epsilon_2)
  = -\frac{1}{18}\( t_1^3+  t_2^3 +t_3^3 \)
  -\frac{1}{6} t_b( t_1^2 + t_2^2 + t_3^2)
  - \frac{1}{6} t_b^2 (t_1 + t_2 + t_3)  \ .
\end{equation}
Keep in mind we use the convention $t_b = t_4$.

We can further decompactify $X$ along the direction of the elliptic
fiber by sending the volume of one of the curves in the vertical
direction to infinity. Let us take the limit\footnote{Since we wish to
  obtain the corresponding 5d gauge theory, we should decompactify the
  vertical curve which intersects with the base
  \cite{DelZotto:2017pti}. Nevertheless since the K\"ahler moduli of
  the three vertical curves appear to be on the equal footing in
  $F_{{\rm 6d},SU(3)}^{{\rm pert}, (0,0)}$, we can choose any of them
  to decompactify, keeping the others intact.}  $t_3\to \infty$ and
after following the same procedure of integrating over finite periods,
we obtain
\begin{equation}\label{eq:F0d5-A2}
  F_{{\rm 5d},SU(3)}^{{\rm pert},(0,0)}(\md t,\epsilon_1,\epsilon_2)
  = -\frac{1}{18}(t_1^3 + t_2^3) - \frac{1}{6} t_b (t_1^2 + t_2^2)
  -\frac{1}{6} t_b^2(t_1 + t_2) + \frac{1}{18}t_b^3 \ .
\end{equation}
This should coincide with the perturbative Nekrasov partition function
for 5d $N=1$ pure SYM with $G=SU(3)$. Combining \eqref{eq:Nek-pert}
and \eqref{eq:dict}, the latter reads
\begin{equation}
  F^{{\rm pert},(0,0),\rm Nek}_{{\rm 5d},SU(3)}
  (\md t,\epsilon_1,\epsilon_2) 
  = - \frac{t_1^3}{3} -\frac{t_1^2 t_2}{2} - \frac{t_1 t_2^2}{2} -
  \frac{t_2^3}{3} - t_m\(\frac{t_1^2}{3} + \frac{t_1t_2}{3} +
  \frac{t_2^2}{3}\) \label{eq:F0Nek-A2} \ .
\end{equation}
To identity the mass parameter $t_m$ in \eqref{eq:F0d5-A2}, we first
write down the curve-divisor intersecton $\md C$-matrix of the 5d
theory
\begin{equation}
  C =
  \begin{pmatrix}
    -2 & 1 \\ 1 & -2 \\ -1 & -1
  \end{pmatrix} \ ,
\end{equation}
which can be obtained by removing in the 6d $\md C$-matrix
\eqref{eq:C-A2} the third row corresponding to $\Sigma_3$ and the
third column corresponding to the divisor $D_3$ containing
$\Sigma_3$. We find the curve $\Sigma_b - \Sigma_1-\Sigma_2$ does not
intersect with any compact divisor. The corresponding mass paramter
for the 5d theory should be
\begin{equation}\label{eq:tmtb-A2}
  t_m = t_b - t_1 - t_2 \ .
\end{equation}
With this identification, it is easy to see that
$F^{{\rm pert},(0,0),\rm Nek}_{{\rm 5d},SU(3)}$ indeed coincides with
$F_{{\rm 5d},SU(3)}^{{\rm pert},(0,0)}$ from decompactification up to
a pure mass parameter term
\begin{equation}\label{eq:F0diffNek-A2}
  F^{{\rm pert},(0,0)}_{{\rm 5d},SU(3)}(\md t,\epsilon_1,\epsilon_2)
  - F^{{\rm pert},(0,0),\rm Nek}_{{\rm 5d},SU(3)}(\md t,\epsilon_1,\epsilon_2)
  = \frac{1}{18}t_m^3 \ .
\end{equation}

We also notice that the 6d and the 5d perturbative prepotentials only
differ by
\begin{equation}\label{eq:F0red-A2}
  F^{\rm (0,0)}_{{\rm 6d},SU(3)}(\md t,\epsilon_1,\epsilon_2)
  - F^{\rm (0,0)}_{{\rm 5d},SU(3)}(\md t,\epsilon_1,\epsilon_2)
  = -\frac{(t_3 + t_b)^3}{18} = -\frac{(\tau + t_m)^3}{18}\ ,
\end{equation}
This implies that the decompactification limit is really obtained by
\begin{equation}\label{eq:declim-A2}
  t_3+t_b = \tau+t_m \to -\infty \ ,\quad Q_\tau Q_m \to 0 \
  ,\quad t_b,t_m\;\text{finite} \ .
\end{equation}
This observation allow us to write down the perturbative contributions
to genus one free energies of the 6d model. By writing down a generic
linear ansatz for $F^{{\rm pert},(1,0)}, F^{{\rm pert},(0,1)}$,
separating out $t_3+t_b$, and demanding the remaining piece coincides
with the second line of \eqref{eq:Nek-pert}, as well as imposing
symmetry between $t_1,t_2,t_3$, we fix
$F^{{\rm pert},(1,0)}, F^{{\rm pert},(0,1)}$ for the 6d $SU(3)$ model
uniquely to be
\begin{align}
  F^{{\rm pert},(1,0)}_{{\rm 6d},SU(3)}
  &(\md t,\epsilon_1,\epsilon_2) = - F^{{\rm pert},(0,1)}_{{\rm 6d},SU(3)}
    (\md t,\epsilon_1,\epsilon_2) \nn
    =& -\frac{t_1}{8} -\frac{t_2}{8} -\frac{t_3}{8} -\frac{t_b}{6}
       \ .\label{eq:F1d6-A2}
\end{align}
The difference from the 5d free energy is
\begin{equation}\label{eq:F1red-A2}
  F^{(1,0)}_{{\rm 6d},SU(3)} - F^{(1,0),\rm Nek}_{{\rm 5d},SU(3)}
  = -\(F^{(0,1)}_{{\rm 6d},SU(3)} - F^{(0,1),\rm Nek}_{{\rm 5d},SU(3)}\)
  = -\frac{1}{8}(t_3+t_b)
  = -\frac{1}{8}(\tau+t_m)\ .
\end{equation}

By specializing \eqref{eq:1-loop}, we get the one-loop partition
function
\begin{align}
  &Z^{\text{1-loop}}_{{\rm 6d},SU(3)} \nn 
  &=
    \PE{-\frac{q_R^{1/2}+q_R^{-1/2}}{\big(q_1^{1/2}-q_1^{-1/2}\big)
    \big(q_2^{1/2}-q_2^{-1/2}\big)}
    \frac{1}{1-Q_\tau}\(Q_1+Q_2+Q_3+Q_1Q_2+Q_1Q_3+Q_2Q_3\)} \ .
\end{align}

\subsubsection{6d $SO(8)$ theory}
\label{sc:input-D4}

\begin{figure}
  \centering
  \includegraphics[width=0.6\linewidth]{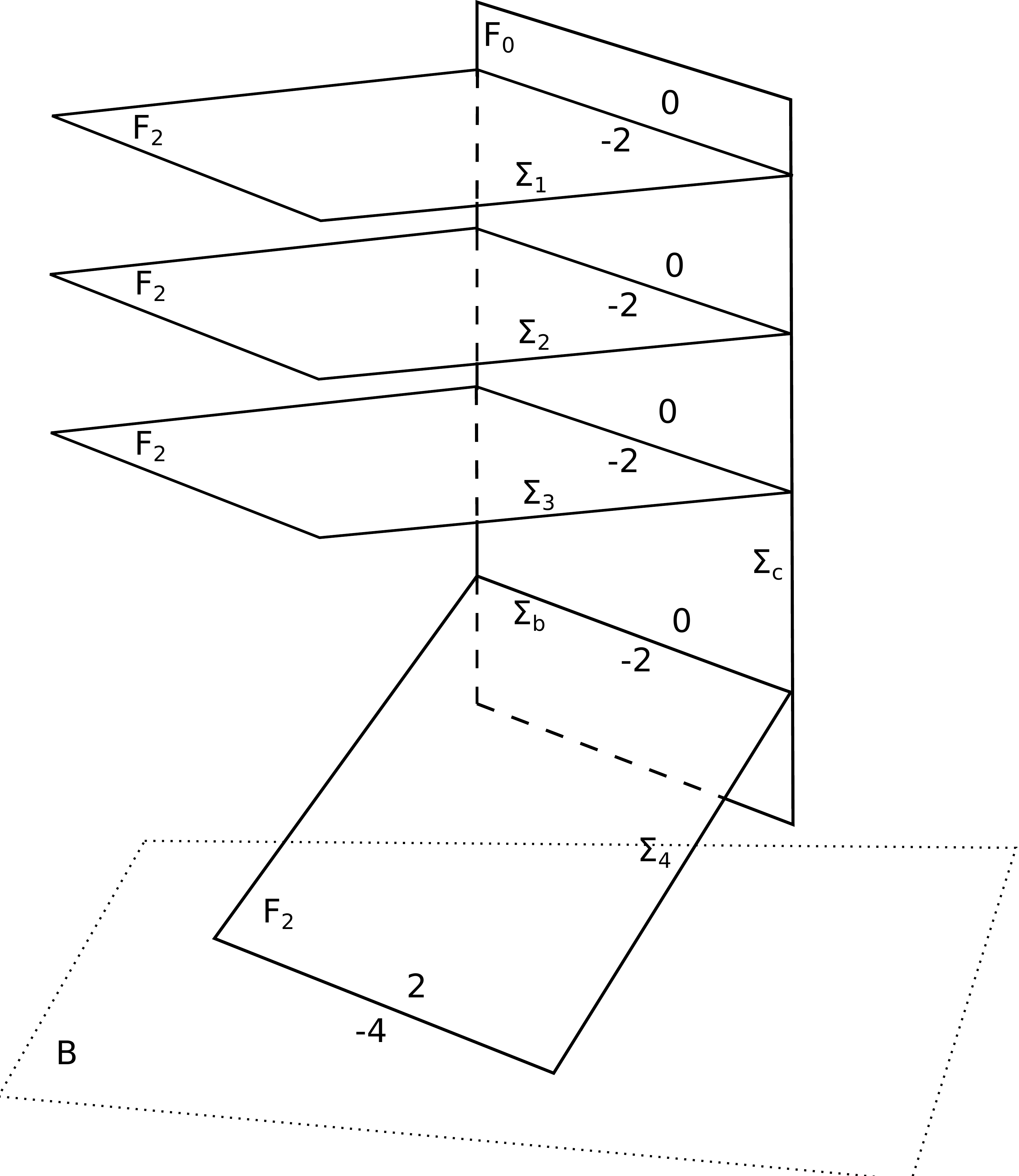}
  \caption{Compact curves and compact divisors in the elliptic
    fibration over $\cO(-4)\to\IP^1$}
  \label{fg:D4}
\end{figure}

The non-compact Calabi-Yau threefold $X$ underlying the 6d $SO(8)$
model on the Omega background is the elliptic fibration over
$\cO(-4)\to\IP^1$ with the singular fiber resolved. As explained in
\cite{DelZotto:2017pti} there are $\nC=6$ compact irreducible curves
and $\nD=5$ compact irreducible divisors. The divisors
$D_1,D_2,D_3,D_4,D_5=D_c$ are Hirzebruch surfaces
$\IF_2,\IF_2,\IF_2,\IF_2,\IF_0$ linking up with each other like the
Dynkin diagram of $\widehat{so}(8)$, where $D_c$ plays the role of the
central node, while $D_4$ lays the role of the affine node and
intersects with the base. $D_1,D_2,D_3,D_4$ intersect with $D_c$ by
the $(-2)$ curves which are all homologously equivalent on $D_c$. We
take this curve to be $\Sigma_6 = \Sigma_b$. The remaining irreducible
curves $\Sigma_i$ ($i=1,\ldots,4$) and $\Sigma_5=\Sigma_c$ are the
$\IP^1$ fibers of $D_i$ ($i=1,\ldots,4$) and $D_c$. This geometry is
illustrated in Fig.~\ref{fg:D4}. The intersection matrix of the curves
$\Sigma_i$ ($i=1,\ldots,6$) and the divisors $D_j$ ($j=1,\ldots,5$) is
\begin{equation}\label{eq:C-D4}
  C = 
  \begin{pmatrix}
    -2 & 0 & 0 & 0 & 1 \\
    0 & -2 & 0 & 0 & 1 \\
    0 & 0 & -2 & 0 & 1 \\
    0 & 0 & 0 & -2 & 1 \\
    1 & 1 & 1 & 1 & -2 \\
    0 & 0 & 0 & 0 & -2
  \end{pmatrix} \ .
\end{equation}
in accord with the general structure \eqref{eq:C-mat}.

Next, with the picture in Fig.~\ref{fg:D4} we can write down the
$\md B$ field
\begin{equation}\label{eq:B-D4}
  \md B = (0,0,0,0,0,0) \ ,
\end{equation}
following the discussion in section~\ref{sc:B}. As in the case of
$SU(3)$ theory, we compute the perturbative prepotential
$F^{{\rm pert},(0,0)}$ by following \cite{Haghighat:2014vxa} and
taking $X$ as the decompactification limit of the compact Calabi-Yau
threefold $\widehat{X}$, the elliptic fibration over $\IF_4$, along
the horizontal direction perpendicular to the $(-4)$ curve in the
base. The compact model $\widehat{X}$ can also be realized as a
hypersurface in a toric variety. Therefore its triple intersection
numbers and thus the perturbative prepotential are readily computable
using usual techniques in toric geometry. We then obtain
$F^{{\rm pert},(0,0)}$ of the non-compact model by integrating over
finite periods in the decompactification limit \cite{Chiang:1999tz}
and we get
\begin{equation}\label{eq:F0d6-D4}
  F_{{\rm 6d},SO(8)}^{(0,0)} = -\frac{1}{6}\( t_1^3 + t_2^3 + t_3^3 + t_4^3\)
  -\frac{1}{4} t_b \(t_1^2 + t_2^2 + t_3^2+ t_4^2\) -\frac{1}{8} t_b^2
  (t_1 + t_2 + t_3 + t_4 + 2 t_c)  \ .
\end{equation}
Keep in mind we use the convention $t_b = t_6, t_c = t_5$.

We further decompactify $X$ along the direction of the elliptic fiber
by
sending $t_4\to -\infty$.
After following the same procedure of integrating over finite periods,
we obtain
\begin{equation}\label{eq:F0d5-D4}
  F^{(0,0)}_{{\rm 5d},SO(8)}(\md t,\epsilon_1,\epsilon_2)
  = -\frac{1}{6}(t_1^3 + t_2^3 +
  t_3^3) - \frac{t_b}{4}(t_1^2 + t_2^2 + t_3^2) - \frac{t_6^2}{8}(t_1 +
  t_2 + t_3 + 2t_c) + \frac{t_b^3}{48} \ .
\end{equation}
This should coincide with the perturbative Nekrasov partition function
for 5d $N=1$ pure SYM with $G=SO(8)$. Combining \eqref{eq:Nek-pert}
and \eqref{eq:dict}, the latter reads
\begin{align}\label{eq:F0Nek-D4}
  &F^{(0,0),\rm Nek}_{{\rm 5d},SO(8)}
  (\md t,\epsilon_1,\epsilon_2)  \nn =
  & - t_1^3 - t_2^3 - t_3^3 - \frac{8 t_c^3}{3} - 4 t_c^2 (t_1 + t_2 + t_3) 
    - 3t_c(t_1^2+ t_2^2 +  t_3^2) - 4 t_c(t_1 t_2+  t_1  t_3 +   t_2 t_3 )\nn
  &  - \frac{3}{2}(t_1^2t_2+t_1t_2^2+t_1^2t_3 +t_1t_3^2+t_1t_3^2+t_2^2 t_3 )
    - 2 t_1 t_2 t_3 \nn
  &  -\frac{t_m}{2}(t_1^2 + t_2^2 +  t_3^2 +2 t_c^2 +
    2t_c(t_1+t_2+t_3)
    + t_1t_2 +  t_1 t_3 +  t_2 t_3  ) \ .
\end{align}
To identity the mass parameter
$t_m$ in \eqref{eq:F0d5-D4}, we first obtain the curve-divisor
intersecton $\md C$-matrix of the 5d theory
\begin{equation}
  C =
  \begin{pmatrix}
    -2 & 0 & 0 & 1 \\
    0 & -2 & 0 & 1 \\
    0 & 0 & -2 & 1 \\
    1 & 1 & 1 & -2 \\
    0 & 0 & 0 & -2
  \end{pmatrix} \ ,
\end{equation}
which is done by removing in the 6d $\md C$-matrix \eqref{eq:C-D4} the
fourth row corresponding to $\Sigma_4$ and the fourth column
corresponding to the divisor $D_4$ containing $\Sigma_4$. We find the
curve $\Sigma_b - 2\Sigma_1-2\Sigma_2-2\Sigma_3-4\Sigma_c$ does not
intersect with any compact divisor. The corresponding mass paramter
for the 5d theory should be
\begin{equation}\label{eq:tmtb-D4}
  t_m = t_b - 2 t_1 - 2 t_2 - 2 t_3 - 4 t_c \ .
\end{equation}
With this identification, it is easy to see that
$F^{{\rm pert},(0,0),\rm Nek}_{{\rm 5d},SO(8)}$
indeed coincides with $F_{{\rm 5d},SO(8)}^{{\rm pert},(0,0)}$
from decompactification up to a pure mass parameter term
\begin{equation}\label{eq:F0diffNek-D4}
  F^{(0,0)}_{{\rm 5d},SO(8)} - F^{(0,0),\rm Nek}_{{\rm 5d},SO(8)} =
  \frac{t_m^3}{48} \ .
\end{equation}

We also notice that the 6d and the 5d perturbative prepotentials only
differ by
\begin{equation}\label{eq:F0red-D4}
  F^{(0,0)}_{{\rm 6d},SO(8)}(\md t,\epsilon_1,\epsilon_2) -
  F^{(0,0)}_{{\rm 5d},SO(8)}(\md t,\epsilon_1,\epsilon_2)
  = -\frac{1}{6}(t_4 + \tfrac{1}{2}t_b)^3
  = -\frac{1}{6}(\tau + \tfrac{1}{2}t_m)^3 \ ,
\end{equation}
This implies that the decompactification limit is really obtained by
\begin{equation}\label{eq:declim-D4}
  t_4+\tfrac{1}{2}t_b = \tau+\tfrac{1}{2} t_m \to -\infty
  \ ,\quad Q_\tau Q_m^{1/2} \to 0 \
  ,\quad t_b,t_m\;\text{finite} \ .
\end{equation}
This observation allow us to write down the perturbative contributions
to genus one free energies of the 6d model. By writing down a generic
linear ansatz for $F^{{\rm pert},(1,0)}, F^{{\rm pert},(0,1)}$,
separating out $t_4+1/2t_b$, and demanding the remaining piece
coincides with the second line of \eqref{eq:Nek-pert}, as well as
imposing symmetry between $t_1,t_2,t_3,t_4$, we fix
$F^{{\rm pert},(1,0)}, F^{{\rm pert},(0,1)}$ for the 6d $SO(8)$ model
uniquely to be
\begin{align}
  F^{(1,0)}_{{\rm 6d},SO(8)}
  &(\md t,\epsilon_1,\epsilon_2) = - F^{(0,1)}_{{\rm 6d},SO(8)}
    (\md t,\epsilon_1,\epsilon_2) \nn
    =& -\frac{t_1}{3} -\frac{t_2}{3} -\frac{t_3}{3} -\frac{t_4}{3}
       -\frac{t_c}{2} - \frac{t_b}{4} \ .\label{eq:F1d6-D4}
\end{align}
The difference from the 5d theory is
\begin{equation}\label{eq:F1red-D4}
  F^{(1,0)}_{{\rm 6d},SO(8)} - F^{(1,0),\rm Nek}_{{\rm 5d},SO(8)}
  = -\(F^{(0,1)}_{{\rm 6d},SO(8)} - F^{(0,1),\rm Nek}_{{\rm 5d},SO(8)}\)
  =-\frac{1}{3}(t_4+\frac{1}{2}t_b)
  = -\frac{1}{3}(\tau+\frac{1}{2}t_m)\ .
\end{equation}

Finally, by specializing \eqref{eq:1-loop}, we get the 1-loop
partition function
\be
\ba
  &Z^{\text{1-loop}}_{{\rm 6d},SO(8)} =
    \text{PE}\Bigg[-\frac{q_R^{1/2}+q_R^{-1/2}}
    {\big(q_1^{1/2}-q_1^{-1/2}\big)\big(q_2^{1/2}-q_2^{-1/2}\big)}
    \frac{1}{1-Q_\tau}\bigg( Q_c^2\sum_{1\leq i<j<k\leq 4} Q_iQ_jQ_k\\
  &+ Q_c \bigg(1+\sum_{i=1}^4 Q_i
    +\sum_{1\leq i<j\leq 4} Q_iQ_j + \sum_{1\leq i<j<k\leq 4}
    Q_iQ_jQ_k + Q_1Q_2Q_3Q_4\bigg)
  + \sum_{i=1}^4 Q_i
    \bigg)\Bigg] \ .
\ea
\ee

\section{Elliptic genera from blowup equations}
\label{sc:expansion}

In this section we put everything together, and demonstrate for the 6d
$SU(3)$ and $SO(8)$ SCFTs the validity of blowup equations order by
order in an expansion in $Q_b$ with the help of the well-known results
of the elliptic genera of these two theories
\cite{Kim:2016foj,DelZotto:2017mee,DelZotto:2017mee}. Then we reverse
the logic and show that the blowup equations can be used to solve the
elliptic genera and BPS invariants, illustrating the power of blowup
equations in the studies of 6d SCFTs

\subsection{Constraint on $\md r$ fields}
\label{sc:r}

We first find a mild condition on the $\md r$ field and argue that
the number of inequivalent and admissble $\md r$ fields satisfying
this condition is finite.

We first rewrite the blowup equation \eqref{eq:geom-blowup} by moving
the unshifted partition function $\hZ(\md t,\epsilon_1,\epsilon_2)$ to
the other side of the equation
\begin{equation}
  \hZ(\md t,\epsilon_1,\epsilon_2)^{-1}
  \sum_{\md n\in\IZ^{g}}(-1)^{|\md n|}
  \hZ(\md t+\epsilon_1\md R,\epsilon_1,\epsilon_2-\epsilon_1)
  \hZ(\md t+\epsilon_2\md R,\epsilon_1-\epsilon_2,\epsilon_2) =
  \Lambda(\tau,\epsilon_1,\epsilon_2) \ ,\label{eq:geom-blowup2}
\end{equation}
where the dependence on $\md r$ is always understood, and
\begin{equation}\label{eq:R}
  \md R = \md C\cdot \md n + {\md r}/2 \ .
\end{equation}
We have also used the fact that $\tau$ is the only mass parameter. When
the l.h.s.\ of the blowup equations are expanded in terms of the
K\"ahler moduli $Q_i = \re^{t_i}$ $(i=1,\ldots,r+1)$,
$Q_{r+2} = Q_b = \re^{t_b}$, the perturbative partition function
$Z^{\rm pert}$ determines the leading order terms. The contributions
of $Z^{\rm pert}$ to the l.h.s\ of \eqref{eq:geom-blowup2} reads
\begin{align}
  &\ \ \ \log
  \Big(    Z^{\rm pert}(\epsilon_1,\epsilon_2-\epsilon_1)
    Z^{\rm pert}(\epsilon_1-\epsilon_2,\epsilon_2)/Z^{\rm pert}(\epsilon_1,\epsilon_2)\Big)  \nn 
  &=
    (\epsilon_1+\epsilon_2) \bigg(-\frac{1}{6}
    \sum_{i,j,k=1}^{r+2}\kappa_{ijk}R_iR_jR_k
    +\sum_{i=1}^{r+2}(b_i^{\rm GV} - b_i^{\rm NS})R_i\bigg) 
    + \sum_{k=1}^{r+2}\bigg(-\frac{1}{2}\sum_{i,j=1}^{r+2}\kappa_{ijk}R_iR_j\bigg) t_k
    \nn 
  & =: f_0(\md n) + \sum_{k=1}^{r+2} f_k(\md n) t_k \ , \label{eq:fk}
\end{align}
where we have used \eqref{eq:bp0}.  For the blowup equations to hold
at the leading order of $Q_k$, we must have
\begin{equation}
  \sum_{\md n \in \cap_{k=1}^g I_k}(-1)^{|\md n|}
  \re^{f_0(\md n)} \re^{f_k(\md n)t_k} = \Lambda(\tau,\epsilon_1,\epsilon_2) \ .
\end{equation}
Here $I_k$ is the set of integral vectors $\md n$ that minimize
$f_k(\md n)$ for true K\"ahler moduli (not mass parameters). The latter
can be written as
\begin{align}
  f_k(\md n)
  &=- \frac{1}{2}\sum_{i,j}\kappa_{ijk}
    \bigg(\sum_{\ell}C_{i\ell}n_\ell+\frac{1}{2}r_i\bigg)
    \bigg(\sum_{m}C_{jm}n_m + \frac{1}{2}r_j\bigg) \nn
  &=
    -\frac{1}{2}\sum_{\ell,m}\bigg(\sum_{i,j}\kappa_{ijk}C_{i\ell}C_{jm}\bigg)
    n_\ell n_m
    -\frac{1}{2} \sum_{\ell}
    \bigg(\sum_{i,j}\kappa_{ijk}r_i C_{j\ell}\bigg) n_\ell
    -\frac{1}{8}\sum_{i,j}\kappa_{ijk}r_ir_j \ .
\end{align}
Note that the functions $f_k(\md n)$ for $k=1,\ldots,r$ and
$f_0(\md n)$ are dependent on the $\md r$ field as well.

In order for the blowup equation to make sense, $f_k(\md n)$ as
functions of the integral vector $\md n$ must have a minimum for any
$k=1,\ldots,r+2$, which allows us to determine the valid $\md r$
fields \cite{Huang:2017mis}. We find that in the case of the $n=3,4$
theories, the valid $\md r$ fields satisfy
\begin{equation}\label{eq:rtau0}
  \sum_{i=1}^{r+1} a_i r_i = 0 \ ,
\end{equation}
in other words, the entry of the $\md r$ field corresponding to $\tau$
is zero.
With this feature we can
write down the blowup equations as identities of Jacobi forms, as we
will see in section~\ref{sc:recursion}, and proceed to prove the
blowup equations order by order in terms of the $Q_b$ expansion using
the modularity argument. We will call the $\md r$ fields satisfying
the constraint \eqref{eq:rtau0} \emph{admissible}.

Interestingly, we notice that \eqref{eq:rtau0} is equivalent to a
slightly stronger condition that $f_b(\md n) = f_{r+2}(\md n)$ in
particular has a minimum for a real vector $\md n$.\footnote{This
  condition is stronger because if a minimal real $\md n$ exists, a
  minimal integral $\md n$ must exist nearby; on the other hand, if a
  minimal integral $\md n$ exists, there can be a non-integral flat
  direction of $f_b(\md n)$.} To see this, let us
recall that $\kappa_{ijk}$ are intersection numbers of
divisors $K_i$ dual to the curve class $\Sigma_i$, satisfying
\begin{equation}
  K_i.\Sigma_j = \delta_{ij}\ .
\end{equation}
In the elliptic Calabi-Yau threefold underlying a 6d minimal SCFT with
a bulk pure gauge theory, the curves $\Sigma_i$ $(i=1,\ldots,r+2)$
have mutual intersection numbers identical to minus the Cartan matrix
of the affine lie algebra
\begin{equation}
  (\Sigma_i.\Sigma_j)_b = -A_{ij} \ ,\quad i,j=1,\ldots,r+1 \ ,
\end{equation}
(the subsript ${}_b$ means restriction to $K_b$) when restricted to
the vertical divisor $K_b$ perpendicular to $\Sigma_b$, which is in
fact the Poincar\'e dual\footnote{We understand that the Poincar\'e is
  only rigorously defined in a compact manifold, while the elliptic
  Calabi-Yau threefold here is non-compact. So we are presenting here
  an argument not a proof. We also checked the validity of
  \eqref{eq:rtau0} for the $SU(3)$, $SO(8)$ as well as some other 6d
  gauge theories.} of $\Sigma_b$. Let $D_i$ $(i=1,\ldots,r+1)$ be the
irreducible compact divisors coming from $\Sigma_i$ fibered over the
$\IP^1$ in the base. We should thus have
\begin{equation}
  D_i. K_b = \Sigma_i \ ,\quad i = 1,\ldots, r+1 \ .
\end{equation}
We note that the two sets of divisors $D_i$ and  $K_j$
should be related by the $\md C$-matrix
\begin{equation}
  D_i = \sum_{\ell=1}^{r+2} K_\ell C_{\ell i} \ ,
\end{equation}
so that\eqref{eq:C-def} still holds.

Let us come back to the discussion of the functions $f_k(\md n)$. Take
the direction $k=r+2 = b$. The coefficients can then be explicitly
evaluated
\begin{align}
  &\sum_{i,j=1}^{r+2} \kappa_{ijb}C_{i\ell}C_{jm} = D_\ell.D_m.K_b
    = (\Sigma_\ell.\Sigma_m)_b
    = -A_{\ell m} \ ,\\
  &\sum_{i,j=1}^{r+2} \kappa_{ijb}r_iC_{j\ell} = \sum_{i=1}^{r+2} r_iK_i.D_\ell.K_b
    = \sum_{i=1}^{r+2} r_i K_i.\Sigma_\ell = r_\ell
    \ ,
\end{align}
and the function $f_b(\md n)$ reads\footnote{This equation and one
  below need slight modification if the bulk gauge group is not of the
  $ADE$ type, as we will see in the companion paper that discusses
  more general cases.}
\begin{equation}\label{eq:fb}
  f_b(\md n) = \frac{1}{2}\sum_{\ell,m=1}^{r+1}A_{\ell m}n_\ell n_m -
  \frac{1}{2}\sum_{\ell=1}^{r+1}r_\ell n_\ell - \frac{1}{8} \sum_{i,j=1}^{r+2}
  \kappa_{ij,b} r_ir_j \ .
\end{equation}
If $f_b(\md n)$ can be minimized for real values of $\md n$, its
derivatives with respect to components of $\md n$ should have a common
zero. These equations are encapsulated in a single linear equation
\begin{equation}
  \md A\cdot \md n = \frac{1}{2} \md r \ .
\end{equation}
For this linear equation to have a solution, the $\md r$ field must be
annihilated by vectors in the (left) kernel of $\md A$. Since $\md A$
is the Cartan of $\widehat{\fg}$, there is only one vector,
\begin{equation}
  \md l = (a_i) \ ,\quad i =1,\ldots,r+1 \ ,
\end{equation}
which annihilates $\md A$ when multiplied from the left. We thus get
$\md l\cdot \md r = 0$, which is the condition \eqref{eq:rtau0}.

Let us now give a counting of inequivalent and admissible
$\md r$ fields. Given the condition \eqref{eq:rB}, we can paramterize
$\md r$ fields by
\begin{equation}\label{eq:rv}
  \md r = \md B+ 2 \md v \ ,\quad \md v\in \IZ^{\nC} \ ,
\end{equation}
and the equivalence condition \eqref{eq:r-equiv} is translated to
\begin{equation}\label{eq:v-equiv}
  \md v - \md v' = \md C\cdot \md n' \ .
\end{equation}
The domain of inequivalent $\md v$-vectors, defined to be the lattice
$\IZ^{\nC}$ modulo the equivalence relation \eqref{eq:v-equiv}, has
only a finite number of points along $\nD$ directions, and extends
freely along the remaining $\nC-\nD$ directions.
In practise, we can always make linear combinations of curve classes
so that the last $\nC-\nD$ rows of the intersection matrix $\md C$ are
empty, i.e.\
\begin{equation}
  \md C = \(\begin{array}{c}
    \md C_{\rm sub} \\  0
  \end{array}\) \ .
\end{equation}
and the $\nD\times\nD$ submatrix $\md C_{\rm sub}$ is of full
rank. The K\"ahler moduli of the curve classes corresponding to the
first $\nD$ rows of $\md C$ are true moduli, while the K\"ahler moduli
of those for the remaining rows are mass parameters. Inequivalent
$\md v$-vectors take the form
\begin{equation}
  \md v = (v_1,\ldots, v_g, *,*,\ldots)
\end{equation}
where $v_i, i=1,\ldots,g$ can only take a finite number of integral
values, while the remaining entries denoted by $*$ can take any value
in $\IZ$. The equivalence condition for the truncated $\md v$-vectors
defined by
\begin{equation}
  \bar{\md v} = (v_1,\ldots,v_g) \ ,
\end{equation}
reads
\begin{equation}
  \bar{\md v} - \bar{\md v}' = \md C_{\rm sub} \cdot \md n' \ .
\end{equation}
Clearly the matrix $\md C_{\rm sub}$ defines a lattice embedding
$\IZ^\nD \hookrightarrow \IZ^\nD$. Therefore the number of
inequivalent truncated $\md v$-vectors is $\det{\md C}_{\rm sub}$.

In the case of 6d minimal SCFT with a pure bulk gauge theory, there is
only one mass parameter $\tau$ corresponding to the volume of elliptic
fiber, and we have seen that the corresponding entry of $\md r$ field
must be zero. Therefore, the number of inequivalent and admissible
$\md r$ fields must be identical with that of inequivalent truncated
$\md v$-vectors, which is $\det\md C_{\rm sub}$. We mentioned in the
previous sections that $\md C$ barring the last row is identified with
the opposite of the affine Cartan matrix. Hence in practice, we can
construct the full rank square submatrix $\md C_{\rm sub}$ of $\md C$
by throwing away the row corresponding to the affine node in the
Dynkin diagram.\footnote{Note the mark associated to the affine node
  is 1. We can construct another full rank $\md C'_{\rm sub}$ by
  throwing away a different row corresponding to a different node. If
  the associated mark $a_i$ is greater than 1, the number of truncated
  $\md v$-vectors might be larger. But we cannot recover integral
  $\md r$ fields from all integral truncated $\md v$-vectors because
  of \eqref{eq:rtau0}. In the end the number of integral $\md r$
  fields is still the determinant of $\md C_{\rm sub}$ constructed
  from throwing away the affine node.} Once all the inequivalent
truncated $\md v$-vectors are found, we can convert them to $\md r$
fields with the help of \eqref{eq:rv} and \eqref{eq:rtau0}. In the 6d
$n=3,4$ minimal SCFTs, we find that all admissible $\md r$ fields give
rise to valid blowup equations.

\subsection{Recursion relations}
\label{sc:recursion}

Here we derive recursion relations of elliptic genera from the blowup
equations. Later when we discuss individual models in
sections~\ref{sc:A2} and \ref{sc:D4}, we will demonstrate the validity of
these recursion relations and then inverse the logic solving elliptic
genera from these relations.

The blowup equations for the partition function of a 6d SCFT can be
written as follows
\begin{equation}
  \sum_{\md n\in \IZ^g} A(\md t,\epsilon_1,\epsilon_2;\md n)
  \hZ^{\rm ell}(\md t+\epsilon_1 \md R,\epsilon_1,\epsilon_2-\epsilon_1)
  \hZ^{\rm ell}(\md t+\epsilon_2 \md R,\epsilon_1-\epsilon_2,\epsilon_2) =
  \Lambda(\tau,\epsilon_1,\epsilon_2)
  \hZ^{\rm ell} (\md t,\epsilon_1,\epsilon_2) \ . \label{eq:recur-0}
\end{equation}
where it is understood that everything depends on the choice of
$\md r$ field.  Here $Z^{\rm ell}$ is the generating function of
elliptic genera
\begin{equation}
  Z^{\rm ell}(\md t,\epsilon_1,\epsilon_2)
  = 1+\sum_{k=1}^\infty Q_b^k Z_k(t_\ell,\epsilon_1,\epsilon_2) \ ,
\end{equation}
where $Z_k$ is proportional to the $k$-string elliptic genus with a
model-dependent prefactor, and it only depends on K\"ahler moduli of
vertical curves $t_\ell$, $\ell=1,\ldots,r+1$. When $Z^{\rm ell}$ is
twisted with $\md t$ shifted to $\md t + \pi\ri \md B$, $Q_b$ is
multiplied with a phase $(-1)^{B_b}$, while $Z_k$ is unchanged, as
$t_\ell$ are volumes of $(-2)$ curves and thus the corresponding
entries of $\md B$ vanish according to the discussion in
Section~\ref{sc:B}.  The function $A$ is given by
\begin{equation}
  A(\md t,\epsilon_1,\epsilon_2;\md n)
  = (-1)^{|\md n|+(k_1+k_2-k)B_b} D^{\text{pert}}(\md t,\epsilon_1,\epsilon_2;\md n)
  D^{\text{1-loop}}(\md t,\epsilon_1,\epsilon_2;\md n)
\end{equation}
including the perturbative contribution
\begin{equation}
  D^{\text{pert}}(\md t,\epsilon_1,\epsilon_2;\md n) =
  Q_b^{f_b(\md n)}
  {\rm exp}\bigg[f_0(\md n)(\epsilon_1+\epsilon_2) +
    \sum_{\ell=1}^{r+1} f_\ell(\md n) t_\ell \bigg] \ ,
\end{equation}
and the one-loop contribution
\begin{equation}
  D^{\text{1-loop}}(t_\ell,\epsilon_1,\epsilon_2;\md n) =
  \frac{\Zloop(t_\ell + \epsilon_1 R_\ell,\epsilon_1, \epsilon_2 -
    \epsilon_1) \Zloop(t_\ell + \epsilon_2 R_\ell, \epsilon_1-
    \epsilon_2, \epsilon_2)}{\Zloop(t_\ell, \epsilon_1,\epsilon_2)} \ ,
\end{equation}
with $f_0, f_\ell$ defined in \eqref{eq:fk}.  We don't have to put hat
over $\Zloop$ because entries of $\md B$ associated to $t_\ell$
($\ell=1,\ldots,r+1$) are all zero. By comparing the coefficients of
$Q_b^k$ on both sides of \eqref{eq:recur-0}, we find the recursion
relation
\begin{align}
  &\Lambda(\tau,\epsilon_1,\epsilon_2)
  Z_k(t_\ell,\epsilon_1,\epsilon_2)\nn=
  &\phantom{=}\sum_{f_b(\md n)+k_1+k_2=k}
    (-1)^{|\md n|+(k_1+k_2-k)B_b} {\rm exp}\bigg[f_0(\md n)(\epsilon_1+\epsilon_2) +
    \sum_{\ell=1}^{r+1} f_\ell(\md n)t_\ell
    +(k_1\epsilon_1+k_2\epsilon_2)R_b(\md n)\bigg]
    \nn   
  &\phantom{==}\times
    D^{\text{1-loop}}(t_\ell,\epsilon_1,\epsilon_2,\md n)
    Z_{k_1}(t_\ell+\epsilon_1 R_\ell, \epsilon_1,\epsilon_2 - \epsilon_1)
    Z_{k_2}(t_\ell+\epsilon_2 R_\ell, \epsilon_1 - \epsilon_2, \epsilon_2)
    \ . \label{eq:recur-1}
\end{align}

The expression above can be simplified due to the following
observation. Given the expression \eqref{eq:fb} of $f_b(\md n)$ and
the condition \eqref{eq:rtau0} on the $\md r$ field,
it is clear that $f_b(\md n)$ is invariant under the shift
\begin{equation}\label{eq:shift}
  \md n \to \md n + m\md a  \ ,\quad m \in \IZ \ ,
\end{equation}
where $\md a = (a_k)$ the vector of marks. Similarly
\begin{equation}
  R_k = \sum_{\ell=1}^{r+1} C_{k,\ell}n_\ell + \frac{1}{2}r_k
  = -A_{k,\ell}n_\ell + \frac{1}{2}r_k
\end{equation}
for $k=1,\ldots,r+1$ is also invariant under the integral shift
\eqref{eq:shift}. Therefore, we could define representatives $\hat{\md
n}$ of $\md n$ by
\begin{equation}
  \md n = \hat{\md n} + m\md a \ , \quad m\in \IZ
\end{equation}
so that no two representatives differ by $\md a \IZ$. Then the
summation in \eqref{eq:recur-1} can be split into two steps
\begin{align}
  &\Lambda(\tau,\epsilon_1,\epsilon_2)
  Z_k(t_\ell,\epsilon_1,\epsilon_2)\nn =
  &\phantom{=}\sum_{f_b(\hat{\md n})+k_1+k_2=k}
    \bigg(\sum_{m\in \IZ} (-1)^{|\md n|+(k_1+k_2-k)B_b}
    {\rm exp}\bigg[f_0(\md n)(\epsilon_1+\epsilon_2) +
    \sum_{\ell=1}^{r+1} f_\ell(\md n)t_\ell +
    (k_1\epsilon_1+k_2\epsilon_2)R_b(\md n)\bigg] \bigg)
    \nn   
  &\phantom{=}\times D^{\text{1-loop}}(t_\ell,\epsilon_1,\epsilon_2,\hat{\md n})
    Z_{k_1}(t_\ell+\epsilon_1 R_\ell(\hat{\md n}),
    \epsilon_1,\epsilon_2 - \epsilon_1)
    Z_{k_2}(t_\ell+\epsilon_2 R_\ell(\hat{\md n}) ,
    \epsilon_1 -\epsilon_2, \epsilon_2) \ ,
    \label{eq:recur-2}
\end{align}
where the summation of $m$ gives a theta function. These are the
equations whose validity we will demonstrate order by order through a
modularity argument in the following example sections.

In the remainder of this section, we illustrate how to derive elliptic
genera from the recursion relation \eqref{eq:recur-2}. We start with
$k=0$. Given the expression \eqref{eq:fb} for $f_b(\md n)$, its value
is already non-negative since the affine Cartan matrix $A_{\ell m}$ is
positive semi-definite. If the minimal value of $f_b(\md n)$ is
greater than zero, the identity \eqref{eq:recur-2} could not hold at
$k=0$ unless $\Lambda = 0$. Thus we should have a vanishing blowup
equation. If the minimal value of $f_b(\md n)$ is zero, there is a
chance that \eqref{eq:recur-2} is satisfied at $k=0$ and we get a
non-vanishing $\Lambda$, which should result in a unity blow up
equation. 
In the latter case, using $Z_0 = 1$, we find the following
expression for $\Lambda$
\begin{equation}\label{eq:Lambda-0}
  \Lambda(\tau,\epsilon_1,\epsilon_2) =
  \sum_{\hat{\md n}\in \hat{I}_b}
  D^{\text{1-loop}}(t_\ell,\epsilon_1,\epsilon_2,\hat{\md n})
  \sum_{m\in \IZ} (-1)^{|\md n|}{\rm exp}\bigg[f_0(\md n)(\epsilon_1+\epsilon_2) +
  \sum_{\ell=1}^{r+1} f_\ell(\md n)t_\ell \bigg]  \ , 
\end{equation}
where $\hat{I}_b$ is the set of representatives $\hat{\md n}$ which
minimize $f_b$ to zero.

Let us now focus on unity blowup equations. As we will see in example
sections, the associated $\md r$ fields have zero entries except for $r_b$:
$\md r = (0,\ldots,0,r_b)$; besides, one can always choose the
representative $\hat{\md n}$ in $\hat{I}_b$ to be the null vector. As
a result, $t_\ell$ are not shifted in $Z_{k_1},Z_{k_2}$ if either
$k_1=k$ or $k_2 = k$. We can thus put the unity recursion relations
for $k\geq 1$ in a more explicit form
\begin{align}
  Z_k(t_\ell,\epsilon_1,\epsilon_2)
  &= Z_k(t_\ell, \epsilon_1,\epsilon_2- \epsilon_1)
    J_{k}^{(1)}(\tau,\epsilon_1,\epsilon_2) 
    +Z_k(t_\ell, \epsilon_1-\epsilon_2,\epsilon_2)
    J_{k}^{(2)}(\tau,\epsilon_1,\epsilon_2)
    + I_k(t_\ell,\epsilon_1,\epsilon_2) \ . \label{eq:recur-3}
\end{align}
The coefficients are
\begin{equation}
  J_{k}^{(1)}(\tau,\epsilon_1,\epsilon_2) 
  = \frac
  {\Lambda_{k}^{(1)}(\tau,\epsilon_1,\epsilon_2) }
  {\Lambda(\tau,\epsilon_1,\epsilon_2)} \ , \quad 
  J_{k}^{(2)}(\tau,\epsilon_1,\epsilon_2) = \frac
  {\Lambda_{k}^{(2)}(\tau,\epsilon_1,\epsilon_2)}
  {\Lambda(\tau,\epsilon_1,\epsilon_2)} \ .
\end{equation}
with
\begin{align}
  \Lambda_{k}^{(1)}(\tau,\epsilon_1,\epsilon_2)
  &= \sum_{\hat{\md n}\in \hat{I}_b}\sum_{m\in \IZ}
    (-1)^{|\md n|}Q_{\tau}^{f_\tau(\md n)} 
    (q_1q_2)^{f_0(\md n)}q_1^{k R_b(\md n)} \\
  \Lambda_{k}^{(2)}(\tau,\epsilon_1,\epsilon_2)
  &= \sum_{\hat{\md n}\in \hat{I}_b}\sum_{m\in \IZ}
    (-1)^{|\md n|}Q_{\tau}^{f_\tau(\md n)} 
    (q_1q_2)^{f_0(\md n)}q_2^{k R_b(\md n)} \ .
\end{align}
$I_k$ is the summation on the r.h.s.\ of \eqref{eq:recur-2} with
$k_1,k_2< k$, and thus are known data in a recursive calculation.

The relations \eqref{eq:recur-3} can be solved to give compact
expressions of $Z_k$, following the procedure in \cite{Keller:2012da}
for a similar calculation for 5d gauge theories. The key observation
is that
$Z_k(t_\ell,\epsilon_1,\epsilon_2),
Z_k(t_\ell,\epsilon_1,\epsilon_2-\epsilon_1),
Z_k(t_\ell,\epsilon_1-\epsilon_2,\epsilon_2)$ do not depend on the
choice of $\md r$ fields. If there are at least three unity $\md r$
fields, we can pick three of them, and write down three equations of
the form \eqref{eq:recur-3}, and combine them into the linear system
\begin{equation}\label{eq:Z-linear}
  \begin{pmatrix}
    -\Lambda(r_1) & \Lambda_k^{(1)}(r_1) & \Lambda_k^{(2)}(r_1) \\
    -\Lambda(r_2) & \Lambda_k^{(1)}(r_2) & \Lambda_k^{(2)}(r_2) \\
    -\Lambda(r_3) & \Lambda_k^{(1)}(r_3) & \Lambda_k^{(2)}(r_3)
  \end{pmatrix} \cdot
  \begin{pmatrix}
    Z_k(\epsilon_1,\epsilon_2) \\ Z_k(\epsilon_1,\epsilon_2 - \epsilon_2) \\
    Z_k(\epsilon_1-\epsilon_2,\epsilon_2)
  \end{pmatrix} = -
  \begin{pmatrix}
   \Lambda(r_1) I_k(r_1) \\ \Lambda(r_2) I_k(r_2) \\ \Lambda(r_3) I_k(r_3)
  \end{pmatrix} \ .
\end{equation}
Here we only write down the most important parameters each function
depends on. If the matrix $M_{\Lambda_k}$ of coefficients
$\Lambda,\Lambda_k^{(1)},\Lambda_k^{(2)}$ on the l.h.s.\ of the linear
system is of full rank, it can inverted to give us the answer of $Z_k$
in terms of $I_k$ and thus lower base degree partition functions. We
will demonsrate that this method also works for 6d theories, except
for the solution of $Z_1$ for the $SU(3)$ model.

\subsection{$SU(3)$ theory}
\label{sc:A2}

\subsubsection{Base degree zero}
\label{sc:b0-A2}

By combining \eqref{eq:C-A2} and \eqref{eq:B-A2} and following
section~\ref{sc:r}, we find all the inequivalent and admissible
$\md r$ fields. By analysing the recursion relation \eqref{eq:recur-2} at
$k=0$ we can divide the resulting blowup equations into unity and
vanishing equations, as listed in Tab.~\ref{tb:r-A2}.

\begin{table}
  \centering
  \begin{tabular}{c *{3}{>{$}c<{$}}}\toprule
    unity & (0,0,0,-1) & (0,0,0,1) & (0,0,0,3) \\\midrule
    \multirow{2}{*}{vanishing} &(2,-2,0,1) & (-2,0,2,1) & (0,2,-2,1) \\
    & (0,-2,2,1) & (2,0,-2,1) & (-2,2,0,1) \\\bottomrule
  \end{tabular}
  \caption{The list of all inequivalent and admissible $\md r$ fields for
    6d $SU(3)$ gauge theory.}\label{tb:r-A2}
\end{table}

For the unity blowup equations, at base degree $k=0$ they reduce to
the computation of $\Lambda$. For all the three $\md r$ fields
$\md r_1,\md r_2,\md r_3$ of unity blowup equations in the first row
of Tab.~\ref{tb:r-A2} there is only one $\hat{\md n} = (0,0,0)$ which
minimize $f_b(\md n)$. Then using \eqref{eq:Lambda-0} we find the
following results for $\Lambda$
\begin{align}
  \Lambda(\md r_1)
  &= \sum_{n\in\IZ} (-1)^n
    Q_\tau^{\frac{3}{2}n^2+\frac{1}{2}n+\frac{1}{24}}(q_1q_2)^{n+\frac{1}{6}}
    = \re^{-\frac{\pi\ri}{6}}\theta_4^{[\tfrac{1}{6}]}(3\tau,\epsilon_1+\epsilon_2)
  \ ,\\
  \Lambda(\md r_2)
  &= \sum_{n\in\IZ}(-1)^n
    Q_\tau^{\frac{3}{2}n^2-\frac{1}{2}n+\frac{1}{24}}(q_1q_2)^{n-\frac{1}{6}}
    = \re^{\frac{\pi\ri}{6}}\theta_4^{[-\tfrac{1}{6}]}(3\tau,\epsilon_1+\epsilon_2)
  \ , \\
  \Lambda(\md r_3)
  &= \sum_{n\in\IZ} (-1)^n
    Q_\tau^{\frac{3}{2}n^2-\tfrac{3}{2}n+\tfrac{3}{8}}(q_1 q_2)^{n-\frac{1}{2}}
    =\re^{\frac{\pi\ri}{2}}\theta_4^{[-\tfrac{1}{2}]}(3\tau,\epsilon_1+\epsilon_2)
  \ .
\end{align}
Here and later in this paper we use the following notation of Jacobi
theta functions with characteristics
\begin{align}
  &\theta_3^{[\alpha]}(\tau,z) =
    \sum_{n\in \IZ}\re^{\tfrac{1}{2}\tau(n+\alpha)^2+ z(n+\alpha)} \
    ,\\
  &\theta_4^{[\alpha]}(\tau,z) =
    \sum_{n\in \IZ}\re^{\tfrac{1}{2}\tau(n+\alpha)^2+ (z+\pi\ri)(n+\alpha)} \ .
\end{align}
Indeed all the $\Lambda$ only depend on $\tau$ and no other K\"ahler
moduli.\footnote{We notice that
  $\Lambda$ are all Jacobi forms of weight 1/2 with respect to the
  modular group acting on $\tau$. This is a subgroup of the monodromy
  group $\Gamma$ of the total modular space, and we thus seem to have
  a contradiction with the claim \cite{Huang:2017mis} that $\Lambda$
  is supposed to have weight zero with respect to $\Gamma$. To
  reconcile them, we recall that we have thrown away the contribution
  of the tensor multiplet to $Z^{\text{1-loop}}$. It is not difficult
  to verify with the help of the identities in Appendix~\ref{sc:app1}
  that if included it contributes to an additional factor
  $\eta(\tau)^{-1}$ to $\Lambda$ reducing the weight of the latter to
  zero.}

As for the vanishing blowup equations, we only have to check one of
them, as their $\md r$ fields are related to each other by $S_3$
symmetry acting on the first three entries. Consider the $\md r$ field
$(-2,2,0,1)$. There are three sets of $\md n$ which minimize
$f_b(\md n)$ represented by
\begin{equation}\label{eq:Ib-A2-vanishing}
  \hat{I}_b=\{(0,0,0), (-1,0,0), (0,1,0)\} \ .
\end{equation}
At the lowest base degree with $k_1=k_2=0$, the elliptic genera contribute
trivially with $Z_0 =1$. The recursion relation only takes
contributions from perturbative and 1-loop partition functions from
the first line of \eqref{eq:recur-2}. Summing over $\hat{\md n}$ in
\eqref{eq:Ib-A2-vanishing}, the recursion relation \eqref{eq:recur-2}
at lowest base degree reads
\begin{equation}\label{su3vanish0}
  \Theta_{(0,0,0)}\theta_{(0,0,0)}+
  \Theta_{(-1,0,0)}\theta_{(-1,0,0)}
  +\Theta_{(0,1,0)}\theta_{(0,1,0)}=0
  \ ,
\end{equation}
where $\Theta$s encapsulate contributions from perturbative partition
functions and have the form
\begin{equation}\label{id1}
\begin{aligned}
  \Theta_{(0,0,0)}&=\sum_{n=-\infty}^{\infty}
  (-1)^n Q_\tau^{\tfrac{3}{2}(n-\tfrac{1}{6})^2}
  Q_1^{n}Q_2^{-n+\tfrac{1}{3}} =
  \re^{\tfrac{\pi\ri}{6}}Q_1^{\tfrac{1}{6}}Q_2^{\tfrac{1}{6}}
  \theta^{[-\tfrac{1}{6}]}_4(3\tau,t_1-t_2)\  ,\\
  \Theta_{(-1,0,0)}&=\sum_{n=-\infty}^{\infty}
  (-1)^n Q_\tau^{\tfrac{3}{2}(n-\tfrac{1}{6})^2}
  Q_1^{-2n+1}Q_2^{-n+\tfrac{1}{3}}=
  \re^{\tfrac{\pi\ri}{6}}Q_1^{\tfrac{2}{3}}Q_2^{\tfrac{1}{6}}
  \theta^{[-\tfrac{1}{6}]}_4(3\tau,-2t_1-t_2)\  ,\\
  \Theta_{(0,1,0)}&=\sum_{n=-\infty}^{\infty}
  (-1)^n Q_\tau^{\tfrac{3}{2}(n-\tfrac{1}{6})^2}
  Q_1^{n}Q_2^{2n+\tfrac{1}{3}}=
  \re^{\tfrac{\pi\ri}{6}}Q_1^{\tfrac{1}{6}}Q_2^{\tfrac{2}{3}}
  \theta^{[-\tfrac{1}{6}]}_4(3\tau,t_1+2t_2)\  ,
\end{aligned}
\end{equation}
while $\theta$s encapsulate contributions from one-loop partition
functions and have the form
\begin{equation}\label{theta0}
\begin{aligned}
  \theta_{(0,0,0)}&=
  \PE{(Q_1+Q_2+Q_1Q_3+Q_2Q_3)\frac{1}{1-Q_\tau}}=
  -\frac{Q_1^{-\tfrac{1}{2}}Q_2^{-\tfrac{1}{2}}Q_\tau^{\tfrac{1}{6}}\eta(\tau)^2}
  {\theta_1(\tau,t_1)\theta_1(\tau,t_2)} \ ,\\
  \theta_{(-1,0,0)}&=
  \PE{(Q_1+Q_3+Q_1Q_2+Q_2Q_3)\frac{1}{1-Q_\tau}}=
  -\frac{Q_1^{-1}Q_2^{-\tfrac{1}{2}}Q_\tau^{\tfrac{1}{6}}\eta(\tau)^2}
  {\theta_1(\tau,t_1)\theta_1(\tau,t_1+t_2)}\ ,\\
  \theta_{(0,1,0)}&=
  \PE{(Q_2+Q_3+Q_1Q_2+Q_1Q_3)\frac{1}{1-Q_\tau}}=
  -\frac{Q_1^{-\tfrac{1}{2}}Q_2^{-1}Q_\tau^{\tfrac{1}{6}}\eta(\tau)^2}
  {\theta_1(\tau,t_2)\theta_1(\tau,t_1+t_2)}\ .
\end{aligned}
\end{equation}
In the derivation of these expressions, the identities in
Appendix~\ref{sc:app1} are very useful. With the reparametrisation
\begin{equation}\label{eq:vA2}
  t_1 = v_1 - v_2 \ ,\quad t_2 = v_2 - v_3
\end{equation}
subject to $v_1 + v_2 + v_3 = 0$, the identity \eqref{su3vanish0} can
be written as
\begin{equation}\label{eq:vand0-A2}
  \theta_4^{[-\tfrac{1}{6}]}(3\tau,-3v_1)\theta_1(\tau,v_2-v_3)
  +\theta_4^{[-\tfrac{1}{6}]}(3\tau,-3v_2)\theta_1(\tau,v_3-v_1)
  +\theta_4^{[-\tfrac{1}{6}]}(3\tau,-3v_3)\theta_1(\tau,v_1-v_2) = 0 \ .
\end{equation}
This last identity can be proved by noticing that each term and
therefore the total sum is a Jacobi form\footnote{Strictly
  speaking, this is a component of a vector-valued Jacobi
  form.\label{fn:vector}} for
$\Gamma(3)$ of weight 1 and index polynomial
\begin{equation}
  \frac{1}{2}(v_1^2+v_2^2+v_3^2-2v_1v_2-2v_2v_3-2v_3v_1) \ ,
\end{equation}
and by verifying that the first few terms in the $Q_\tau$ expansion
vanish, which we have checked up to very high orders.

\subsubsection{Modularity at generic base degree}
\label{sc:modularity-A2}

Here we given an argument for the validity of the recursion relation
\eqref{eq:recur-2} by demonstrating that both sides of the equation
\eqref{eq:recur-2} are multivariate meromorphic Jacobi forms of the
same weight and index polynomial at any base degree $k$. Once this is
achieved, after multiplying both side of \eqref{eq:recur-2} with the
common denominator, we get an identity of multivariate weak Jacobi
forms of the same weight and index, which can then be proved by
plugging in the expression of $Z_k$ given in \cite{Kim:2016foj} and
comparing the first few terms in the $Q_\tau$ expansion.

Consider a blowup equation with $\md r = (r_1,r_2,r_3,r_b)$ subject to
the condition $r_1+r_2+r_3 = 0$. The perturbative contribution to the
recursion \eqref{eq:recur-2} is
\begin{align}
  D^{\text{pert},'}:=
  &{\rm exp} \bigg[f_0(\md n)(\epsilon_1+\epsilon_2)
    +\sum_{\ell=1}^{r+1}f_\ell(\md n)t_\ell
    +(k_1\epsilon_1+k_2\epsilon_2)R_b(\md n)\bigg] \nn =
  &Q_1^{f_1(\md n)}Q_2^{f_2(\md n)} Q_3^{f_3(\md n)}
    (q_1q_2)^{f_0(\md n)} \Big(q_1^{k_1}q_2^{k_2}\Big)^{R_b(\md n)}
\end{align}
where
\begin{align}
  f_\ell(\md n) =
  &\frac{3}{2}\bigg(n_\ell-\frac{r_\ell+r_b}{6}\bigg)^2 \ ,\quad
    \ell=1,2,3 \\
  f_0(\md n) =
  &\frac{\sum_{\ell=1}^3 r_\ell^3}{144}
    +r_b\bigg(-\frac{1}{6}+\frac{\sum_{\ell=1}^3r_\ell^2}{48}\bigg)
    +\frac{n_1+n_2+n_3}{3}
    - \frac{\sum_{\ell=1}^3 r_\ell^2n_\ell}{8}
    -\frac{r_b}{4}\sum_{\ell=1}^3 r_\ell n_\ell \nn
  &  +\frac{3}{4}\sum_{\ell=1}^3r_\ell n_\ell^2 +
    \frac{r_b}{2}
    \bigg(\sum_{\ell=1}^3 n_\ell^2 -\sum_{1\leq \ell<m\leq 3}n_\ell n_m\bigg)
    -\frac{4}{3}\sum_{\ell=1}^3 n_\ell^3 +\frac{1}{2}\sum_{\ell\neq m}
    n_\ell^2n_m +n_1n_2n_3\
    ,\\
  R_b =
  &-n_1-n_2-n_3 + \frac{r_b}{2} \ .
\end{align}
In addition
\begin{equation}
  f_b(\md n) = \frac{1}{24}\sum_{\ell=1}^3r_\ell^2
  - \frac{1}{2}\sum_{\ell=1}^3 r_\ell n_\ell
  -\sum_{1\leq\ell<m\leq 3}n_\ell n_m + \sum_{\ell=1}^3 n_\ell^2 \ .
\end{equation}
Following the discussin in section~\ref{sc:recursion}, we can split
$\md n$ to a representative $\hat{\md n}$, which we uniquely fix by
setting $n_3 = 0$, and $(m,m,m)$, i.e.
\begin{equation}
  \md n = (n_1,n_2,0) + (m,m,m) \ .
\end{equation}
Then $D^{\text{pert},'}$ can be written as
\begin{align}
  &D^{\text{pert},'} \nn
  &=Q_\tau^{\frac{3}{2}\(m-\frac{r_3+r_b}{6}\)^2}
    \Big(Q_1^{n_1+\frac{r_3-r_1}{6}}Q_2^{n_2+\frac{r_3-r_2}{6}}\Big)^{3\(m-\frac{r_3+r_b}{6}\)}
    \(q_1 q_2\)^{\(m-\frac{r_3+r_b}{6}\)(1-3f_b(\hat{\md n}))}
    \big(q_1^{k_1}q_2^{k_2}\big)^{-3\(m-\frac{r_3+r_b}{6}\)}\nn
  &\times Q_1^{\frac{3}{2}\(n_1+\frac{r_3-r_1}{6}\)^2}
    Q_2^{\frac{3}{2}\(n_2+\frac{r_3-r_2}{6}\)^2}
    \(q_1q_2\)^{f_0(\hat{\md n})+\frac{r_3+r_b}{6}(1-3f_b(\hat{\md n}))}
    \big(q_1^{k_1}q_2^{k_2}\big)^{-n_1-n_2-\frac{r_3}{2}}
    \ . \label{eq:mod-pert-A2}
\end{align}

For the contributions from vector multiplets, using the notation
(\ref{Blfunc}), we have
\begin{align}
  D^{\text{1-loop}}=
  &\text{PE}\Big[-(
    Bl_{(0,1/2,-2n_1+n_2+\frac{r_1}{2})}(q_1,q_2)Q_1
    +Bl_{(0,1/2,2n_1-n_2+\frac{r_2+r_3}{2})}(q_1,q_2)Q_2Q_3\nn +
  &Bl_{(0,1/2,n_1-2n_2+\frac{r_2}{2})}(q_1,q_2)Q_2
    +Bl_{(0,1/2,-n_1+2n_2+\frac{r_1+r_3}{2})}(q_1,q_2)Q_1Q_3\nn +
  &Bl_{(0,1/2,n_1+n_2+\frac{r_3}{2})}(q_1,q_2)Q_3
    +Bl_{(0,1/2,-n_1-n_2+\frac{r_1+r_2}{2})}(q_1,q_2)Q_1Q_2)/(1-Q_\tau)\Big]
    \ ,
\end{align}
where we used the property that
$D^{\text{1-loop}}(\md n) = D^{\text{1-loop}}(\hat{\md n})$ and set
$n_3=0$. Then noticing $\Qtau=Q_1Q_2Q_3$ and using the notation
(\ref{brevedef}), we have
\begin{align}
  D^{\text{1-loop}}=
  &T_{-2n_1+n_2+\frac{r_1}{2}}(t_1)\,
    T_{n_1-2n_2+\frac{r_2}{2}}(t_2)\,
    T_{-n_1-n_2+\frac{r_1+r_2}{2}}(t_1+t_2)\nn =
  &(-1)^{f_b(\hat{\md n})}Q_\tau^{\frac{f_b(\hat{\md n})}{2}}
    Q_1^{-f_b(\hat{\md n})-\frac{3}{2}\(n_1+\frac{r_3-r_1}{6}\)^2}
    Q_2^{-f_b(\hat{\md n})-\frac{3}{2}\(n_2+\frac{r_3-r_2}{6}\)^2}
    \(q_1q_2\)^{d_R(\mf a_2)}\nn
  & \times\breve{\theta}_{-2n_1+n_2+\frac{r_1}{2}}(t_1)\,
    \breve{\theta}_{n_1-2n_2+\frac{r_2}{2}}(t_2)\,
    \breve{\theta}_{-n_1-n_2+\frac{r_1+r_2}{2}}(t_1+t_2) \ ,
    \label{eq:mod-1loop-A2}
\end{align}
where
\begin{align}
  d_R(\mf a_2)
  &= \frac{4n_1^3 + 4n_2^3}{3} -\frac{n_1n_2(n_1+n_2)}{2}
    - r_1\frac{5n_1^2-2n_1n_2+2n_2^2}{4}
    - r_2\frac{2n_1^2-2n_1n_2+5n_2^2}{4}\nn
  & -\frac{n_1+n_2}{3} + \frac{3r_1^2n_1+3r_2^2n_2}{8}
    + \frac{r_1r_2(n_1+n_2)}{4}
    +\frac{r_1+r_2}{6} -\frac{r_1^3+r_2^3}{24}
    -\frac{r_1r_2(r_1+r_2)}{16} \ .
\end{align}
In Appenidx \ref{sc:app1}, we show $\breve{\theta}_R(t)$ is a
meromorphic Jacobi form of weight 0, and thus so is
$D^{\text{1-loop}}$ up to the prefactor.

Finally, given the relation between $Z_k$ and the $k$-string elliptic
genus $\IE_k$ for the $SU(3)$ theory
\begin{align}
  Z_k(t_{\ell},\epsilon_1,\epsilon_2) =
  \bigg(\frac{Q_\tau^{1/2}}{Q_1Q_2}\bigg)^k
  \IE_k(t_{\ell},\epsilon_1,\epsilon_2) \ .
\end{align}
we also have
\begin{equation}\label{eq:mod-ell-A2}
  \begin{aligned}
    Z_{k_1}(t_\ell +\epsilon_1R_\ell,\epsilon_1,\epsilon_2-\epsilon_1)
    =
    &q_1^{k_1\(n_1+n_2-\frac{r_1+r_2}{2}\)}\bigg(\frac{Q_\tau^{1/2}}{Q_1Q_2}\bigg)^{k_1}
    \IE_{k_1}(t_\ell+\epsilon_1R_\ell,
    \epsilon_1,\epsilon_2-\epsilon_1)
    \ ,\\
    Z_{k_2}(t_\ell +\epsilon_2R_\ell,\epsilon_1-\epsilon_2,\epsilon_2)
    =
    &q_2^{k_2\(n_1+n_2-\frac{r_1+r_2}{2}\)}\bigg(\frac{Q_\tau^{1/2}}{Q_1Q_2}\bigg)^{k_2}
    \IE_{k_2}(t_\ell+\epsilon_2R_\ell,
    \epsilon_1-\epsilon_2,\epsilon_2) \ ,
  \end{aligned}
\end{equation}
for the last two factors.

Combining
\eqref{eq:mod-pert-A2}, \eqref{eq:mod-pert-A2} and \eqref{eq:mod-pert-A2}
all together, we find the following expression for the r.h.s.\ of the
recursion relation \eqref{eq:recur-2} in terms of (meromorphic) Jacobi
forms
\begin{align}
  \text{r.h.s.}
  &=\re^{\frac{\pi\ri (r_3+r_b)}{6}}\bigg(\frac{Q_\tau^{1/2}}{Q_1Q_2}\bigg)^k
    \sum_{f_b(\hat{\md n})+k_1+k_2=k}
    (-1)^{n_1+n_2} \nn
  &\times \theta_4^{\[-(r_3+r_b)/6\]}
    \(3\tau,3\((n_1+\tfrac{r_3-r_1}{6})t_1+(n_2+\tfrac{r_3-r_2}{6})t_2\)
    +(1-3(k-k_2))\epsilon_1+(1-3(k-k_1))\epsilon_2\)\nn
  &\times \breve{\theta}_{-2n_1+n_2+\frac{r_1}{2}}(t_1)
    \breve{\theta}_{n_1-2n_2+\frac{r_2}{2}}(t_2)
    \breve{\theta}_{-n_1-n_2+\frac{r_1+r_2}{2}}(t_1+t_2)\nn
  &\times \IE_{k_1}(t_\ell+\epsilon_1R_\ell,\epsilon_1,\epsilon_2-\epsilon_1)
    \IE_{k_2}(t_\ell+\epsilon_2R_\ell,\epsilon_1-\epsilon_2,\epsilon_2)
    \ . \label{eq:rhs-urec-A2}
\end{align}
Using the fact that the elliptic genus
$\IE_k(t_\ell,\epsilon_1,\epsilon_2)$ is a meromorphic Jacobi form of
weight zero and index polynomial
\cite{DelZotto:2016pvm,DelZotto:2017mee}
\begin{equation}
  \text{Ind}(\IE_k)=
  -\frac{k}{2}(\epsilon_1^2+\epsilon_2^2)+\frac{3k(k-1)}{2}\epsilon_1\epsilon_2
  - \frac{3k}{2}(\md a,\md a)_{\mf a_2}\ ,
\end{equation}
where
\begin{equation}
  (\md a,\md a)_{\mf a_2} = \frac{2}{3}(t_1^2 +t_1t_2 + t_2^2) =
  v_1^2+v_2^2+v_3^2 \ ,
\end{equation}
It is not difficult to show that up to the common prefactor every term
in \eqref{eq:rhs-urec-A2} is a meromorphic Jacobi
form\footnote{The same as footnote \ref{fn:vector}.} for
$\Gamma(3)$ of weight 1/2 and index polynomial
\begin{equation}\label{eq:ind-A2}
  \text{Ind}(\text{r.h.s.}) = \frac{3k-1}{6}(\epsilon_1^2+\epsilon_2^2)
  +\frac{(3k-2)(3k-1)}{6} \epsilon_1\epsilon_2
  -\frac{3k}{2}(\md a,\md a)_{\mf a_2} \ ,
\end{equation}
which is indepdent of the summation indices $n_1,n_2,k_1,k_2$, and
thus so is the total sum.

On the other hand, if the blowup equation is of
vanishing type, the l.h.s.\ of \eqref{eq:recur-2} vanish; if the
blowup equation is of unity type, we have $\md r=(0,0,0,r_b)$, and
after plugging in the expression of $\Lambda$, we find the l.h.s.\ of
\eqref{eq:recur-2} to be
\begin{equation}
  \text{l.h.s.}
  = \re^{\frac{\pi\ri r_b}{6}}\bigg(\frac{Q_\tau^{1/2}}{Q_1Q_2}\bigg)^k
  \theta_4^{[-\frac{r_b}{6}]}(3\tau,\epsilon_1+\epsilon_2)
  \IE_k(t_\ell,\epsilon_1,\epsilon_2) \ , \label{eq:lhs-urec-A2}
\end{equation}
which is also a meromorphic Jacobi form of the same weight and the
same index \eqref{eq:ind-A2} up to the same prefactor. In both cases,
after multiplied with a common denominator, the recursion relations
\eqref{eq:recur-2} can be cast as identities of (weak) Weyl invariant
Jacobi forms of identical weights and indices. As the ring of Jacobi
forms is finitely generated, these identities can be proved by
checking that when the \emph{correct} $\md r$ (Tab.~\ref{tb:r-A2}) are
plugged in the first few terms in $Q_\tau$ expansion are correct.
For instance when $k=0$ we find \eqref{eq:rhs-urec-A2}
indeed reduces to the computation of $\Lambda$ in the unity cases and
the identity \eqref{eq:vand0-A2} in the vanishing cases. When $k=1$
and with a unity $\md r$ plugged in, the recursion relations reduce to
\begin{align}
  \theta_4^{[-\tfrac{r_b}{6}]}&(3\tau,-2\eq+\et)\IE_1(\md v,\eq,\et-\eq)
  +\theta_4^{[-\tfrac{r_b}{6}]}(3\tau,\eq-2\et)\IE_1(\md v,\eq-\et,\et)
  \nn - \,
  &\theta_4^{[-\tfrac{r_b}{6}]}(3\tau,\eq+\et)\IE_1(\md v,\eq,\et)
    +I_1^{[-\tfrac{r_b}{6}]}(\eq,\et)
   = 0
  \ , \label{eq:ureck1-A2}
\end{align}
in which
\begin{equation}
  I_1^{[-\tfrac{r_b}{6}]}(\eq,\et)=
  -\sum_{i\neq j\neq k}
  \frac{\theta_4^{[-\tfrac{r_b}{6}]}(3\tau,3v_{ij}-2\eq-2\et)\,\eta^6}
  {\theta_1(v_{ij})\theta_1(v_{ij}-\eq)\theta_1(v_{ij}-\et)
    \theta_1(v_{ij}-\eq-\et)\theta_1(v_{ik})\theta_1(v_{jk})}
  \ .
\end{equation}
We have used the variables $\md v=(v_1,v_2,v_3)$ for K\"ahler moduli
defined in \eqref{eq:vA2} to make the Weyl symmetry of $SU(3)$ more
transparent, and we define $v_{ij} = v_i-v_j$. Here and in the
following we suppress the modular parameter of a theta function if it
is simply $\tau$. One can readily verify by using the expression
(3.51) of $\IE_1$ in \cite{Kim:2016foj} and by the first few terms in
the $Q_\tau$ expansion that \eqref{eq:ureck1-A2} is valid if $r_b$ is
odd but not if $r_b$ is even. Similarly when the vanishing
$\md r = (-2,2,0,1)$ is plugged in, we find the identity
\begin{align}
  &\sum_{i=1}^3
    \frac{\theta_4^{[-\frac{1}{6}]}(3\tau,-3v_i-3\eq)\, \eta^2}
    {\prod_{j\neq i}\theta_1(v_{ij})}
    \IE_{1}({\md v}+\eq{\md R_v^{\{i\}}},\eq,\et-\eq)\nn +
  &\sum_{i=1}^3
    \frac{\theta_4^{[-\frac{1}{6}]}(3\tau,-3v_i-3\et)\, \eta^2}
    {\prod_{j\neq i}\theta_1(v_{ij})}
    \IE_{1}({\md v}+\et{\md R_v^{\{i\}}},\eq-\et,\et)\nn -
  &\sum_{i=1}^3
    \frac{\theta_4^{[-\frac{1}{6}]}(3\tau,6v_i-3\eq-3\et)\,\eta^8}
    {\prod_{j\neq i}
    \theta_1(v_{ij})\theta_1(v_{ij}-\eq)\theta_1(v_{ij}-\et)\theta_1(v_{ij}-\eq-\et)}
    = 0
    , \label{eq:Z1con}
\end{align}
with the $\md R$ fields
\begin{equation}\label{Rv1}
  \ba
  \md {R}_v^{\{1\}}&=\Big(\ \frac{2}{3},-\frac{1}{3},-\frac{1}{3}\Big)\ ,\\
  \md {R}_v^{\{2\}}&=\Big(-\frac{1}{3},\frac{2}{3},-\frac{1}{3}\Big)\ ,\\
  \md {R}_v^{\{3\}}&=\Big(-\frac{1}{3},-\frac{1}{3},\frac{2}{3}\Big)\ ,
  \ea
\end{equation}
which can also be explicitly verified. The cases of the other
vanishing $\md r$ fields can be obtained by permutations of the
components $r_1,r_2,r_3$ of the $\md r$ field. We have also checked
the cases of $k=2$. Identities for $k > 2$ can be checked in a similar
manner.

\subsubsection{Recursion formula for elliptic genera}
\label{sc:recursion-A2}

We would like to inverse the logic and illustrate here that it is
possible to solve $\IE_k$ $(k\geq 2)$ for the $SU(3)$ theory from the
recursion relations \eqref{eq:recur-3} following the argument at the
end of section~\ref{sc:recursion}.  In the case of 6d $SU(3)$ theory,
there are three unity $\md r$ fields. The corresponding
$\Lambda_k^{(1)},\Lambda_k^{(2)}$ are
\begin{itemize}
\item $\md r_1 = (0,0,0,-1)$
\begin{align}
  \Lambda_k^{(1)}(\md r_1)
  &= \sum_{n\in\IZ} (-1)^n
    Q_\tau^{\frac{3}{2}n^2+\frac{1}{2}n+\frac{1}{24}}(q_1^{-3k+1}q_2)^{n+\frac{1}{6}}
    = \re^{-\frac{\pi\ri}{6}}\theta_4^{[\tfrac{1}{6}]}(3\tau,-(3k-1)\epsilon_1+\epsilon_2)
    \ ,\nn
  \Lambda_k^{(2)}(\md r_1)
  &= \sum_{n\in\IZ} (-1)^n
    Q_\tau^{\frac{3}{2}n^2+\frac{1}{2}n+\frac{1}{24}}(q_1q_2^{-3k+1})^{n+\frac{1}{6}}
    = \re^{-\frac{\pi\ri}{6}}\theta_4^{[\tfrac{1}{6}]}(3\tau,\epsilon_1-(3k-1)\epsilon_2)
    \ .
\end{align}
\item $\md r_2 = (0,0,0,1)$
\begin{align}
  \Lambda_k^{(1)}(\md r_2)
  &= \sum_{n\in\IZ} (-1)^n
    Q_\tau^{\frac{3}{2}n^2-\frac{1}{2}n+\frac{1}{24}}(q_1^{-3k+1}q_2)^{n-\frac{1}{6}}
    = \re^{\frac{\pi\ri}{6}}\theta_4^{[-\tfrac{1}{6}]}(3\tau,-(3k-1)\epsilon_1+\epsilon_2)
    \ , \nn
  \Lambda_k^{(2)}(\md r_2)
  &= \sum_{n\in\IZ} (-1)^n
    Q_\tau^{\frac{3}{2}n^2-\frac{1}{2}n+\frac{1}{24}}(q_1 q_2^{-3k+1})^{n-\frac{1}{6}}
    = \re^{\frac{\pi\ri}{6}}\theta_4^{[-\tfrac{1}{6}]}(3\tau,\epsilon_1-(3k-1)\epsilon_2)
    \ .
\end{align}
\item $\md r_3 = (0,0,0,3)$
\begin{align}
  \Lambda_k^{(1)}(\md r_3)
  &= \sum_{n\in\IZ} (-1)^n
    Q_\tau^{\frac{3}{2}n^2-\tfrac{3}{2}n+\tfrac{3}{8}}(q_1^{-3k+1} q_2)^{n-\frac{1}{2}}
    =\re^{\frac{\pi\ri}{2}} \theta_4^{[-\tfrac{1}{2}]}(3\tau,-(3k-1)\epsilon_1+\epsilon_2)
    \ ,\nn
  \Lambda_k^{(2)}(\md r_3)
  &= \sum_{n\in\IZ} (-1)^n
    Q_\tau^{\frac{3}{2}n^2-\tfrac{3}{2}n+\tfrac{3}{8}}(q_1 q_2^{-3k+1})^{n-\frac{1}{2}}
    =\re^{\frac{\pi\ri}{2}} \theta_4^{[-\tfrac{1}{2}]}(3\tau,\epsilon_1-(3k-1)\epsilon_2)
    \ .
\end{align}
\end{itemize}
Surprisingly, we find that at base degree one the matrix $M_{\Lambda_1}$ is
actually not of full-rank. Therefore one cannot invert $M_{\Lambda_1}$
to solve $Z_1$ from the recursion relation.

That $\det M_{\Lambda_1} = 0$ may have something to do with the
curious coincidence that while the characteristics of the theta
functions enjoy a $\IZ_3$ symmetry connected to the gauge group
$SU(3)$, the elliptic parameters of theta functions enjoy some $S_3$
symmetry. Note that the three types of theta functions
\begin{equation}
  \theta_4^{[\tfrac{1}{6}]} \ ,\quad
  \theta_4^{[-\tfrac{1}{6}]} \ ,\quad
  \theta_4^{[-\tfrac{1}{2}]} \ .
\end{equation}
are invariant under the shift of the upper characteristic
$\alpha \to \alpha -1/3$ because
$\theta_4^{[\alpha]} =
\theta_4^{[\alpha+1]}$. On the other
hand, the three elliptic parameters for each theta function
\begin{equation}
  \epsilon_1+\epsilon_2 \ ,\quad -2\epsilon_1+\epsilon_2
  \ ,\quad \epsilon_1-2\epsilon_2
\end{equation}
sum up to zero, and enjoy a $S_3$ symmetry.

On the other hand, $\det M_{\Lambda_k}$ does not vanish
at base degrees $k>1$. For instance the leading order contribution in
the $Q_\tau$ expansion is
\begin{align}
  \det M_{\Lambda_k} =
  &-(q_1q_2)^{-\frac{3k+1}{2}}
    ((q_1q_2)^{\frac{k}{2}}- (q_1q_2)^{\frac{1}{2}})
    ((q_1q_2)^{\frac{k}{2}}+(q_1q_2)^{\frac{1}{2}})\nn
  &\times(q_1^k+q_2^k +q_1^{2k}+q_2^{2k}+q_1^{2k}q_2^k+q_1^kq_2^{2k})
    Q_\tau^{11/24} + \cO(Q_\tau^{11/24+1}) \ .
\end{align}
We can thus obtain compact expressions of $Z_k$ from the
linear equation \eqref{eq:Z-linear} by inverting $M_{\Lambda_k}$. We
do not give explicit formulas for $Z_k$ here as they are quite
lengthy, and the results of $Z_k$ are already well known
in the literature \cite{Kim:2016foj,Gu:2017ccq}. Instead we will
compute and list the BPS invariants in the next subsection, which also
serves as another check on the blowup equations.

\subsubsection{Solving refined BPS invariants}
\label{sc:BPS-A2}

Among the recursion relations those from unity blowup equations
\eqref{eq:recur-3} are most useful as it is rather easy to solve them
and obtain compact formulas of ellptic genera; recursion relations
from vanishing blowup equations are rather complicated and it is
difficult to get headway with them.

Another way to solve the blowup equations \eqref{eq:geom-blowup} is to
expand them in terms of all K\"ahler moduli $Q_i$ and extract
equations of refined BPS invariants. There are two advantages to this
method: one can equally easily extract equations from vanishing blowup
equations and thus increase the number of available constraint
equations; one can in fact start without the input of
$Z^{\text{1-loop}}$ but with only the truly perturbative data: the
$\md C$-matrix, the $\md B$-field, and $Z^{\rm pert}$.

We have succeeded to exploit this method to great effect. We have used
the equations extracted from the blowup equations associated to the
following $\md r$ fields
\begin{equation}
  (0,0,0,1),(0,0,0,3),(-2,2,0,1) + \text{permutations of}\; r_1,r_2,r_3
  \ ,
\end{equation}
and computed the BPS invariants up to total degree
$d_1+d_2+d_3+d_b = 7$. They are listed in Tab.~\ref{tb:A2-BPS}. These
BPS invariants respect the permuation symmetry of the degrees
$d_1,d_2,d_3$. Therefore we only list the non-vanishing invariants
with $d_1\leq d_2 \leq d_3$ and omit the invariants which can be
obtained by permuting $d_1,d_2,d_3$. All the other curve classes which
are not listed have vanishing BPS invariants. These invariants agree
with the results in the existing literature. In this way we have not
only demonstrated the validity of the generalized blowup equations but
also shown the power of the blowup equations as a computational
tool. We expect that BPS invariants of higher degrees can also be
computed with enough time.

\subsection{$SO(8)$ theory}
\label{sc:D4}

\subsubsection{Base degree zero}
\label{sc:b0-D4}

Following the same analysis as in the $SU(3)$ theory, we can find all
the inequivalent and admissible $\md r$ fields and divide the
resulting blowup equations into unity and vanishing equations. The
results are listed in Tab.~\ref{tb:r-D4}.

\begin{table}
  \centering
  \begin{tabular}{c *{4}{>{$}c<{$}}}\toprule
    unity & (0,0,0,0,0,-2) & (0,0,0,0,0,0) & (0,0,0,0,0,2) & (0,0,0,0,0,4) \\\midrule
    \multirow{4}{*}{vanishing} & (-2,0,0,2,0,0) & (0,-2,0,2,0,0) & (0,0,-2,2,0,0) &\\
          & (-2,2,0,0,0,0) & (-2,0,2,0,0,0) & (0,-2,2,0,0,0) &\\
          & (-2,-2,0,0,2,2) & (-2,0,-2,0,2,2) &  (0,-2,-2,0,2,2) & \\
          & (-2,0,0,-2,2,2) & (0,-2,0,-2,2,2) & (0,0,-2,-2,2,2) & \\\bottomrule
  \end{tabular}
  \caption{The list of all inequivalent and admissible $\md r$ fields for
    6d $SO(8)$ gauge theory.}\label{tb:r-D4}
\end{table}

For the unity blowup equations, at base degree $k=0$ they reduce to the
computation of $\Lambda$. We find the following results for the four
$\md r$ fields in the first row of Tab.~\ref{tb:r-D4}
\begin{align}
  \Lambda(\md r_1)
  &= \sum_{n\in\IZ} 
    Q_\tau^{2n^2+n+\tfrac{1}{8}}(q_1 q_2)^{2n+\frac{1}{2}}
    =\theta_3^{[\tfrac{1}{4}]}(4\tau,2\epsilon_1+2\epsilon_2)
    \ ,\\
  \Lambda(\md r_2)
  &= \sum_{n\in\IZ} 
    Q_\tau^{2n^2}(q_1 q_2)^{2n}
    =\theta_3^{[0]}(4\tau,2\epsilon_1+2\epsilon_2)
    \ ,\\
  \Lambda(\md r_3)
  &= \sum_{n\in\IZ} 
    Q_\tau^{2n^2-n+\tfrac{1}{8}}(q_1 q_2)^{2n-\frac{1}{2}}
    =\theta_3^{[-\tfrac{1}{4}]}(4\tau,2\epsilon_1+2\epsilon_2)
    \ ,\\
  \Lambda(\md r_4)
  &= \sum_{n\in\IZ} 
    Q_\tau^{2n^2-2n+\tfrac{1}{2}}(q_1 q_2)^{2n-1}
    =\theta_3^{[-\tfrac{1}{2}]}(4\tau,2\epsilon_1+2\epsilon_2)
    \ .
\end{align}
Notice that indeed all the $\Lambda$ only depend on $\tau$ and no
other K\"ahler moduli.


As for the vanishing blowup equations, we check two of them with
$\md r$ fields
\begin{equation}
  (-2,2,0,0,0,0) \ ,\quad (-2,-2,0,0,2,2)
\end{equation}
while the other $\md r$ fields could be obtained by acting $S_4$ symmetry
on the first four entries.

In the case of $\md r = (-2,2,0,0,0,0)$, there are eight sets of $\md
n$ which minimize $f_b(\md n)$ and they are represented by
\begin{align}
  \hat{I}_b = \{
  &(-1,0,0,0,0), (0,1,0,0,0), (-1,1,0,0,0), (0,0,0,0,0),\nn
  &(0,1,0,0,1),(0,1,1,1,1),(0,1,0,1,1),(0,1,1,0,1)\} \ .
\end{align}
Summing over them, the lowest order blowup equation is
\begin{equation}
  \sum_{\hat{\md n}\in \hat{I}_b}(-1)^{|\md n|}\Theta_{\md n}
  \theta_{\md n} = 0
\end{equation}
where
\begin{equation}
  \begin{aligned}
    \Theta_{(-1,0,0,0,0)} = \Theta_{(0,1,0,0,0)} = &\sum_{n\in\IZ}
    Q_\tau^{2n^2}Q_1^{2n+\tfrac{1}{2}}Q_2^{2n+\tfrac{1}{2}} =
    Q_1^{\tfrac{1}{2}}Q_2^{\tfrac{1}{2}}\theta_3(4\tau,2t_1+2t_2) \ ,\\
    \Theta_{(-1,1,0,0,0)}=\Theta_{(0,0,0,0,0)} = &\sum_{n\in\IZ}
    Q_\tau^{2n^2}Q_1^{2n+\tfrac{1}{2}}Q_2^{-2n+\tfrac{1}{2}} =
    Q_1^{\tfrac{1}{2}}Q_2^{\tfrac{1}{2}}\theta_3(4\tau,2t_1-2t_2) \ ,\\
    \Theta_{(0,1,0,0,1)}=\Theta_{(0,1,1,1,1)} = &\sum_{n\in\IZ}
    Q_\tau^{2(n+\tfrac{1}{2})^2}Q_3^{2n+\tfrac{3}{2}}Q_4^{2n+\tfrac{3}{2}}
    =
    Q_3^{\tfrac{1}{2}}Q_4^{\tfrac{1}{2}}\theta_3^{[\tfrac{1}{2}]}(4\tau,2t_3+2t_4)
    \  ,\\
    \Theta_{(0,1,0,1,1)}=\Theta_{(0,1,1,0,1)} = &\sum_{n\in\IZ}
    Q_\tau^{2(n+\tfrac{1}{2})^2}Q_3^{2n+\tfrac{3}{2}}Q_4^{-2n-\tfrac{1}{2}}
    =
    Q_3^{\tfrac{1}{2}}Q_4^{\tfrac{1}{2}}\theta_3^{[\tfrac{1}{2}]}(4\tau,2t_3-2t_4)
    \ .
  \end{aligned}
\end{equation}
and
\begin{equation}
\begin{aligned}
  \theta_{(-1,0,0,0,0)} &= \theta_{(0,1,0,0,0)}\\ =
  &-\frac{Q_1^{-\tfrac{1}{2}}Q_2^{-\tfrac{1}{2}}\eta^6}
  {\theta_1(t_1)\theta_1(t_2)\theta_4(\tfrac{-t_1-t_2-t_3-t_4}{2})
    \theta_4(\tfrac{-t_1-t_2+t_3-t_4}{2})
    \theta_4(\tfrac{-t_1-t_2-t_3+t_4}{2})
    \theta_4(\tfrac{+t_1+t_2-t_3-t_4}{2})} \ ,\\
  \theta_{(-1,1,0,0,0)} &= \theta_{(0,0,0,0,0)}\\ =
  &-\frac{Q_1^{-\tfrac{1}{2}}Q_2^{-\tfrac{1}{2}}\eta^6}
  {\theta_1(t_1)\theta_1(t_2)\theta_4(\tfrac{+t_1-t_2-t_3-t_4}{2})
    \theta_4(\tfrac{-t_1+t_2-t_3-t_4}{2})
    \theta_4(\tfrac{+t_1-t_2+t_3-t_4}{2})
    \theta_4(\tfrac{+t_1-t_2-t_3+t_4}{2})} \ ,\\
  \theta_{(0,1,0,0,1)} &= \theta_{(0,1,1,1,1)}\\ =
  &-\frac{Q_3^{-\tfrac{1}{2}}Q_4^{-\tfrac{1}{2}}\eta^6}
  {\theta_1(t_3)\theta_1(t_4)\theta_4(\tfrac{-t_1-t_2-t_3-t_4}{2})
    \theta_4(\tfrac{+t_1-t_2-t_3-t_4}{2})
    \theta_4(\tfrac{-t_1+t_2-t_3-t_4}{2})
    \theta_4(\tfrac{+t_1+t_2-t_3-t_4}{2})} \ ,\\
  \theta_{(0,1,0,1,1)} &= \theta_{(0,1,1,0,1)}\\ =
  &-\frac{Q_3^{-\tfrac{1}{2}}Q_4^{-\tfrac{1}{2}}\eta^6}
  {\theta_1(t_3)\theta_1(t_4)\theta_4(\tfrac{-t_1-t_2+t_3-t_4}{2})
    \theta_4(\tfrac{-t_1-t_2-t_3+t_4}{2})
    \theta_4(\tfrac{+t_1-t_2+t_3-t_4}{2})
    \theta_4(\tfrac{+t_1-t_2-t_3+t_4}{2})} \ .
\end{aligned}
\end{equation}
It is equivalent to the identity
\begin{align}
  0 = -
  &\theta_3(4\tau,2t_1+2t_2)\theta_1(t_3)\theta_1(t_4)
    \theta_4(\tfrac{+t_1-t_2-t_3-t_4}{2})
    \theta_4(\tfrac{-t_1+t_2-t_3-t_4}{2})
    \theta_4(\tfrac{+t_1-t_2+t_3-t_4}{2})
    \theta_4(\tfrac{+t_1-t_2-t_3+t_4}{2})\nn + 
  &\theta_3(4\tau,2t_1-2t_2)\theta_1(t_3)\theta_1(t_4)
    \theta_4(\tfrac{-t_1-t_2-t_3-t_4}{2})
    \theta_4(\tfrac{-t_1-t_2+t_3-t_4}{2})
    \theta_4(\tfrac{-t_1-t_2-t_3+t_4}{2})
    \theta_4(\tfrac{+t_1+t_2-t_3-t_4}{2})\nn + 
  &\theta_3^{[1/2]}(4\tau,2t_3+2t_4)\theta_1(t_1)\theta_1(t_2)
    \theta_4(\tfrac{-t_1-t_2+t_3-t_4}{2})
    \theta_4(\tfrac{-t_1-t_2-t_3+t_4}{2})
    \theta_4(\tfrac{+t_1-t_2+t_3-t_4}{2})
    \theta_4(\tfrac{+t_1-t_2-t_3+t_4}{2})\nn - 
  &\theta_3^{[1/2]}(4\tau,2t_3-2t_4)\theta_1(t_1)\theta_1(t_2)
    \theta_4(\tfrac{-t_1-t_2-t_3-t_4}{2})
    \theta_4(\tfrac{+t_1-t_2-t_3-t_4}{2})
    \theta_4(\tfrac{-t_1+t_2-t_3-t_4}{2})
    \theta_4(\tfrac{+t_1+t_2-t_3-t_4}{2})
    \ .  \label{eq:vand0-D4-1}
\end{align}
It can be proved by noticing that each summand and thus the total sum
is a Jacobi form for $\Gamma(4)$ of weight
$7/2$ and index polynomial
\begin{equation}
  t_1^2+t_2^2+t_3^2+t_4^2 \ ,
\end{equation}
and that the first few terms in $Q_\tau$ expansion vanish,
which we checked up to very high orders.

In the case of $\md r = (-2,-2,0,0,2,2)$, there are eight sets of $\md
n$ which minimize $f_b(\md n)$ and they are represented by
\begin{align}
  \hat{I}_b = \{
  &(-1,-1,0,0,0),(0,0,0,0,0),(-1,0,0,0,0),(0,-1,0,0,0),\nn
  &(0,0,1,1,1),(0,0,0,0,1),(0,0,1,0,1),(0,0,0,1,1)\} \ .
\end{align}
Summing over them, we get for the lowest order blowup equation
\begin{equation}
  \sum_{\hat{\md n}\in \hat{I}_b}(-1)^{|\md n|}\Theta_{\md n}
  \theta_{\md n} = 0
\end{equation}
where
\begin{equation}
  \begin{array}{ll}
    \Theta_{(-1,-1,0,0,0)} =
    Q_1^{\tfrac{1}{2}}Q_2^{\tfrac{1}{2}}\theta_3^{[\tfrac{1}{4}]}(4\tau,2t_1+2t_2)
    \ , \quad
    &\Theta_{(0,0,0,0,0)}=
      Q_1^{\tfrac{1}{2}}Q_2^{\tfrac{1}{2}}\theta_3^{[-\tfrac{1}{4}]}(4\tau,2t_1+2t_2)
      \ ,\\
    \Theta_{(-1,0,0,0,0)}  =
    Q_1^{\tfrac{1}{2}}Q_2^{\tfrac{1}{2}}\theta_3^{[\tfrac{1}{4}]}(4\tau,2t_1-2t_2)
    \ , \quad
    &\Theta_{(0,-1,0,0,0)} =
      Q_1^{\tfrac{1}{2}}Q_2^{\tfrac{1}{2}}\theta_3^{[-\tfrac{1}{4}]}(4\tau,2t_1-2t_2)
      \ ,\\
    \Theta_{(0,0,1,1,1)} = 
    Q_3^{\tfrac{1}{2}}Q_4^{\tfrac{1}{2}}\theta_3^{[\tfrac{1}{4}]}(4\tau,2t_3+2t_4)
    \ , \quad
    &\Theta_{(0,0,0,0,1)} = 
      Q_3^{\tfrac{1}{2}}Q_4^{\tfrac{1}{2}}\theta_3^{[-\tfrac{1}{4}]}(4\tau,2t_3+2t_4)
      \ ,\\
    \Theta_{(0,0,1,0,1)} = 
    Q_3^{\tfrac{1}{2}}Q_4^{\tfrac{1}{2}}\theta_3^{[\tfrac{1}{4}]}(4\tau,2t_3-2t_4)
    \ , \quad
    &\Theta_{(0,0,0,1,1)} =
      Q_3^{\tfrac{1}{2}}Q_4^{\tfrac{1}{2}}\theta_3^{[-\tfrac{1}{4}]}(4\tau,2t_3-2t_4)
      \ .
  \end{array}
\end{equation}
and
\begin{equation}
\begin{aligned}
  \theta_{(-1,-1,0,0,0)} &= \theta_{(0,0,0,0,0)}\\ =
  &-\frac{Q_1^{-\tfrac{1}{2}}Q_2^{-\tfrac{1}{2}}\eta^6}
  {\theta_1(t_1)\theta_1(t_2)\theta_4(\tfrac{-t_1-t_2-t_3-t_4}{2})
    \theta_4(\tfrac{-t_1-t_2+t_3-t_4}{2})
    \theta_4(\tfrac{-t_1-t_2-t_3+t_4}{2})
    \theta_4(\tfrac{+t_1+t_2-t_3-t_4}{2})} \ ,\\
  \theta_{(-1,0,0,0,0)} &= \theta_{(0,-1,0,0,0)}\\ =
  &-\frac{Q_1^{-\tfrac{1}{2}}Q_2^{-\tfrac{1}{2}}\eta^6}
  {\theta_1(t_1)\theta_1(t_2)\theta_4(\tfrac{+t_1-t_2-t_3-t_4}{2})
    \theta_4(\tfrac{-t_1+t_2-t_3-t_4}{2})
    \theta_4(\tfrac{+t_1-t_2-t_3+t_4}{2})
    \theta_4(\tfrac{+t_1-t_2+t_3-t_4}{2})} \ ,\\
  \theta_{(0,0,1,1,1)} &= \theta_{(0,0,0,0,1)}\\ =
  &-\frac{Q_3^{-\tfrac{1}{2}}Q_4^{-\tfrac{1}{2}}\eta^6}
  {\theta_1(t_3)\theta_1(t_4)\theta_4(\tfrac{-t_1-t_2-t_3-t_4}{2})
    \theta_4(\tfrac{+t_1-t_2-t_3-t_4}{2})
    \theta_4(\tfrac{-t_1+t_2-t_3-t_4}{2})
    \theta_4(\tfrac{+t_1+t_2-t_3-t_4}{2})} \ ,\\
  \theta_{(0,0,1,0,1)} &= \theta_{(0,0,0,1,1)}\\ =
  &-\frac{Q_3^{-\tfrac{1}{2}}Q_4^{-\tfrac{1}{2}}\eta^6}
  {\theta_1(t_3)\theta_1(t_4)\theta_4(\tfrac{-t_1-t_2+t_3-t_4}{2})
    \theta_4(\tfrac{-t_1-t_2-t_3+t_4}{2})
    \theta_4(\tfrac{+t_1-t_2-t_3+t_4}{2})
    \theta_4(\tfrac{+t_1-t_2+t_3-t_4}{2})} \ .
\end{aligned}
\end{equation}
It is equivalent to
\begin{align}
  0 = -
  &\(\theta_3^{[\tfrac{1}{4}]}(4\tau,2t_1+2t_2)
    +\theta_3^{[-\tfrac{1}{4}]}(4\tau,2t_1+2t_2)\)
    \theta_1(t_3)\theta_1(t_4) \nn
  &\phantom{=}\times
    \theta_4\(\tfrac{+t_1-t_2-t_3-t_4}{2}\)
    \theta_4\(\tfrac{-t_1+t_2-t_3-t_4}{2}\)
    \theta_4\(\tfrac{+t_1-t_2+t_3-t_4}{2}\)
    \theta_4\(\tfrac{+t_1-t_2-t_3+t_4}{2}\)\nn +
  &\(\theta_3^{[\tfrac{1}{4}]}(4\tau,2t_1-2t_2)
    +\theta_3^{[-\tfrac{1}{4}]}(4\tau,2t_1-2t_2)\)
    \theta_1(t_3)\theta_1(t_4)\nn
  &\phantom{=}\times
    \theta_4\(\tfrac{-t_1-t_2-t_3-t_4}{2}\)
    \theta_4\(\tfrac{-t_1-t_2+t_3-t_4}{2}\)
    \theta_4\(\tfrac{-t_1-t_2-t_3+t_4}{2}\)
    \theta_4\(\tfrac{+t_1+t_2-t_3-t_4}{2}\)\nn +
  &\(\theta_3^{[\tfrac{1}{4}]}(4\tau,2t_3+2t_4)
    +\theta_3^{[-\tfrac{1}{4}]}(4\tau,2t_3+2t_4)\)
    \theta_1(t_1)\theta_1(t_2)\nn
  &\phantom{=}\times
    \theta_4\(\tfrac{-t_1-t_2+t_3-t_4}{2}\)
    \theta_4\(\tfrac{-t_1-t_2-t_3+t_4}{2}\)
    \theta_4\(\tfrac{+t_1-t_2+t_3-t_4}{2}\)
    \theta_4\(\tfrac{+t_1-t_2-t_3+t_4}{2}\)\nn -
  &\(\theta_3^{[\tfrac{1}{4}]}(4\tau,2t_3-2t_4)
    +\theta_3^{[-\tfrac{1}{4}]}(4\tau,2t_3-2t_4)\)
    \theta_1(t_1)\theta_1(t_2)\nn
  &\phantom{=}\times
    \theta_4\(\tfrac{-t_1-t_2-t_3-t_4}{2}\)
    \theta_4\(\tfrac{+t_1-t_2-t_3-t_4}{2}\)
    \theta_4\(\tfrac{-t_1+t_2-t_3-t_4}{2}\)
    \theta_4\(\tfrac{+t_1+t_2-t_3-t_4}{2}\)
    \ ,  \label{eq:vand0-D4-2}
\end{align}
which can be similarly proved.

Note that in order to write recursion relations in terms of
appropriate Jacobi forms, we need to absorb one of the five K\"ahler
moduli $t_1,t_2,t_3,t_4,t_c$ completely into $\tau$. It is canonical
to absorb $t_4$ associated to the affine node as one goes down from
the affine Lie algebra to the simple Lie algebra. Here we choose to
absorb $t_c$ so that the symmetry between $t_1,t_2,t_3,t_4$ still
survives.

\subsubsection{Modularity at generic base degree}
\label{sc:modularity-D4}

Following the example of the $SU(3)$ theory, we show here that both
sides of the recursion relation \eqref{eq:recur-2} for the $SO(8)$
theory are meromorphic Jacobi forms of the same weight and index
polynomial at any base degree $k$. When this is established, one can
multiply both sides with the common denominator and obtain an identity
of weak Jacobi forms and proceeds to prove it by comparing the first
few terms in the $Q_\tau$ expansin.

Consider a blowup equation with $\md r=(r_1,r_2,r_3,r_4,r_c,r_b)$
subject to the condition $r_1+r_2+r_3+r_4+2r_c = 0$. The perturbative
contribution is
\begin{align}
  D^{\rm pert,'}:=
  &{\rm exp}\bigg[f_0(\md n)(\epsilon_1+\epsilon_2)
    +\sum_{\ell=1}^{r+1}f_\ell(\md n)t_\ell
    + (k_1\epsilon_1+k_2\epsilon_2)R_b(\md n)\bigg] \nn=
  & Q_1^{f_1(\md n)}Q_2^{f_2(\md n)}Q_3^{f_3(\md n)}Q_4^{f_4(\md n)}Q_c^{f_c(\md n)}
    \(q_1q_2\)^{f_0(\md n)}\big(q_1^{k_1}q_2^{k_2}\big)^{R_b(\md n)} \ ,
\end{align}
where
\begin{align}
  f_\ell(\md n)=
  & 2\Big(n_\ell - \frac{r_\ell+r_b/2}{4}\Big)^2 \ ,\quad \ell = 1,2,3,4 \nn
    f_c(\md n)=
  & \Big(n_c - \frac{r_b}{4}\Big)^2 \ ,\nn
    f_0(\md n)=
  &-\frac{4}{3}\sum_{i=1}^5n_i^3 + n_c^2\sum_{\ell=1}^4 n_\ell 
    +\sum_{\ell=1}^4 \(r_\ell+\frac{r_b}{2}\)n_\ell^2
    -\frac{r_bn_c}{2}\sum_{\ell=1}^4 n_\ell
    +\frac{r_b+r_c}{2}n_c^2 \nn
  & +\frac{\sum_{i=1}^5 n_i}{3}
    -\frac{\sum_{\ell=1}^4r_\ell^2n_\ell}{4}
    -\frac{r_b\sum_{\ell=1}^4r_\ell n_\ell}{4}
    -\frac{r_br_cn_c}{4}\nn
  & +\frac{\sum_{\ell=1}^4 r_\ell^3}{48}
    +\frac{r_b\sum_{\ell=1}^4r_\ell^2}{32}
    +\frac{r_b^2\sum_{\ell=1}^4 r_\ell}{64}
    +\frac{r_b^2r_c}{32} +\frac{r_c}{6}-\frac{r_b}{4} \ ,\nn
    R_b(\md n) =
  &- 2n_c + \frac{r_b}{2} \ ,
\end{align}
with the notation $r_5 = r_c$.
In addition
\begin{align}
  f_b(\md n) =
  & \frac{\sum_{\ell=1}^4 r_\ell^2}{16}
    - \frac{\sum_{\ell=1}^5 r_i n_i }{2}
    +\sum_{i=1}^5 n_i^2 - n_c\sum_{\ell=1}^4 n_\ell \ .
\end{align}
We split $\md n$ to a representative $\hat{\md n}$, which we fix
uniquely by setting $n_4 = 0$, and $(m,m,m,m,2m)$, i.e.
\begin{equation}
  \md n = (n_1,n_2,n_3,0,n_c) + (m,m,m,m,2m) \ .
\end{equation}
Then separating $m$-dependent and -independent parts, $D^{\rm pert,'}$
can be written as
\begin{align}
  D^{\rm pert,'} 
  &=Q_\tau^{2\(m-\frac{r_4+r_b/2}{4}\)^2}
    \prod_{\ell=1}^3
    Q_\ell^{4(m-\frac{r_4+r_b/2}{4})(n_\ell+\frac{r_4-r_\ell}{4})}
    Q_c^{4(m-\frac{r_4+r_b/2}{4})(n_c+\frac{r_4}{2})}\nn
  &\times \(q_1q_2\)^{(m-\frac{r_4+r_b/2}{4})(2-4f_b(\hat{\md n}))}
    \big(q_1^{k_1}q_2^{k_2}\big)^{-4(m-\frac{r_4+r_b/2}{4})}
    \nn
  &\times
    \prod_{\ell=1}^3
    Q_\ell^{2(n_\ell+\frac{r_4-r_\ell}{4})^2}Q_c^{(n_c+\frac{r_4}{2})^2}
    (q_1q_2)^{f_0(\hat{\md n})+\frac{1}{4}(r_4+r_b/2)(2-4f_b(\hat{\md n}))}
    \big(q_1^{k_1}q_2^{k_2}\big)^{-2n_c-r_4}
    \ . \label{eq:mod-pert-D4}
\end{align}
The contribution of 1-loop partition function is
\begin{align}
  D^{\text{1-loop}} =
  &T_{-2n_1+n_c+\frac{r_1}{2}}(t_1)
    T_{-2n_2+n_c+\frac{r_2}{2}}(t_2)
    T_{-2n_3+n_c+\frac{r_3}{2}}(t_3)
    T_{n_1+n_2+n_3-2n_c+\frac{r_c}{2}}(t_c)\nn\times 
  &T_{-n_1+n_2+n_3-n_c+\frac{r_1+r_c}{2}}(t_1+t_c)
    T_{n_1-n_2+n_3-n_c+\frac{r_2+r_c}{2}}(t_2+t_c)
    T_{n_1+n_2-n_3-n_c+\frac{r_3+r_c}{2}}(t_3+t_c)\nn\times 
  &T_{-n_1-n_2+n_3+\frac{r_1+r_2+r_c}{2}}(t_1+t_2+t_c)
    T_{-n_1+n_2-n_3+\frac{r_1+r_3+r_c}{2}}(t_1+t_3+t_c)\nn\times
  &T_{n_1-n_2-n_3+\frac{r_2+r_3+r_c}{2}}(t_2+t_3+t_c)
    T_{-n_1-n_2-n_3+n_c+\frac{r_1+r_2+r_3+r_c}{2}}(t_1+t_2+t_3+t_c)\nn\times 
  & T_{-n_c++\frac{r_1+r_2+r_3+2r_c}{2}}(t_1+t_2+t_3+2t_c) \nn =
  &Q_\tau^{f_b(\hat{\md n})}
    \prod_{\ell=1}^3Q_\ell^{-2f_b(\hat{\md n})-2\(n_\ell+\frac{r_4-r_\ell}{4}\)^2}
    Q_c^{-4f_b(\hat{\md n})-\(n_c+\frac{r_4}{2}\)^2}\(q_1q_2\)^{d_R(\mf d_4)}\nn\times 
  &\breve{\theta}_{-2n_1+n_c+\frac{r_1}{2}}(t_1)
    \breve{\theta}_{-2n_2+n_c+\frac{r_2}{2}}(t_2)
    \breve{\theta}_{-2n_3+n_c+\frac{r_3}{2}}(t_3)
    \breve{\theta}_{n_1+n_2+n_3-2n_c+\frac{r_c}{2}}(t_c)\nn\times 
  &\breve{\theta}_{-n_1+n_2+n_3-n_c+\frac{r_1+r_c}{2}}(t_1+t_c)
    \breve{\theta}_{n_1-n_2+n_3-n_c+\frac{r_2+r_c}{2}}(t_2+t_c)
    \breve{\theta}_{n_1+n_2-n_3-n_c+\frac{r_3+r_c}{2}}(t_3+t_c)\nn\times 
  &\breve{\theta}_{-n_1-n_2+n_3+\frac{r_1+r_2+r_c}{2}}(t_1+t_2+t_c)
    \breve{\theta}_{-n_1+n_2-n_3+\frac{r_1+r_3+r_c}{2}}(t_1+t_3+t_c)\nn\times
  &\breve{\theta}_{n_1-n_2-n_3+\frac{r_2+r_3+r_c}{2}}(t_2+t_3+t_c)
    \breve{\theta}_{-n_1-n_2-n_3+n_c+\frac{r_1+r_2+r_3+r_c}{2}}(t_1+t_2+t_3+t_c)\nn\times 
  &\breve{\theta}_{-n_c+\frac{r_1+r_2+r_3+2r_c}{2}}(t_1+t_2+t_3+2t_c)
    \ ,\label{eq:mod-1loop-D4}
\end{align}
where
\begin{align}
  d_R(\mf d_4)
  &=\frac{4}{3}\sum_{\ell=1}^3n_\ell^3 -n_c^2\sum_{\ell=1}^3 n_\ell +
    \frac{4}{3}n_c^3
    +\sum_{\ell=1}^3 (r_4-r_\ell) n_\ell^2 -r_4
    n_c\sum_{\ell=1}^3n_\ell + \(r_4-\frac{r_c}{2}\)n_c^2\nn
  &+\sum_{\ell=1}^3\Big(-\frac{1}{3}+\frac{r_\ell^2}{4}-\frac{r_4r_\ell}{2}\Big)n_\ell
    +\Big(-\frac{1}{3}-\frac{r_4r_c}{2}\Big)n_c \nn
  &-\frac{r_c}{6}-\frac{r_4}{2}
    -\frac{\sum_{\ell=1}^3r_\ell^3-r_4^3}{16}-\frac{\sum_{\ell=1}^3
    r_\ell^2-r_4^2}{8}r_c-\frac{r_c^2\sum_{\ell=1}^3 r_\ell}{4}
    -\frac{r_c^3}{3}+\frac{r_1 r_2r_3}{8} \ .
\end{align}
Here $T_R(t)$ and $\breve{\theta}_R(t)$ are defined in
Appendix~\ref{sc:app1}; in particular, $\breve{\theta}_R(t)$ is a
Jacobi form of weight 0 and index given by \eqref{indbreve}.

Finally, $Z_k$ is related to the $k$-string elliptic genus $\IE_k$ for
the $SO(8)$ theory by
\begin{equation}
  Z_k (t_\ell,\epsilon_1,\epsilon_2)=
  \bigg(\frac{Q_\tau^{1/2}}{Q_1Q_2Q_3Q_c^2}\bigg)^{2k}
  \IE_k(t_\ell,\epsilon_1,\epsilon_2) \ ,
\end{equation}
and the latter is a meromorphic Jacobi form of weight 0 and index
\cite{DelZotto:2016pvm,DelZotto:2017mee}
\begin{equation}
  \text{Ind}(\IE_k)
  =-k(\epsilon_1^2+\epsilon_2^2) + k(2k-3)\epsilon_1\epsilon_2 - 2k(\md
  a,\md a)_{\mf d_4} \ ,
\end{equation}
with the invariant bilinear form normalized as\footnote{Recall
  $t_i = \vev{\boldsymbol{\alpha}_i,\md a}$.}
\begin{align}
  (\md a,\md a)_{\mf d_4} =
  &t_1^2 + t_2^2 + t_3^2 +
    t_1t_2+t_2t_3+t_3t_1 +2t_c(t_1+t_2+t_3)+2t_c^2 \ .
\end{align}
We also have
\begin{equation}\label{eq:mod-ell-D4}
  \begin{aligned}
    Z_{k_1}(t_\ell +\epsilon_1R_\ell,
    \epsilon_1,\epsilon_2-\epsilon_1)=
    &q_1^{k_1(2n_c+r_4)}
    \bigg(\frac{Q_\tau^{1/2}}{Q_1Q_2Q_3Q_c^2}\bigg)^{2k_1}
    \IE_{k_1}(t_\ell +\epsilon_1R_\ell,
    \epsilon_1,\epsilon_2-\epsilon_1)
    \ ,\\
    Z_{k_2}(;t_\ell +\epsilon_2R_\ell,
    \epsilon_1-\epsilon_2,\epsilon_2) =
    &q_2^{k_2(2n_c+r_4)}\bigg(\frac{Q_\tau^{1/2}}{Q_1Q_2Q_3Q_c^2}\bigg)^{2k_2}
    \IE_{k_2}(t_\ell +\epsilon_2R_\ell,
    \epsilon_1-\epsilon_2,\epsilon_2) \ ,
  \end{aligned}
\end{equation}

Combining
\eqref{eq:mod-pert-D4}, \eqref{eq:mod-1loop-D4} and \eqref{eq:mod-ell-D4}
all together, we get for the r.h.s.\ of the recursion relation
\eqref{eq:recur-2} 
\begin{align}
  \text{r.h.s.}=
  &\bigg(\frac{Q_\tau^{1/2}}{Q_1Q_2Q_3Q_c^2}\bigg)^{2k}
    \sum_{f_b(\hat{\md n})+k_1+k_2=k} (-1)^{n_1+n_2+n_3+n_5}\nn\times 
  &\theta_3^{\[-\frac{2r_4+r_b}{8}\]}\bigg(4\tau,4\sum_{\ell=1}^3
    \(n_\ell+\tfrac{r_4-r_\ell}{4}\) t_\ell +
    4\(n_c+\tfrac{r_4}{2}\)t_c
    + (2-4(k-k_2))\epsilon_1+(2-4(k-k_1))\epsilon_2\bigg)\nn\times 
  &\breve{\theta}_{-2n_1+n_c+\frac{r_1}{2}}(t_1)
    \breve{\theta}_{-2n_2+n_c+\frac{r_2}{2}}(t_2)
    \breve{\theta}_{-2n_3+n_c+\frac{r_3}{2}}(t_3)
    \breve{\theta}_{n_1+n_2+n_3-2n_c+\frac{r_c}{2}}(t_c)\nn\times 
  &\breve{\theta}_{-n_1+n_2+n_3-n_c+\frac{r_1+r_c}{2}}(t_1+t_c)
    \breve{\theta}_{n_1-n_2+n_3-n_c+\frac{r_2+r_c}{2}}(t_2+t_c)
    \breve{\theta}_{n_1+n_2-n_3-n_c+\frac{r_3+r_c}{2}}(t_3+t_c)\nn\times 
  &\breve{\theta}_{-n_1-n_2+n_3+\frac{r_1+r_2+r_c}{2}}(t_1+t_2+t_c)
    \breve{\theta}_{-n_1+n_2-n_3+\frac{r_1+r_3+r_c}{2}}(t_1+t_3+t_c)
    \breve{\theta}_{n_1-n_2-n_3+\frac{r_2+r_3+r_c}{2}}(t_2+t_3+t_c)\nn\times 
  &\breve{\theta}_{-n_1-n_2-n_3+n_c+\frac{r_1+r_2+r_3+r_c}{2}}(t_1+t_2+t_3+t_c)
    \breve{\theta}_{-n_c+\frac{r_1+r_2+r_3+2r_c}{2}}(t_1+t_2+t_3+2t_c) \nn\times 
  &\IE_{k_1}(t_\ell+\epsilon_1R_\ell,\epsilon_1,\epsilon_2-\epsilon_1)
    \IE_{k_2}(t_\ell+\epsilon_2R_\ell,\epsilon_1-\epsilon_2,\epsilon_2)
    \ . \label{eq:rhs-urec-D4}
\end{align}
Up to the common prefactor, each summand happens to be a meromorphic
Jacobi form\footnote{The same as footnote \ref{fn:vector}.}
for $\Gamma{4}$ of the same weight 1/2 and index polynomial
\begin{equation}\label{eq:ind-D4}
  \text{Ind}(\text{r.h.s.}) = \frac{-2k+1}{2}(\epsilon_1^2+\epsilon_2^2) +
  (k-1)(2k-1)\epsilon_1\epsilon_2 - 2k(\md a,\md a)_{\mf d_4} \ ,
\end{equation}
which is independent of the summation indices $n_i,k_1,k_2$ and thus
so is the total sum.

On the other hand, if the blowup equation is of
vanishing type, the l.h.s.\ of \eqref{eq:recur-2} vanish; if the
blowup equation is of unity type, we have $\md r=(0,0,0,0,0,r_b)$, and
after plugging in the expression of $\Lambda$, we find the l.h.s.\ of
\eqref{eq:recur-2} to be
\begin{equation}\label{eq:lhs-urec-D4}
  \text{l.h.s.}=\bigg(\frac{Q_\tau^{1/2}}{Q_1Q_2Q_3Q_c^2}\bigg)^{2k}
  \theta_3^{\[-\frac{r_b}{8}\]}(4\tau,2\epsilon_1+2\epsilon_2)
  \IE(t_\ell,\epsilon_1,\epsilon_2) \ ,
\end{equation}
which is a meromorphic Jacobi form of the same weight and the same
index \eqref{eq:ind-D4}. In both cases, after multiplied with a common
denominator, the recursion relations \eqref{eq:recur-2} become
identities of (weak) Weyl invariant Jacobi forms of identical weights
and indices. As in the case of $SU(3)$ theory, these identities can be
proved by checking that when the \emph{correct} $\md r$
(Tab.~\ref{tb:r-D4}) are plugged in the first few terms in $Q_\tau$
expansion are correct. For instance, when $k=0$ we find
\eqref{eq:rhs-urec-D4} indeed reduces to the computation of $\Lambda$
in the unity cases, and the identities \eqref{eq:vand0-D4-1},
\eqref{eq:vand0-D4-2} in the vanishing cases. When $k=1$, let us first
reparametrise the K\"ahler moduli by
\begin{equation}
  m_i = \vev{\md e_i,\avec}
\end{equation}
with the standard basis $\{\md e_i\}$ of $\IR^4$, in which the root
lattice of $SO(8)$ is embedded, so that the Weyl symmetry of $SO(8)$
is more transparent. The variables $m_i$ are related to $t_i$ by
\begin{equation}
  \begin{cases}
    &t_1= m_1 - m_2 \\
    &t_2 = m_3 - m_4 \\
    &t_3 = m_3 + m_4 \\
    &t_c = m_2 - m_3 \\
    &t_4 = \tau - m_1 -m_2
  \end{cases} \ .
\end{equation}
In the case of unity equations, $f_b(\hat{\md n})$ can only be 0 or
1. It has one set of solution $\hat{\md n} = (0,0,0,0,0)$ in the
former case, and
24 sets of solutions in the latter case, which are
\begin{gather}
  \hat{I}_b^{(1)} = \{(-1,0,0,0,0),\ (1,0,0,0,0),\ (0,0,0,0,1),\
  (1,0,0,0,1),\ (1,1,0,0,1),\nn
  \ (1,1,1,0,1),\ (1,1,1,1,1) 
  +  \text{permutations of first four entries}\} \ .
\end{gather}
Then the recursion relations
\eqref{eq:rhs-urec-D4},\eqref{eq:lhs-urec-D4} become
\begin{align}
  \theta_3^{[-\frac{r_b}{8}]}
  &(4\tau,-2\eq+2\et)\IE_1(\md v, \eq,\et-\eq)
    +\theta_3^{[-\frac{r_b}{8}]}(4\tau,2\eq-2\et)\IE_1(\md v,
    \eq-\et,\et)
    +I^{[-\frac{r_b}{8}]} \nn =\,
  &\theta_3^{[-\frac{r_b}{8}]}(4\tau,2\eq+2\et)\IE_1(\md v, \eq,\et)
    \ , \label{eq:unityd1-D4}
\end{align}
where
\begin{align}
  I_1^{[-\frac{r_b}{8}]}=
  -\sum_{i<j}
  \sum_{\displaystyle \substack{r=\pm 1\\s=\pm 1}}
  &\frac{\theta_3^{[-\frac{r_b}{8}]}\(4\tau,4(rm_i+sm_j)-2\eq-2\et\)\ \eta^4}
    {\theta_1(rm_i+sm_j)\theta_1(rm_i+sm_j-\eq)\theta_1(rm_i+sm_j-\et)
    \theta_1(rm_i+sm_j-\eq-\et)}\nn
  &\times \prod_{\displaystyle \substack{k\neq i\\ k\neq j}}
  \frac{\eta^4}{\theta_1(m_i\pm m_k)\theta_1(m_j\pm m_k)}\ .
\end{align}
Here we define $\theta_1(m_i\pm m_k)$ as
$\theta_1(m_i+m_k)\theta_1(m_i- m_k)$. One can readily verify by the
first terms in the $Q_\tau$ expansion and using the expression of
$\IE_1$ in (3.24) of \cite{Haghighat:2014vxa}\footnote{We need to
  multiply the expression in \cite{Haghighat:2014vxa} by a factor of
  two.} that the identity \label{eq:unityd1-D4} holds only when $r_b$
is even. The vanishing equations at $k=1$ and the cases of $k \geq 2$
can be checked in a similar manner.

\subsubsection{Recursion formula for elliptic genera}
\label{sc:recursion-D4}

In the case of 6d $SO(8)$ theory, there are four unity $\md r$ fields. The
corresponding $\Lambda_k^{(1)}, \Lambda_k^{(2)}$ are
\begin{itemize}
\item $\md r_1 = (0,0,0,0,0,-2)$
\begin{align}
  \Lambda_k^{(1)}(\md r_1)
  &= \sum_{n\in\IZ} 
    Q_\tau^{2n^2+n+\tfrac{1}{8}}(q_1^{-2k+1} q_2)^{2n+\frac{1}{2}}
    =\theta_3^{[\tfrac{1}{4}]}(4\tau,-(4k-2)\epsilon_1+2\epsilon_2)
  \ ,\\
  \Lambda_k^{(2)}(\md r_1)
  &= \sum_{n\in\IZ} 
    Q_\tau^{2n^2+n+\tfrac{1}{8}}(q_1 q_2^{-2k+1})^{2n+\frac{1}{2}}
    =\theta_3^{[\tfrac{1}{4}]}(4\tau,2\epsilon_1-(4k-2)\epsilon_2)
  \ .
\end{align}
\item $\md r_2 = (0,0,0,0,0,0)$
\begin{align}
  \Lambda_k^{(1)}(\md r_2)
  &= \sum_{n\in\IZ} 
    Q_\tau^{2n^2}(q_1^{-2k+1} q_2)^{2n}
    =\theta_3(4\tau,-(4k-2)\epsilon_1+2\epsilon_2)
  \ ,\\
  \Lambda_k^{(2)}(\md r_2)
  &= \sum_{n\in\IZ} 
    Q_\tau^{2n^2}(q_1 q_2^{-2k+1})^{2n}
    =\theta_3(4\tau,2\epsilon_1-(4k-2)\epsilon_2)
  \ .
\end{align}
\item $\md r_3 = (0,0,0,0,0,2)$
\begin{align}
  \Lambda_k^{(1)}(\md r_3)
  &= \sum_{n\in\IZ} 
    Q_\tau^{2n^2-n+\tfrac{1}{8}}(q_1^{-2k+1} q_2)^{2n-\frac{1}{2}}
    =\theta_3^{[-\tfrac{1}{4}]}(4\tau,-(4k-2)\epsilon_1+2\epsilon_2)
  \ ,\\
  \Lambda_k^{(2)}(\md r_3)
  &= \sum_{n\in\IZ} 
    Q_\tau^{2n^2-n+\tfrac{1}{8}}(q_1 q_2^{-2k+1})^{2n-\frac{1}{2}}
    =\theta_3^{[-\tfrac{1}{4}]}(4\tau,2\epsilon_1-(4k-2)\epsilon_2)
  \ .
\end{align}
\item $\md r_4 = (0,0,0,0,0,4)$
\begin{align}
  \Lambda_k^{(1)}(\md r_4)
  &= \sum_{n\in\IZ} 
    Q_\tau^{2n^2-2n+\tfrac{1}{2}}(q_1^{-2k+1} q_2)^{2n-1}
    =\theta_3^{[-\tfrac{1}{2}]}(4\tau,-(4k-2)\epsilon_1+2\epsilon_2)
  \ ,\\
  \Lambda_k^{(2)}(\md r_4)
  &= \sum_{n\in\IZ} 
    Q_\tau^{2n^2-2n+\tfrac{1}{2}}(q_1 q_2^{-2k+1})^{2n-1}
    =\theta_3^{[-\tfrac{1}{2}]}(4\tau,2\epsilon_1-(4k-2)\epsilon_2)
  \ .
\end{align}
\end{itemize}
In this case, the matrix $M_{\Lambda_k}$
constructed out of any three of the four unity $\md r$ fields at any
base degree have non-vanishing determinant and is thus of full
rank. For instance when $\md r_1,\md r_2,\md r_3$ are used, the
leading order contribution to $\det M_{\Lambda_k}$ is
\begin{align}
  \det M_{\Lambda_k} =
  (q_1q_2)^{-k}(q_1^k+q_2^k+q_1^{2k}+q_2^{2k}+q_1^{2k}q_2^k+q_1^kq_2^{2k})
  Q_\tau^{1/4} + \cO(Q_\tau^{5/4}) \ .
\end{align}
One can therefore invert $M_{\Lambda_k}$ to solve for $Z_k$ or $\IE_k$ from the
recursion relation.

For instance, using the identity \eqref{eq:unityd1-D4} with the
unity $\br$ fields $(0,0,0,0,0,2)$, $(0,0,0,0,0,0)$, and
$(0,0,0,0,0,-2)$, we obtain the following expression for the one
string elliptic genus:
\begin{equation}
  \IE_1=\frac{\Delta^{[-\frac{1}{4}]}I_1^{[-\frac{1}{4}]}
    +\Delta^{[0]}I_1^{[0]}+\Delta^{[\frac{1}{4}]}I_1^{[\frac{1}{4}]}}{\Delta}\
  ,
\end{equation}
where $\Delta$ is the determinant of the matrix
\begin{equation}
  \left(\begin{array}{ccc} 
          \theta_3^{[-{1\over 4}]}(4\tau,-2\eq+2\et)\quad
          &\quad \theta_3^{[-{1\over 4}]}(4\tau,2\eq-2\et)
          &\quad \theta_3^{[-{1\over 4}]}(4\tau,2\eq+2\et) \\
          \theta_3^{[0]}(4\tau,-2\eq+2\et) \quad
          &\quad \theta_3^{[0]}(4\tau,2\eq-2\et)
          &\quad \theta_3^{[0]}(4\tau,2\eq+2\et) \\
          \theta_3^{[{1\over 4}]}(4\tau,-2\eq+2\et)\quad
          &\quad \theta_3^{[{1\over 4}]}(4\tau,2\eq-2\et)
          &\quad \theta_3^{[{1\over 4}]}(4\tau,2\eq+2\et) \\
      \end{array}\right),
\end{equation}
and $\Delta^{[\alpha]}$ is the minor of
$\theta_3^{[\alpha]}(4\tau,2\eq+2\et)$. Here $\Delta$ only has poles
at $\eq=0$, $\et=0$ and $\eq-\et=0$. It is a Jacobi form of weight
$3/2$ and index
$(3\epsilon_1^2-2\epsilon_1\epsilon_2+3\epsilon_2^2)/2$. The leading
order is $\Qtau^{9/4}$. The expressions of the elliptic genera for
higher numbers of strings can be similarly written down, although they
are much more lengthy.

Before ending this subsection, let us mention an interesting
phenomenon. In the case of $SO(8)$, the four theta functions
\begin{equation}
  \theta_3^{[\tfrac{1}{4}]}\ ,\quad
  \theta_3^{[0]}\ ,\quad
  \theta_3^{[-\tfrac{1}{4}]} \ ,\quad
  \theta_3^{[-\tfrac{1}{2}]} \ ,
\end{equation}
enjoy a cyclic $\IZ_4$ symmetry, as they are invariant under the shift
of the upper characteristic $\alpha \to \alpha - 1/4$.
The matrix
\begin{equation}
  \begin{pmatrix}
    \theta_3^{[\tfrac{1}{4}]}(\tau,z_1) &
    \theta_3^{[\tfrac{1}{4}]}(\tau,z_2) &
    \theta_3^{[\tfrac{1}{4}]}(\tau,z_3)
    & \theta_3^{[\tfrac{1}{4}]}(\tau,z_4) \\
    \theta_3(\tau,z_1) & \theta_3(\tau,z_2) & \theta_3(\tau,z_3)
    & \theta_3(\tau,z_4) \\
    \theta_3^{[-\tfrac{1}{4}]}(\tau,z_1) &
    \theta_3^{[-\tfrac{1}{4}]}(\tau,z_2) &
    \theta_3^{[-\tfrac{1}{4}]}(\tau,z_3)
    & \theta_3^{[-\tfrac{1}{4}]}(\tau,z_4) \\
    \theta_3^{[-\tfrac{1}{2}]}(\tau,z_1) &
    \theta_3^{[-\tfrac{1}{2}]}(\tau,z_2) &
    \theta_3^{[-\tfrac{1}{2}]}(\tau,z_3)
    & \theta_3^{[-\tfrac{1}{2}]}(\tau,z_4) \\
  \end{pmatrix} \ ,\quad \text{with }\; z_1+z_2+z_3+z_4 = 0 \ ,
\end{equation}
which has $S_4$ symmetry amongst the elliptic parameters and thus is
an analogue of the matrix $M_{\Lambda_1}$ of $SU(3)$, has a vanishing
determinant.

\subsubsection{Solving refined BPS invariants}
\label{sc:BPS-D4}

Here we compute the BPS invariants from the equations extracted from
the exansion of the blowup equations with respect to all K\"ahler
moduli. We used the blowup equations associated with the following
$\md r$ fields
\begin{equation}
  (0,0,0,0,0,2),(-2,-2,0,0,2,2),(-2,2,0,0,0,0) + \text{permutations
    of}\; r_1,r_2,r_3,r_4\ .
\end{equation}
We managed to compute all the BPS invariants up to total degree of
$d_1+d_2+d_3+d_4+d_c+d_b = 5$ and list them in
Tab.~\ref{tb:D4-BPS}. They satisfy the obvious permutation symmetry of
$d_1,d_2,d_3,d_4$. Therefore we only list the non-vanishing invariants
with $d_1\leq d_2\leq d_3\leq d_4$ and omit those which can obtained
by permuting these degrees. The other curve classes that are not
listed in the table all have vanishing BPS invariants. These
invariants agree with the results in the literature.

Note that in addition there seems to be a curious symmetry
between $d_c,d_b$ if both are nonzero
\begin{equation}
  N^{d_1,d_2,d_3,d_4,d_c,d_b}_{j_L,j_R} =
  N^{d_1,d_2,d_3,d_4,d_b,d_c}_{j_L,j_R} \ ,\quad d_c,d_b \neq 0 \ .
\end{equation}
We trace this symmetry to fiber-base duality of D-type theories
\cite{Haghighat:2018dwe}. This can be understood as follows. Starting
with an affine $D_4$ base, the central fiber (i.e. the fiber over the
central node in $\widehat{D}_4$) is of affine $SU(2)$ type. Switching
the role of fiber and base, one obtains an affine $SU(2)$ base
consisting of a $(-4)$ and a $(-1)$-curve with the fiber over the $(-4)$
curve of affine $D_4$ type. Decompactifying the $-1$ one arrives
exactly at our present setup. Thus what we have effectively done, when
switching off all non-central nodes of $\widehat{D}_4$ is to swap the
central node with the base curve. This is a remnant of the actual
exact duality where the $(-1)$ curve has finite size.

We believe that BPS invariants of higher degree can be computed
similarly given more computing time and that the same symmetry will
hold.


\section{Reduction to blowup equations for 5d theories}
\label{sc:reduction}

We demonstrate here that the blowup equations for 6d gauge theories
could be dimensinally reduced to the blowup equations for 5d gauge
theories.

We have seen in sections~\ref{sc:input-A2}, \ref{sc:input-D4} that the
perturbative free energy of the 6d theory is reduced to that of the 5d
theory through the limit
\begin{equation}\label{eq:5dlimit}
  \limprime :\; \tau + c\, t_m \to - \infty \ ,\quad t_m \;\text{finite} \ ,
\end{equation}
with a model-dependent constant $c$.  When applied to $Z^{\rm inst}$
this limit is equivalent to keeping finite terms in the limit
$Q_\tau = \re^{\tau} \to 0$. The 6d one-loop partition function given
by \eqref{eq:1-loop} becomes
\begin{align}
  Z^{\text{1-loop}} \to
  & {\rm PE}\Bigg[-\frac{q_R+q_R^{-1}}
    {\big(q_1^{1/2}-q_1^{-1/2}\big)\big(q_2^{1/2}-q_2^{-1/2}\big)}
    \sum_{\alpha\in\Delta_+}\re^{-\alpha\cdot a}\Bigg] \nn =
  & \prod_{i,j=1}^\infty\prod_{\alvec\in\Delta_+}
    \Big(1-q_1^{i+1}q_2^{j+1}\re^{\aroot}\Big)
    \Big(1-q_1^iq_2^j \re^{\aroot}\Big) \ . \label{eq:1-loop-5d}
\end{align}
Using the formal identity
\begin{equation}
  \prod_{n=0}^\infty(1-x q^n) = \prod_{n=1}^\infty(1-x q^{-n})^{-1} \ ,
\end{equation}
The last line of \eqref{eq:1-loop-5d} could be written as 
\begin{equation}
  \prod_{i,j=0}^\infty\prod_{\alvec\in\Delta_+}
  \(1-q^i t^{j+1}\re^{\aroot}\)^{-1}
  \(1-q^{i+1}t^j \re^{\aroot}\)^{-1}
\end{equation}
with
\begin{equation}
  q = \re^{-\epsilon_1} \ ,t = \re^{\epsilon_2}
\end{equation}
which is precisely the 1-loop partition function of a 5d pure SYM
theory \cite{Hayashi:2017jze}.

Furthermore, the partition function component $Z_k$ of the 6d gauge
theory is identified with the $k$-string elliptic genus
by\footnote{This relation coincides with that in
  \cite{DelZotto:2017mee} when $n=3,4$. The discrepany for $n>4$ is
  due to that $Q_b$ defined in \cite{DelZotto:2017mee} is no longer
  the volume of a Mori cone generator.}
\begin{equation}
  Z_k (\md t,\epsilon_1,\epsilon_2)=
  \(\frac{\prod_{i=0}^{\lfloor\frac{n-3}{2}\rfloor}Q_{r+1-i}^{n-2-2i}}
  {Q_{\tau}^{\tfrac{n-2}{2}}}\)^{k}
  \IE_k(\tau,\md m,\epsilon_1,\epsilon_2) \ .
\end{equation}
When reduced to 5d gauge theory by sending $Q_\tau$ to 0, we recover
5d gauge instanton partiton functions $Z_k^{\rm 5d}$ by
\cite{DelZotto:2017mee,Kim:2016foj}\footnote{The discussion in
  section~5.4 of \cite{DelZotto:2017mee} is slightly inaccurate.}
\begin{equation}\label{eq:dict-6d5d}
  \left\{\begin{aligned}
      &\frac{\prod_{i=0}^{\lfloor\frac{n-3}{2}\rfloor}Q_{r+1-i}^{n-2-2i}}
      {Q_{\tau}^{n-2}} Q_b = Q_\tau^{-(n-2)/2}Q_{\rm ell}\to \mf q \\
      &Q_{\tau}^{k (n-2)/2} \IE_k \to Z_k^{\rm inst}
  \end{aligned}\right.
\end{equation}
such that
\begin{equation}
  1 + \sum_{k=1}^\infty Q_b^k \, Z_k \to
  1 + \sum_{k=1}^\infty \mf q^k\, Z_k^{\rm inst} \ .
\end{equation}
Here $\fq$ is the gauge instanton fugacity related to the 5d mass
paramter $t_m$ by $\fq = \re^{t_m}$. The first line in the dictionary
\eqref{eq:dict-6d5d} is then consistent with the observation
\eqref{eq:tmtb-A2}, \eqref{eq:tmtb-D4}.

These observations allow us to conclude that we can obtain the full
partition function of the 5d pure SYM theory from the partition
function of the 6d gauge theory throug the operation
\begin{equation}\label{eq:Zred}
  Z_{{\rm 5d}}(\md t,\epsilon_1,\epsilon_2)
  = \limprime Z^{\rm dec}(\tau,t_m,\epsilon_1,\epsilon_2)^{-1}
  Z_{{\rm 6d}}(\md T,\epsilon_1,\epsilon_2) \ ,
\end{equation}
where $Z^{\rm dec}(\tau,t_m\epsilon_1,\epsilon_2)$ is the component
that runs off in the limit \eqref{eq:5dlimit}, which is the
exponential of the extra piece in the perturbative free energy given
by \eqref{eq:F0red-A2},\eqref{eq:F1red-A2} combined for $SU(3)$ and
\eqref{eq:F0red-D4},\eqref{eq:F1red-D4} combined for $SO(8)$ theories
respectively. Here we use $\md T$ for K\"ahler moduli in 6d instead of
$\md t$ to stress that there is one more K\"ahler modulus in 6d
theories. We make it explicit that $Z^{\rm dec}$ only depends on
$\tau,t_m$ and no other K\"ahler moduli. Besides, we are free to twist
the partition functions in the sense of \eqref{eq:hF} and put hats
over $Z_{\rm 5d}$, $Z_{\rm 6d}$.

Then by multiplying both sides of the blowup equations for the 6d
theory with an inverse power of $Z^{\rm dec}$ and taking the limit
$\limprime$, we get
\begin{align}
  \limprime
  &\Lambda_{\rm 6d}(\tau,\epsilon_1,\epsilon_2)
  \hZ_{\rm 5d}(\md t,\epsilon_1,\epsilon_2) \nn=
  & \limprime\sum_{\md n\in \IZ^{r+1}} (-1)^{|\md n|}
    B^{\rm dec}(\tau,t_m,\epsilon_{1,2};\md n)
    \hZ_{\rm 5d}(\md t+\epsilon_1 \md R,\epsilon_1,\epsilon_2-\epsilon_1)
    \hZ_{\rm 5d}(\md t+\epsilon_2 \md R,\epsilon_1-\epsilon_2,\epsilon_2)
    \label{eq:blowup-red}
\end{align}
where we have defined
\begin{equation}
  B^{\rm dec}(\tau,t_m,\epsilon_{1,2};\md n)
  = \frac{Z^{\rm dec}(\tau,t_m+\epsilon_1 R_{t_m},\epsilon_1,\epsilon_2-\epsilon_1)
    Z^{\rm dec}(\tau,t_m+\epsilon_2 R_{t_m},\epsilon_1-\epsilon_2,\epsilon_2)}
  {Z^{\rm dec}(\tau,t_m,\epsilon_1,\epsilon_2)} \ .
\end{equation}
As we will illustrate by the examples of $SU(3)$ and $SO(8)$ theories,
if we expand $\Lambda_{\rm 6d}$ and $B^{\rm dec}$ in
terms of $Q_{\tau}$, and keep only the coefficients of the lowest
power on both sides of
\eqref{eq:blowup-red}, we get the blowup equations for the 5d gauge theory
\begin{equation}\label{eq:blowup-5d}
  \Lambda_{\rm 5d}(t_m,\epsilon_{1,2})
  \hZ_{\rm 5d}(\md t,\epsilon_{1,2})
  = \sum_{\md n \in \IZ^{r}} (-1)^{|\md n|}
  \hZ_{\rm 5d}(\md t+\epsilon_1 \md R,\epsilon_1,\epsilon_2-\epsilon_1)
  \hZ_{\rm 5d}(\md t+\epsilon_2 \md R,\epsilon_1-\epsilon_2,\epsilon_2)
  \ .
\end{equation}
Note that the dimension of $\md n$ is reduced by 1 in the 5d blowup
equations.

\subsection{$SU(3)$ model}

In the case of 6d $SU(3)$ model, we have concretely
\begin{equation}
  \limprime_{SU(3)}  =\lim_{\tau+t_m \to -\infty}  = \lim_{t_3+t_b
    \to -\infty}\ ,\quad t_m,t_b\; \text{finite}\ .
\end{equation}
We take the Nekrasove partition function to be the partition function
of the 5d gauge theory. $Z^{\rm dec}$ should include an extra piece
from \eqref{eq:F0diffNek-A2}, and it reads
\begin{equation}
  Z^{\rm dec}_{SU(3)}(\md t,\epsilon_1,\epsilon_2)
  = \exp\(-\frac{(\tau+t_m)^3-t_m^3}{18\epsilon_1\epsilon_2}
  - \frac{(\epsilon_1^2+\epsilon_2^2+3\epsilon_1\epsilon_2)(\tau+t_m)}
  {8\epsilon_1\epsilon_2}\)
\end{equation}
Then 
\begin{equation}
  B^{\rm dec}_{SU(3)} =
  {\rm
  exp}\[\tau\(\frac{3}{2}n_3^2-\frac{1}{2}n_3(r_3+r_4)+\frac{1}{24}(r_3+r_4)^2\)
    + \ldots\]
\end{equation}
where the remaining pieces in the ellipses are linear in $t_m$ and
$\epsilon_R$ and depend only on $n_3$ but not on $n_1,n_2$. Clearing,
depending on the value of $r_3+r_4$, there are only one or two
integral values of $n_3$ which minimize the power of $Q_\tau$. If we
only keep the minimal power of $Q_\tau$ in \eqref{eq:blowup-red},
although we still sum $n_1,n_2$ over all integers, we only sum $n_3$
over one or two values. This is the reason the dimension of the
summation index vector $\md n$ is reduced by 1 in the 5d blowup
equations. In the case where $n_3$ can take two values (this happens
when $r_3+r_4 = 6k+3$, $k\in\IZ$), one 6d blowup equation splits to
two 5d blowup equations.

Note that given the definition of $\md R$ in \eqref{eq:R}, when we sum
$\md n\in \IZ$ in the blowup equations, we are effectively summing
$2\md R$ over all the $\md r$ fields in the same equivalence
class. Therefore when dimensionally reducing blowup equations, we can
get the equivalence classes of $\md r$ fields for the 5d theory simply
by fixing the value of $n_3$, as we prescribed above, and deleting the
entry $r_3$ associated to $t_3$.

This procedure also gives us immediately a way to compute
$\Lambda_{\rm 5d}$ from $\Lambda_{\rm 6d}$, which are given in
section~\ref{sc:b0-A2}
\begin{equation}\label{eq:lambda5d-A2}
  \Lambda_{\rm 5d}^{(n_3^{\rm min})}(t_m)
  = \lim_{\tau+t_m\to -\infty} \Lambda_{\rm 6d}(\tau)
  B^{\rm dec}(\tau,t_m,\epsilon_{1,2};n_3^{\rm min})^{-1}\ ,
\end{equation}
where $n_3^{\rm min}$ is a value of $n_3$ that minimizes the power of
$Q_\tau$.

We list the 5d $\md r$ fields reduced from 6d $\md r$ fields as well
as the corresponding $\Lambda_{\rm 5d}$ in Tab.~\ref{tb:red-A2}. They
are consistent with a direct compute with 5d blowup equations. Note
that $\Lambda_{\rm 5d}$ for the 5d unity $\md r$ fields $(2,-2,-3)$,
$(2,-2,3)$, $(-2,2,-3)$, $(-2,2,3)$ cannot be derived by
\eqref{eq:lambda5d-A2} though, and they are instead
computed from 5d blowup equataions. They are also notably pairwise
identical as they should since they descend from the same 6d vanishing
blowup equations pairwise. Note that the last entry of 5d $\md r$
field is $r_m$ related to $r_b$ by
\begin{equation}
  r_m = r_b - r_1 - r_2 \ .
\end{equation}

\begin{table}
  \centering \renewcommand*{\arraystretch}{1.2}
  \begin{tabular}{*{3}{>{$}c<{$}}}\toprule
    (r_1,r_2,r_3,r_b) & (r_1,r_2,r_m) & \Lambda_{\rm 5d} \\\midrule
    \ucol{(0,0,0,-1)} & \ucol{(0,0,-1)} & q_R^{1/12}\\
    \ucol{(0,0,0,1)}  & \ucol{(0,0,1)} & q_R^{-1/12}\\
    \ucol{(0,0,0,3)}  & \ucol{(0,0,-3)},
                        \ucol{(0,0,3)}
                                      & q_R^{1/4}, q_R^{-1/4}\\
    \vcol{(-2,0,2,1)} & \ucol{(2,-2,-3)}, \ucol{(2,-2,3)}
                                      & q_R^{-3/4}Q_m^{1/3}, q_R^{-3/4}Q_m^{1/3}\\
    \vcol{(0,-2,2,1)} & \ucol{(-2,2,-3)}, \ucol{(-2,2,3)}
                                      & q_R^{3/4}Q_m^{1/3}, q_R^{-3/4}Q_m^{1/3}\\
    \vcol{(-2,2,0,1)} & \vcol{(-2,2,1)} & 0 \\
    \vcol{(2,-2,0,1)} & \vcol{(2,-2,1)} & 0 \\
    \vcol{(2,0,-2,1)} & \vcol{(-2,2,-1)} & 0 \\
    \vcol{(0,2,-2,1)} & \vcol{(2,-2,-1)} & 0 \\
    \bottomrule
  \end{tabular}

  \caption{Reduction of 6d $\md r$ fields to 5d $\md r$ fields and corresponding
  $\Lambda_{\rm 5d}$ for the $SU(3)$ model. Unity and vanishing
  $\md r$ fields are colored in blue and green respectively.}\label{tb:red-A2}
\end{table}

\subsection{$SO(8)$ model}

In the case of $SO(8)$ model, we have
\begin{equation}
  \limprime_{SO(8)} = \lim_{\tau+\tfrac{1}{2}t_m \to -\infty}
  = \lim_{t_4+\tfrac{1}{2}t_b \to -\infty}\ ,\quad
  t_m,t_b\;\text{finite} \ ,
\end{equation}
as well as
\begin{equation}
  Z^{\rm dec}_{SO(8)}(\md t,\epsilon_1,\epsilon_2)
  = \exp\(-\frac{(2\tau+t_m)^3 - t_m^3}{48\epsilon_1\epsilon_2}
  - \frac{(\epsilon_1^2+\epsilon_2^2+3\epsilon_1\epsilon_2)(2\tau+t_m)}
  {6\epsilon_1\epsilon_2}\) \ ,
\end{equation}
which leads to 
\begin{equation}
  B^{\rm dec}_{SO(8)} =
   {\rm
  exp}\[\tau\( 2n_4^2-\frac{2r_4+r_6}{2}n_4+\frac{4r_4^2+r_6^2+4r_4r_6}{32}\)
    + \ldots\] \ .
\end{equation}
The analysis is completely analogous as in the case of the $SU(3)$
theory. The splitting of 6d $\md r$ fields to pairs of 5d $\md r$
fields happens when $2r_4+r_6 = 8k+4$, $k\in\IZ$. We list the
resulting 5d $\md r$ fields and the corresponding $\Lambda_{\rm 5d}$
in Tab.~\ref{tb:red-D4}. Note that the last entry of 5d $\md r$ field
is $r_m$ related to $r_b$ by
\begin{equation}
  r_m = r_b - 2r_1 - 2r_2 - 2r_3 - 4r_c \ .
\end{equation}

\begin{table}
  \centering
  \renewcommand*{\arraystretch}{1.2}
  \begin{tabular}{*{3}{>{$}c<{$}}}\toprule
    (r_1,r_2,r_3,r_4,r_c,r_b) & (r_1,r_2,r_3,r_c,r_m) & \Lambda_{\rm 5d}\\\midrule
    \ucol{(0,0,0,0,0,-2)}  & \ucol{(0,0,0,0,-2)} & q_R^{1/3}\\
    \ucol{(0,0,0,0,0,0)}   & \ucol{(0,0,0,0,0)} & 1 \\
    \ucol{(0,0,0,0,0,2)}   & \ucol{(0,0,0,0,2)} & q_R^{-1/3}\\
    \ucol{(0,0,0,0,0,4)}   & \ucol{(0,0,0,0,4)},\ucol{(0,0,0,2,-4)}
                                                      & q_R^{-2/3},q_R^{2/3}\\
    \vcol{(-2,0,0,2,0,0)}  & \vcol{(-2,0,0,0,4)},\vcol{(-2,0,0,2,-4)}
                                                      & 0, 0\\
    \vcol{(0,-2,0,2,0,0)}  & \vcol{(0,-2,0,0,4)},\vcol{(0,-2,0,2,-4)} & 0,0\\
    \vcol{(0,0,-2,2,0,0)}  & \vcol{(0,0,-2,0,4)},\vcol{(0,0,-2,2,-4)} & 0,0\\
    \vcol{(-2,2,0,0,0,0)}  & \vcol{(-2,2,0,0,0)} & 0\\
    \vcol{(-2,0,2,0,0,0)}  & \vcol{(-2,0,2,0,0)} & 0\\
    \vcol{(0,-2,2,0,0,0)}  & \vcol{(0,-2,2,0,0)} & 0\\  
    \vcol{(-2,-2,0,0,2,2)} & \vcol{(-2,-2,0,2,2)} & 0\\
    \vcol{(-2,0,-2,0,2,2)} & \vcol{(-2,0,-2,2,2)} & 0\\
    \vcol{(0,-2,-2,0,2,2)} & \vcol{(0,-2,-2,2,2)} & 0\\    
    \vcol{(-2,0,0,-2,2,2)} & \vcol{(-2,0,0,2,-2)} & 0\\
    \vcol{(0,-2,0,-2,2,2)} & \vcol{(0,-2,0,2,-2)} & 0\\
    \vcol{(0,0,-2,-2,2,2)} & \vcol{(0,0,-2,2,-2)} & 0\\\bottomrule
  \end{tabular}
  \caption{Reduction of 6d $\md r$ fields to 5d $\md r$ fields and the
    corresponding $\Lambda_{\rm 5d}$ for the $SO(8)$ model. Unity and
    vanishing $\md r$ fields are colored in blue and green
    respectively.}\label{tb:red-D4}
\end{table}

\section{Conclusion and discussion}
\label{sc:conclusion}

In this paper, we consider the $n=3,4$ minimal 6d SCFTs in the tensor
branch. These theories are obtained by F-theory compactification on
non-compact elliptic Calabi-Yau threefolds. We demonstrate that the
elliptic genera of these theories, which encode the refined BPS
invariants of the underlying Calabi-Yau threefolds, satisfy the
generalized blowup equations. Furthermore, we illustrate that the
generalized blowup equations can be used to solve the elliptic genera
as well as the refined BPS invariants.

We emphasize here that the generalized blowup equations is an
extremely powerful tool for computing the BPS invariants of
non-compact Calabi-Yau threefolds. All the currently existing
techniques for computing BPS invariants in local geometry, being well
established and very powerful, have their limitations in terms of
accessible geometries. The topological vertex
\cite{Aganagic:2003db,Iqbal:2007ii,Iqbal:2012mt} is only applicable
for toric geometries or generalizations thereof. The holomorphic
anomaly equations \cite{Bershadsky:1993ta,Bershadsky:1993cx} and the
topological recursion
\cite{Eynard:2007kz,Eynard:2008we,Bouchard:2007ys} are useful only if
the mirror geometry is known, and in particular if the number of
compact divisors $g$\footnote{If the mirror geometry can be reduced to
  a curve, this number is equal to the genus of the curve.} in the
original geometry is low ($g \leq 2$). The modular bootstrap
\cite{Huang:2015sta,Gu:2017ccq,DelZotto:2016pvm,DelZotto:2017mee,Kim:2018gak}
only works if the Calabi-Yau is elliptic and is most efficient if
there is no singular elliptic fiber (see also \cite{Duan:2018sqe}). On
the other hand, the generalized blowup equations are more versatile
than any of these individual methods. They have been applied in toric
geometries \cite{Huang:2017mis}, elliptic geometries (which are
non-toric), and the cases where the number of compact divisors is
greater than two (this paper). Up to this moment, there does not seem
to be any restriction on the type of non-compact Calabi-Yau threefolds
for which the generalized blowup equations are applicable.

Nevertheless, we have to point out that why the generalized blowup
equations work still remains a mystery. The only case where the blowup
equations have a rigorous mathematical proof is when applied on the
$X_{N,m}$ geometries \cite{Nakajima:2009qjc}. A better mathematical or
physical understanding of the generalized blowup equations would be
extremely desirable. For example, what is the relation between blowup
equations and refined holomorphic anomaly equations? To anwser this
question requires a non-holomorphic version of blowup equations. And
what is the relation between blowup equations and refined topological
vertex? For 6d SCFTs, this may involves the recently proposed elliptic
topological vertex \cite{Foda:2018jwz}. Furthermore the moduli space
of the topological string theory usually contains both geometric and
non-geometric phases. In this paper we only work deep in the geometric
phase around the large volume limit. It is an interesting problem to
study the blowup equations in the other phases of the moduli space as
in \cite{Huang:2017mis}. Finally one could certainly push the
computation of the BPS invariants for an almost infinite range of
non-compact Calabi-Yau threefolds. The easiest targets and the most
similar to what are studied here are those for the remaining cases of
minimal 6d SCFTs, the results of which we will report in companion
papers in the near future.

\section*{Acknowledgement}

We would like to thank Giulio Bonelli, Min-xin Huang, Joonho Kim,
Albrecht Klemm, Amir-Kian Kashani-Poor, Alessandro Tanzini and
Rui-Dong Zhu for valuable discussions. JG is supported by the European
Research Council (Programme ``Ideas'' ERC-2012-AdG 320769
AdS-CFT-solvable) during the early stage of the project, and by the
Fonds National Suisse, subsidiary 200020-175539 (project ``Quantum
mechanics, geometry and strings''), during the late stage of the
project.

\appendix

\section{Useful identities}
\label{sc:app1}

We collect some identities which are useful in the main text of the
paper.

Using the triple product formula of $\theta_1$
\begin{equation}
  \theta_1(\tau,z)=
  \ri
  Q_\tau^{\frac{1}{12}}Q_z^{-\frac{1}{2}}\eta(\tau)\prod_{n=1}^{\infty}
  \(1-Q_zQ_\tau^{n-1}\)\(1-\frac{Q_\tau^n}{Q_z}\),
\end{equation}
we can simplify the following plethystic exponentials which often
appear in the evaluation of vector multiplet contributions to the
one-loop partition function
\begin{equation}
  \PE{\frac{Q}{1-Q_\tau}}=\prod_{n=0}^{\infty}\frac{1}{1-QQ_\tau^n},
\end{equation}
and
\begin{equation}\label{petheta}
  \PE{\(Q_z+{Q_\tau\over Q_z}\)\(\frac{1}{1-Q_\tau}\)}
  =\frac{\ri Q_\tau^{\frac{1}{12}}Q_z^{-\frac{1}{2}}\eta(\tau)}{\theta_1(\tau,z)}.
\end{equation}

In the following, we would like to present some elementary but useful
formulas when dealing with blowup equations. Denote
\begin{equation}
  f_{(j_L,j_R)}(q_1,q_2)=
  \frac{\chi_{j_L}(q_L)\chi_{j_R}(q_R)}{\Big(q_1^{1/2}-q_1^{-1/2}\Big)\Big(q_2^{1/2}-q_2^{-1/2}\Big)}
\end{equation}
which is the spin-related prefactor in the contribution to the one-loop
partition function of a multiplet with spin $(j_L,j_R)$ (see for
instance \eqref{eq:multiplets}). It satisfies the relations
\begin{gather}
  f_{(j_L,j_R)}(q_1^{-1},q_2^{-1})=f_{(j_L,j_R)}(q_1,q_2)=f_{(j_L,j_R)}(q_2,q_1)
  \ ,\\
  f_{(j_L,j_R)}(q_1^{-1},q_2)=f_{(j_L,j_R)}(q_1,q_2^{-1})=f_{(j_R,j_L)}(q_1,q_2)
  \ .
\end{gather}
In the blowup equation this prefactor contributes by
\begin{equation}\label{Blfunc}
  Bl_{(j_L,j_R,R)}(q_1,q_2) =
  f_{(j_L,j_R)}(q_1,q_2/q_1)q_1^R+f_{(j_L,j_R)}(q_1/q_2,q_2)q_2^R-f_{(j_L,j_R)}(q_1,q_2)\
  ,,
\end{equation}
where $R = \md R\cdot \md d \in \tfrac{1}{2}\IZ$ is the entry of $\md R$
associated to the K\"ahler modulus $\md Q^\md d$ multiplying this
prefactor. The checkerboard pattern \eqref{eq:checker} translates to
the condition
\begin{equation}
  2j_L+2j_R+1\equiv 2 R \ (\rm{mod}\, 2) \ .
\end{equation}
It has been argued from the $\eq,\et$ expansion of refined free energy
and blowup equations \cite{Huang:2017mis} that under this condition
the apparent denominator of $Bl_{(j_L,j_R,R)}(q_1,q_2)$ can always be
factored out so that
\begin{equation}\label{eq:B-finseries}
 Bl_{(j_L,j_R,R)}(q_1,q_2) = \text{finite series in} \; q_1,q_2 \ .
\end{equation}
We call \eqref{eq:B-finseries} \emph{fundamental identities}. Note
that since
\begin{equation}\label{minusR}
  Bl_{(j_L,j_R,-R)}(q_1,q_2)=Bl_{(j_L,j_R,R)}(q_1^{-1},q_2^{-1}) \ ,
\end{equation}
 we only need to consider the cases with  $R\geq 0$.

In the following, we present some frequently used instances of the fundamental
identities for small spins.
\begin{itemize}
\item For $(j_L,j_R)=(0,0)$, $R$ should be half integers. Then

  \begin{equation}
    Bl_{(0,0,R)}(q_1,q_2)=  -\sum_{\substack{m,n\geq 0\\m+n\leq R-3/2}}
    q_1^{m+1/2}q_2^{n+1/2}\ , \quad R \geq 1/2 \ .
  \end{equation}
\item For $(j_L,j_R)=(1/2,0)$, $R$ should be integers. Then
  \begin{equation}
    Bl_{(1/2,0,R)}(q_1,q_2)=\begin{cases}
        -\sum\limits_{\substack{m,n\geq 0\\1\leq m+n\leq R}}q_1^{m}q_2^{n}
        -\sum\limits_{\substack{m,n\geq 0\\m+n\leq R-3}}q_1^{m+1}q_2^{n+1},\  
        &R\geq 1 \ , \\
        -1 \ , \quad &R = 0 \ .
    \end{cases}
  \end{equation}
\item For $(j_L,j_R)=(0,1/2)$, $R$ should be integers. Then
  \begin{equation}\label{0half}
    Bl_{(0,1/2,R)}(q_1,q_2)=
    -\sum_{\substack{m,n\geq 0\\m+n\leq R-1}}q_1^mq_2^n
    -\sum_{\substack{m,n\geq 0\\m+n\leq R-2}}q_1^{m+1}q_2^{n+1}\ ,\quad
    R\geq 0 \ .
\end{equation}  
\end{itemize}

As we have seen in the main text, the contribution of vector multiplets
can always be factorized as products of
\begin{equation}\label{brevedef}
  T_R(z)=
  \PE{-\(Bl_{(0,1/2,R)}(q_1,q_2)Q_z+Bl_{(0,1/2,-R)}(q_1,q_2){Q_\tau\over Q_z}\)\({1\over {1-Q_\tau}}\)}.
\end{equation}
Using (\ref{0half}) and (\ref{petheta}) and assuming $R\geq 0$, it can
be written as
\begin{align}
  T_R(z)=
  &\prod_{\substack{m,n\geq 0\\m+n\leq R-1}} \frac{\ri
  Q_\tau^{1/12}\eta\(Q_zq_1^mq_2^n\)^{-1/2}}
  {\theta_1(z+m\eq+n\et)}\prod_{\substack{m,n\geq 0\\m+n\leq R-2}}
  \frac{\ri Q_\tau^{1/12}\eta\(Q_zq_1^{m+1}q_2^{n+1}\)^{-1/2}}
  {\theta_1(z+(m+1)\eq+(n+1)\et)}\nn
  =\,
  &\(\ri Q_\tau^{1/12}
    Q_z^{-1/2}\)^{R^2}(q_1q_2)^{-\frac{(R-1)R(R+1)}{6}}
    \breve{\theta}_R(z) \ ,
\end{align}
where
\begin{equation}
  \breve{\theta}_R(z) =\prod_{\substack{m,n\geq 0\\m+n\leq R-1}}
  \frac{\eta}{\theta_1(z+m\eq+n\et)}
  \prod_{\substack{m,n\geq 0\\m+n\leq R-2}}
  \frac{\eta}{\theta_1(z+(m+1)\eq+(n+1)\et)} \ .
\end{equation}
In the case of $R<0$ we can use the above expression for $-R$ with
$\epsilon_{1,2}$ replaced by $-\epsilon_{1,2}$ or equivalently
$q_{1,2}$ replaced by $1/q_{1,2}$. In both cases,
$\breve{\theta}_R(z)$ is a multivariate Jacobi form of weight zero and
index quadratic form
\begin{equation}\label{indbreve}
  \mathrm{Ind}_{\breve{\theta}}^R(z)=
  -{R^2z^2\over 2}
  -\frac{(R-1)R(R+1)}{3}z(\eq+\et)
  -\frac{(R-1)R^2(R+1)}{12}(\epsilon_1^2+\epsilon_1\epsilon_2+\epsilon_2^2)
  \ .
\end{equation}

\pagebreak

\section{Refined BPS invariants}
\begin{longtable} {c p{11.5cm}}
   \caption{BPS invariants of 6d $n=3$ minimal model} \label{tb:A2-BPS}\\
  \toprule
  $\mathbf{d}=(d_1,d_2,d_3,d_b)$
  &$\oplus N^{\mathbf{d}}_{j_L,j_R}(j_L,j_R)$ \\
  \toprule\endfirsthead
  \toprule
  $\mathbf{d}=(d_1,d_2,d_3,d_b)$
  &$\oplus N^{\mathbf{d}}_{j_L,j_R}(j_L,j_R)$ \\
  \toprule
  \endhead
  \midrule\multicolumn{2}{r}{\emph{Continued on the next page}}\\\endfoot\endlastfoot
  (0, 0, 1, 0) & $(0,1/2)$ \\ \hline 
  (0, 1, 1, 0) & $(0,1/2)$ \\ \hline 
  (1, 1, 2, 0) & $(0,1/2)$ \\ \hline 
  (1, 2, 2, 0) & $(0,1/2)$ \\ \hline 
  (2, 2, 3, 0) & $(0,1/2)$ \\ \hline 
  (0, 0, 0, 1) & $(0,0)$ \\ \hline 
  (0, 0, 1, 1) & $(0,1)$ \\ \hline 
  (0, 1, 1, 1) & $ (0,0)\oplus(0,1)$ \\ \hline 
  (0, 0, 2, 1) & $(0,2)$ \\ \hline 
  (1, 1, 1, 1) & $3 (0,0)\oplus3(0,1)\oplus(1/2,1/2)$ \\ \hline 
  (0, 1, 2, 1) & $ (0,1)\oplus(0,2)$ \\ \hline 
  (0, 0, 3, 1) & $(0,3)$ \\ \hline 
  (1, 1, 2, 1) & $2 (0,0)\oplus4(0,1)\oplus2(0,2)\oplus(1/2,1/2)\oplus(1/2,3/2)$ \\ \hline 
  (0, 2, 2, 1) & $ (0,0)\oplus(0,1)\oplus(0,2)$ \\ \hline 
  (0, 1, 3, 1) & $ (0,2)\oplus(0,3)$ \\ \hline 
  (0, 0, 4, 1) & $(0,4)$ \\ \hline 
  (1, 2, 2, 1) & $4 (0,0)\oplus7(0,1)\oplus3(0,2)\oplus2(1/2,1/2)\oplus(1/2,3/2)$ \\ 
  \hline 
  (1, 1, 3, 1) & $2 (0,1)\oplus4(0,2)\oplus2(0,3)\oplus(1/2,3/2)\oplus(1/2,5/2)$ \\ \hline 
  (0, 2, 3, 1) & $ (0,1)\oplus(0,2)\oplus(0,3)$ \\ \hline 
  (0, 1, 4, 1) & $ (0,3)\oplus(0,4)$ \\ \hline 
  (0, 0, 5, 1) & $(0,5)$ \\ \hline 
  (2, 2, 2, 1) & $13(0,0)\oplus15(0,1)\oplus6(0,2)\oplus7(1/2,1/2)\oplus3(1/2,3/2)\oplus(1,1)$ \\ \hline 
  (1, 2, 3, 1) & $2 (0,0)\oplus6(0,1)\oplus6(0,2)\oplus2(0,3)\oplus(1/2,1/2)\oplus2(1/2,3/2)\oplus(1/2,5/2)$ \\ \hline 
  (1, 1, 4, 1) & $2 (0,2)\oplus4(0,3)\oplus2(0,4)\oplus(1/2,5/2)\oplus(1/2,7/2)$ \\ \hline 
  (0, 3, 3, 1) & $ (0,0)\oplus(0,1)\oplus(0,2)\oplus(0,3)$ \\ \hline 
  (0, 2, 4, 1) & $ (0,2)\oplus(0,3)\oplus(0,4)$ \\ \hline 
  (0, 1, 5, 1) & $ (0,4)\oplus(0,5)$ \\ \hline 
  (0, 0, 6, 1) & $(0,6)$ \\ \hline 
  (0, 0, 2, 2) & $(0,5/2)$ \\ \hline 
  (1, 1, 1, 2) & $2 (0,1/2)\oplus(0,3/2)$ \\ \hline 
  (0, 1, 2, 2) & $ (0,3/2)\oplus(0,5/2)$ \\ \hline 
  (0, 0, 3, 2) & $ (0,5/2)\oplus(0,7/2)\oplus(1/2,4)$ \\ \hline 
  (1, 1, 2, 2) & $3 (0,1/2)\oplus5(0,3/2)\oplus3(0,5/2)\oplus(1/2,1)\oplus(1/2,2)$ \\ 
  \hline 
  (0, 2, 2, 2) & $2 (0,1/2)\oplus2(0,3/2)\oplus2(0,5/2)\oplus(0,7/2)$ 
  \\ \hline 
  (0, 1, 3, 2) & $ (0,3/2)\oplus3(0,5/2)\oplus2(0,7/2)\oplus(1/2,3)\oplus(1/2,4)$ \\ 
  \hline 
  (0, 0, 4, 2) & $ (0,5/2)\oplus(0,7/2)\oplus2(0,9/2)\oplus(1/2,4)\oplus(1/2,5)\oplus(1,11/2)$ \\ 
  (1, 2, 2, 2) & $12 (0,1/2)\oplus14(0,3/2)\oplus8(0,5/2)\oplus2(0,7/2)\oplus2(1/2,0)\oplus4(1/2,1)\oplus3(1/2,2)\oplus(1/2,3)$ \\ \hline 
  (1, 1, 3, 2) & $2 (0,1/2)\oplus9(0,3/2)\oplus13(0,5/2)\oplus6(0,7/2)\oplus(1/2,1)\oplus4(1/2,2)\oplus5(1/2,3)\oplus2(1/2,4)\oplus(1,5/2)\oplus(1,7/2)$ 
  \\ \hline 
  (0, 2, 3, 2) & $2 (0,1/2)\oplus4(0,3/2)\oplus5(0,5/2)\oplus3(0,7/2)\oplus(0,9/2)\oplus(1/2,2)\oplus(1/2,3)\oplus(1/2,4)$ \\ \hline 
  (0, 1, 4, 2) & $ (0,3/2)\oplus3(0,5/2)\oplus5(0,7/2)\oplus3(0,9/2)\oplus(1/2,3)\oplus3(1/2,4)\oplus2(1/2,5)\oplus(1,9/2)\oplus(1,11/2)$ \\ \hline 
  (0, 0, 5, 2) & $ (0,5/2)\oplus(0,7/2)\oplus2(0,9/2)\oplus2(0,11/2)\oplus(1/2,4)\oplus( 1/2,5)\oplus2(1/2,6)\oplus(1,11/2)\oplus(1,13/2)\oplus(3/2,7)$ \\ 
  \hline 
  (0, 0, 3, 3) & $ (0,3)\oplus(1/2,9/2)$ \\ \hline 
  (1, 1, 2, 3) & $ (0,1)\oplus2(0,2)\oplus(0,3)$ \\ \hline 
  (0, 2, 2, 3) & $ (0,0)\oplus(0,1)\oplus(0,2)\oplus(0,3)\oplus(0,4)$ 
  \\ \hline 
  (0, 1, 3, 3) & $ (0,2)\oplus2(0,3)\oplus(0,4)\oplus(1/2,7/2)\oplus(1/2,9/2)$ \\ \hline 
  (0, 0, 4, 3) & $ (0,2)\oplus(0,3)\oplus2(0,4)\oplus(0,5)\oplus(0,6)\oplus(1/2,7/2)\oplus2(1/2,9/2)\oplus2(1/2,11/2)\oplus(1,5)\oplus(1,6)\oplus(3/2,13/2)$
  \\ \bottomrule
\end{longtable} 

\begin{longtable}{c p{11.5cm}}
  \caption{BPS invariants of 6d $n=4$ minimal model}\label{tb:D4-BPS}\\
  \toprule
  $\mathbf{d}=(d_1,d_2,d_3,d_4,d_c,d_b)$
  &$\oplus N^{\mathbf{d}}_{j_L,j_R}(j_L,j_R)$ \\
  \toprule\endhead
  \midrule \multicolumn{2}{r}{\emph{Continued on the next page}}\\
  \endfoot
  \endlastfoot
  (0, 0, 0, 1, 0, 0) & $(0,1/2)$ \\ \hline 
  (0, 0, 0, 0, 1, 0) & $(0,1/2)$ \\ \hline 
  (0, 0, 0, 1, 1, 0) & $(0,1/2)$ \\ \hline 
  (0, 0, 1, 1, 1, 0) & $(0,1/2)$ \\ \hline 
  (0, 1, 1, 1, 1, 0) & $(0,1/2)$ \\ \hline 
  (1, 1, 1, 1, 1, 0) & $(0,1/2)$ \\ \hline 
  (0, 1, 1, 1, 2, 0) & $(0,1/2)$ \\ \hline 
  (0, 0, 0, 0, 0, 1) & $(0,1/2)$ \\ \hline 
  (0, 0, 0, 1, 0, 1) & $(0,1/2)$ \\ \hline 
  (0, 0, 0, 2, 0, 1) & $(0,3/2)$ \\ \hline 
  (0, 0, 0, 3, 0, 1) & $(0,5/2)$ \\ \hline 
  (0, 0, 0, 4, 0, 1) & $(0,7/2)$ \\ \hline 
  (0, 0, 0, 0, 1, 1) & $(0,3/2)$ \\ \hline 
  (0, 0, 0, 1, 1, 1) & $ (0,1/2)\oplus(0,3/2)$ \\ \hline 
  (0, 0, 1, 1, 1, 1) & $2 (0,1/2)\oplus(0,3/2)$ \\ \hline 
  (0, 0, 0, 2, 1, 1) & $ (0,1/2)\oplus(0,3/2)$ \\ \hline 
  (0, 1, 1, 1, 1, 1) & $3 (0,1/2)\oplus(0,3/2)$ \\ \hline 
  (0, 0, 1, 2, 1, 1) & $ (0,1/2)\oplus(0,3/2)$ \\ \hline 
  (0, 0, 0, 3, 1, 1) & $ (0,3/2)\oplus(0,5/2)$ \\ \hline 
  (0, 0, 0, 0, 2, 1) & $(0,5/2)$ \\ 
  (0, 0, 0, 1, 2, 1) & $ (0,3/2)\oplus(0,5/2)$ \\ \hline 
  (0, 0, 1, 1, 2, 1) & $ (0,1/2)\oplus2(0,3/2)\oplus(0,5/2)$ \\ \hline 
  (0, 0, 0, 2, 2, 1) & $ (0,1/2)\oplus(0,3/2)\oplus(0,5/2)$ \\ \hline 
  (0, 0, 0, 0, 3, 1) & $(0,7/2)$ \\ \hline 
  (0, 0, 0, 1, 3, 1) & $ (0,5/2)\oplus(0,7/2)$ \\ \hline 
  (0, 0, 0, 0, 4, 1) & $(0,9/2)$ \\ \hline 
  (0, 0, 0, 3, 0, 2) & $(0,5/2)$ \\ \hline 
  (0, 0, 0, 0, 1, 2) & $(0,5/2)$ \\ \hline 
  (0, 0, 0, 1, 1, 2) & $ (0,3/2)\oplus(0,5/2)$ \\ \hline 
  (0, 0, 1, 1, 1, 2) & $ (0,1/2)\oplus2(0,3/2)\oplus(0,5/2)$ \\ \hline 
  (0, 0, 0, 2, 1, 2) & $ (0,1/2)\oplus(0,3/2)\oplus(0,5/2)$ \\ \hline 
  (0, 0, 0, 0, 2, 2) & $ (0,5/2)\oplus(0,7/2)\oplus(1/2,4)$ \\ \hline 
  (0, 0, 0, 1, 2, 2) & $ (0,3/2)\oplus3(0,5/2)\oplus2(0,7/2)\oplus(1/2,3)\oplus(1/2,4)$ \\ \hline 
  (0, 0, 0, 0, 3, 2) & $ (0,5/2)\oplus(0,7/2)\oplus2(0,9/2)\oplus(1/2,4)\oplus(1/2,5)\oplus(1,11/2)$ \\ \hline 
  (0, 0, 0, 0, 1, 3) & $(0,7/2)$ \\ \hline 
  (0, 0, 0, 1, 1, 3) & $ (0,5/2)\oplus(0,7/2)$ \\ \hline 
  (0, 0, 0, 0, 2, 3) & $ (0,5/2)\oplus(0,7/2)\oplus2(0,9/2)\oplus(1/2,4)\oplus(1/2,5)\oplus(1,11/2)$ \\ \hline 
  (0, 0, 0, 0, 1, 4) & $(0,9/2)$ \\ \bottomrule 
\end{longtable}

\printindex

\bibliographystyle{JHEP}
\bibliography{blowupscft}

\end{document}